\documentclass[useAMS,usenatbib]{mn2e}

\usepackage{amssymb}
\usepackage{times}
\usepackage{graphicx}
\usepackage{epstopdf}
\usepackage{subfigure}

\newcommand{\beq}{\begin{equation}}
\newcommand{\eeq}{\end{equation}}

\newcommand{\bfd}{\mathbf{d}}
\newcommand{\bfr}{\mathbf{r}}
\newcommand{\bfP}{\mathbf{P}}
\newcommand{\bfV}{\mathbf{V}}

\def\gs{\mathrel{\lower0.6ex\hbox{$\buildrel {\textstyle >}\over{\scriptstyle \sim}$}}}
\def\ls{\mathrel{\lower0.6ex\hbox{$\buildrel {\textstyle <}\over{\scriptstyle \sim}$}}}
\newcommand{\simgt}{\lower.5ex\hbox{$\; \buildrel > \over \sim \;$}}
\newcommand{\simlt}{\lower.5ex\hbox{$\; \buildrel < \over \sim \;$}}

\newcommand{\aap}{A\&A}
\newcommand{\apj}{ApJ}
\newcommand{\apjl}{ApJ}
\newcommand{\apjs}{ApJS}
\newcommand{\aj}{AJ}

\newcommand{\mnras}{MNRAS}

\begin{document}

\title[Multi-probe of A1689]{Mass, shape and thermal properties of A1689 by a multi-wavelength X-ray, lensing and Sunyaev-Zel'dovich analysis}
\author[M. Sereno et al.]{
Mauro Sereno$^{1,2}$\thanks{E-mail: mauro.sereno@polito.it (MS)}, Stefano Ettori$^{3,4}$, Keiichi Umetsu$^{5}$ and Alessandro Baldi$^{6}$
\\
$^1$Dipartimento di Scienza Applicata e Tecnologia, Politecnico di Torino, corso Duca degli Abruzzi 24, I-10129 Torino, Italia\\
$^2$INFN, Sezione di Torino, via Pietro Giuria 1, I-10125, Torino, Italia\\
$^3$INAF, Osservatorio Astronomico di Bologna, via Ranzani 1, I-40127 Bologna, Italia\\
$^4$INFN, Sezione di Bologna, viale Berti Pichat 6/2, I-40127 Bologna, Italia\\
$^5$Institute of Astronomy and Astrophysics, Academia Sinica, P. O. Box 23-141, Taipei 10617, Taiwan\\
$^6$Dipartimento di Astronomia, Universit\`a di Bologna, via Ranzani 1, IÐ40127, Bologna, Italia
}


\maketitle

\begin{abstract}
Knowledge of mass and concentration of galaxy clusters is crucial to understand their formation and evolution. Unbiased estimates require the understanding of the shape and orientation of the halo as well as its equilibrium status. We propose a novel method to determine the intrinsic properties of galaxy clusters from a multi-wavelength data set spanning from X-ray spectroscopic and photometric data to gravitational lensing to the Sunyaev-Zel'dovich effect (SZe). The method relies on two quite non informative geometrical assumptions: the distributions of total matter or gas are approximately ellipsoidal and co-aligned; they have different, constant axial ratios but share the same degree of triaxiality.  Weak and strong lensing probe the features of the total mass distribution in the plane of the sky. X-ray data measure size and orientation of the gas in the plane of the sky. Comparison with the SZ amplitude fixes the elongation of the gas along the line of sight. These constraints are deprojected thanks to Bayesian inference. The mass distribution is described as a Navarro-Frenk-White halo with arbitrary orientation, gas density and temperature are modelled with parametric profiles. We applied the method to Abell 1689. Independently of the priors, the cluster is massive, $M_{200} = (1.3 \pm 0.2)\times 10^{15}M_\odot$, and over-concentrated, $c_{200} =8\pm1$, but still consistent with theoretical predictions. The total matter is triaxial (minor to major axis ratio $\sim 0.5 \pm 0.1$ exploiting priors from $N$-body simulations) with the major axis nearly orientated along the line of sight. The gas is rounder (minor to major axis ratio $\sim 0.6 \pm 0.1$ ) and deviates from hydrostatic equilibrium. The contribution of non-thermal pressure is $\sim$20--50 per cent in inner regions, $\ls 300~\mathrm{kpc}$, and $\sim 25\pm 5$ per cent at $\sim 1.5~\mathrm{Mpc}$. This picture of A1689 was obtained with a small number of assumptions and in a single framework suitable to application to a large variety of clusters.
\end{abstract}


\section{Introduction}

Clusters of galaxies, the most recent bound structures to form in the Universe, are excellent laboratories for precision astronomy \citep{voi05}. Their use in cosmological tests relies on accurate measurements of their mass and concentration \citep{men+al10,ras+al12}. Assessing the equilibrium status is also crucial in determining evolution and mechanisms of interaction of baryons and dark matter \citep{le+su03,kaz+al04}. An unbiased look at the cluster properties must take into account their shape and orientation too \citep{ogu+al05}. The intrinsic form shows how material aggregates from large-scale perturbations \citep{wes94,ji+su02}. On the other hand, estimations of mass, inner matter density slope and concentration may be biased if derived under the assumption of spherical symmetry \citep{gav05,men+al10,ras+al12}.

Assessing the intrinsic shape and orientation is also critical when comparing observations with theoretical predictions. The slope of the concentration-mass $c(M)$ relation of galaxy clusters is consistent with $N$-body simulations, though the normalization factor is higher \citep{co+na07,ett+al10}. Galaxy clusters selected according to their gravitational lensing strength or X-ray flux may form biased samples \citep{men+al11}. Triaxial halos are more efficient lenses than their more spherical counterparts \citep{og+bl09} with the strongest lenses in the Universe expected to be a highly biased population preferentially orientated along the line of sight. In fact, the over-concentration problem is less prominent in analyses accounting for triaxiality \citep{ogu+al05,se+zi12}.

We can test the intrinsic three dimensional form of a population of astronomical objects with statistical approaches  \citep{hub26,noe79,bin80,bi+de81,fa+vi91,det+al95,moh+al95,bas+al00,coo00,th+ch01,al+ry02,ryd96,pli+al04,paz+al06,kaw10}. Clusters of galaxy can be observed with very heterogeneous data-sets at very different wave-lengths from X-ray surface brightness and spectral observations of the intra-cluster medium (ICM), to gravitational lensing (GL) observations of the total mass distribution to the Sunyaev-Zel'dovich effect (SZe) in the radio-band. This enables us to tackle the structure of a cluster with a multi-probe approach \citep{zar+al98,reb00,dor+al01,pu+ba06}. Only a few works have tried to infer shape or orientation of single objects. Combined use of X-ray and SZe data enables to constrain the shape of the ICM without any assumption regarding equilibrium. \citet{def+al05} and \citet{ser+al06} first studied a sample of 25 clusters finding signs of a quite general triaxial morphology. \citet{ma+ch11} constrained the minimum line-of-sight extent of the hot plasma of the Bullet cluster with a model-independent technique. 

Lensing observations offer an alternative way to study triaxiality. Projected mass distributions from either weak or strong lensing can be deprojected exploiting some a priori assumptions on the intrinsic shapes \citep{ogu+al05,cor+al09,mor+al11,ser+al10,ser+al10b,se+um11}.

Abell 1689 (A1689) is a very luminous cluster at redshift $z=0.183$ \citep{bro+al05,lim+al07}. The mass distribution in the inner $\ls 300~\mathrm{kpc}$ regions of the cluster has been accurately determined by strong lensing analyses that favoured a quite concentrated mass distribution  \citep{bro+al05,hal+al06,lim+al07,coe+al10}. On the larger virial scale, different weak lensing analyses suggest somewhat different degrees of concentration \citep{ume+bro08,ume+al09,cor+al09,lim+al07}. A further puzzle is the conflict between X-ray and lensing analyses, with lensing masses exceeding estimates derived under the hypothesis of hydrostatic equilibrium by 30-40 per cent in the inner regions \citep{pen+al09}. 

A1689 has been object of a number of triaxial analyses \citep{ogu+al05,cor+al09}. \citet{se+um11} analyzed weak and strong lensing data and found evidence for a mildly triaxial lens (minor to major axis ratio $\sim 0.5 \pm 0.2$) with the major axis nearly aligned with the line of sight. The halo was over-concentrated but still consistent with theoretical predictions. \citet{pen+al09} recognized that a prolate configuration, aligned with the line of sight and with an axis ratio of $\sim 0.6$, could solve the central mass discrepancy between lensing and X-ray mass estimates. \citet{mor+al11} combined lensing and X-ray data assuming the cluster to be aligned with the line of sight and fixing the relation between gas and matter shape. They found an axial ratio for the matter distribution of $\simeq 0.5$. \citet{ser+al12} combined X-ray and SZe data to model the gas shape and orientation by using Bayesian inference. Their analysis favoured a mildly triaxial gas distribution with a minor to major axis ratio of $0.70 \pm 0.15$, preferentially elongated along the line of sight, as expected for massive lensing clusters.

Here, we propose a method to infer the intrinsic shape and orientation of clusters based on weak and strong lensing (WL and SL) observations plus deep X-ray and SZe observations. Lensing gives a picture of the total projected mass distribution without any assumption on the equilibrium status of the cluster. X-ray plus SZe observations fix the size of the gas distribution along the line of sight and in the plane of the sky \citep{ser07,ser+al12}. The method exploits Bayesian inference to study a number of variables larger than the number of observational constraints. This enables us to investigate both intrinsic shape and orientation. The version of the method detailed in the present paper joins and integrate the lensing analysis in \citet{se+um11} with the X-ray plus SZe investigation of \citet{ser+al12}.

The paper is organised as follows. In Sec.~\ref{sec_tria}, we discuss how triaxial ellipsoids project and the relation between total matter and gas distributions. Sec.~\ref{sec_x_sz} summarizes the X-ray plus SZe analysis. Secs.~\ref{sec_wl} and \ref{sec_sl} are devoted to the WL and SL parts, respectively. In Sec.~\ref{sec_depr}, we show how to combine the different data-sets to infer the intrinsic properties. Results are discussed in Sec.~\ref{sec_resu}. The hydrodynamical status of the cluster is wrote about in Sec.~\ref{sec_he}. Comparisons with previous works are listed in Sec.~\ref{sec_comp}. Final considerations are contained in Sec.~\ref{sec_conc}.

Throughout the paper, we assume a flat $\Lambda$CDM cosmology with density parameters $\Omega_\mathrm{M}=0.3$, $\Omega_{\Lambda}=0.7$ and Hubble constant $H_0=100h~\mathrm{km~s}^{-1}\mathrm{Mpc}^{-1}$, $h=0.7 \pm 0.014$ \citep{kom+al11}. At the A1689 distance, $1\arcsec$ corresponds to $2.15~\mathrm{kpc}/h$
($=3.08~\mathrm{kpc}$).

\section{Triaxial halos}
\label{sec_tria}

In this section we describe how we model the total matter and the gas distribution and how we relate them.

\subsection{Matter profile}

The ellipsoidal Navarro-Frenk-White (NFW ) density profile,
\begin{equation}
\label{nfw1}
	\rho_\mathrm{NFW}=\frac{\rho_\mathrm{s}}{(\zeta/r_\mathrm{s})(1+\zeta/r_\mathrm{s})^2},
\end{equation}
provides a very good fit to the density distribution of dark matter halos in high resolution $N$-body simulations \citep{nfw96,nav+al97,ji+su02}. In Eq.~(\ref{nfw1}), $\zeta$ is the ellipsoidal radius. The shape of the halo is determined by the axial ratios, which we denote as $q_1$ (minor to major axial ratio) and $q_2$ (intermediate to major axial ratio). The eccentricity is $e_i=\sqrt{1-q_i^2}$. The orientation of the halo is fixed by three Euler's angles, $\theta, \varphi$ and $\psi$, with $\vartheta$ quantifying the inclination of the major axis with respect to the line of sight. 

The NFW density profile can be described by two parameters, the concentration and the mass. By definition, $r_{200}$ is such that the mean density contained within an ellipsoid of semi-major axis $r_{200}$ is 200 times the critical density at the halo redshift, $\rho_\mathrm{cr}$ \citep{cor+al09,ser+al10,se+um11}. Then, the corresponding concentration is $c_{200} \equiv r_{200}/ r_\mathrm{s}$. $M_{200}$ is the mass within the ellipsoid of semi-major axis $r_{200}$, $M_{200}=(800\pi/3)q_1q_2 r_{200}^3 \rho_\mathrm{cr}$.

The projection into the sky of the ellipsoidal 3D NFW halo is an elliptical 2D profile \citep{sta77,ser07,ser+al10b}. The convergence $\kappa$, i.e., the surface mass density in units of the critical density for lensing, $\Sigma_\mathrm{cr}=(c^2\,D_\mathrm{s})/(4\pi G\,D_\mathrm{d}\,D_\mathrm{ds})$, where $D_\mathrm{s}$, $D_\mathrm{d}$ and $D_\mathrm{ds}$ are the source, the lens and the lens-source angular diameter distances respectively, can be written as
\beq
\kappa_\mathrm{NFW}(x)=\frac{2 \kappa_\mathrm{s}}{1-x^2}\left[ \frac{1}{\sqrt{1-x^2}} \mathrm{arccosh}\left(\frac{1}{x}\right) -1\right];
\eeq
$x$ is the dimensionless elliptical radius,
\beq
x \equiv \xi /r_\mathrm{sP}, \ \ \xi= [x_1^2 +x_2^2/(1-\epsilon)^2)]^{1/2},
\eeq
where $\epsilon$ is the ellipticity and $x_1$ and $x_2$ are the abscissa and the ordinate in the plane of the sky oriented along the the ellipse axes, respectively. 

The strength $\kappa_\mathrm{s}$ and the projected length scale $r_\mathrm{sP}$ are related to mass and concentration and depend on the shape and orientation parameters too. Explicit formulae can be found in \citet{ser+al10}. 

Since we can only observe projected maps, the problem of determining the 3D orientation and shape of halos is intrinsically degenerate. Ellipsoids project into ellipses \citep{sta77}. Even with an ideal multi-probe data-set without noise, we can only measure three observable quantities which help us to constrain the five unknown intrinsic properties (two axial ratios and three orientation angles) of the ellipsoidal halo \citep{ser07}. The three measurable quantities are the ellipticity $\epsilon$, the orientation $\theta_\epsilon$ and the elongation $e_\Delta$. 

The parameters $\epsilon$ and $\theta_\epsilon$ characterise the projected ellipse in the plane of the sky. The ellipticity $\epsilon$ gives a measurement of the width in the plane of the sky. It is defined as $1-b_\mathrm{p}/a_\mathrm{p}$, where $b_\mathrm{p}$ and $a_\mathrm{p}$ are the minor and major axis of the projected ellipse. $\theta_\epsilon$ measures the orientation in the plane of the sky of this ellipse.

$e_\Delta$ quantifies the extent of the cluster along the line of sight. It is the ratio between $a_\mathrm{p}$ and the size of the ellipsoid along the line of sight \citep[ see figure~1]{ser+al12}. The smaller $e_\Delta$, the larger the elongation along the line of sight. If $e_\Delta < 1$, the cluster is more elongated along the line of sight than wide in the plane of the sky.

These three observable quantities depend on the intrinsic axial ratios and on the Euler's angle \citep{bin80,ser07}.

\subsection{ICM}

Observations \citep{kaw10} and theory \citep{bu+hu12} suggest that also the density of the intra-cluster medium (ICM) is nearly constant on a family of similar, concentric, coaxial ellipsoids. Modelling both the gas and the matter distribution as ellipsoids with constant eccentricity is formally wrong in haloes in hydrostatic equilibrium. If the cluster of galaxies is in hydrostatic equilibrium the gas distribution traces the gravitational potential. Given an ellipsoidal gas density, the gravitational potential is ellipsoidal too and can turn unphysical for extreme axial ratios, with negative density regions or unlikely configurations. On the other hand, even in hydrostatic equilibrium, the ellipsoidal approximation for the gas is suitable in the inner regions or when small eccentricities are considered. 

If the potential is ellipsoidal, the matter distribution that originates it can not be ellipsoidal. However, dark matter haloes formed in cosmological simulations typically have radially varying shapes too \citep{kaz+al04}, which might be compatible with an ellipsoidal potential.

If the matter halo isodensity surfaces are triaxial ellipsoids, the isodensity surfaces of the intracluster gas are well approximated as triaxial ellipsoids with eccentricities slowly varying with the radius \citep{le+su03,le+su04}. The ratio of eccentricities of gas ($e^\mathrm{ICM}$) and matter ($e^\mathrm{Mat}$) is nearly constant up to the length scale, with $e^\mathrm{ICM}_i/e^\mathrm{Mat}_i \simeq 0.7$ for $i=1,2$. Furthermore, the variation in eccentricity is usually smaller than the observational error on the measured ellipticity of the X-ray surface brightness map.

A further complication is that microphysical processes such as radiative cooling, turbulence or feedback mechanisms strongly affect the shape of the baryonic component and make the ICM shape more triaxial and with a distinctly oblate shape towards the central cluster regions compared to the underlying dark matter potential shape \citep{lau+al11}. The gas traces the underlying gravitational potential better outside the core, even if radiative processes can make the ICM shape rounder. These mechanisms can be effective so that assuming that the overall triaxiality of the gas is strictly due to the underlying shape of the dark matter potential can be misleading.

Finally, density and shape of the gas distribution have a strong correlation, whereas temperature is essentially uncorrelated \citep{sam+al12}. Then the proper modelling of a varying eccentricity would require an independent measurements of the density profile.

Under these circumstances, we can make two simplifying but quite non-informative working hypotheses to relate gas and total matter distributions. Firstly, we conservatively model both the total matter density and the gas distribution as coaligned ellipsoids with fixed, but different, axial ratios. Secondly, we set the two distributions to have the same triaxiality parameter ${\cal T}(=e_2^2/e_1^2)$. If two distributions have the same triaxialities, then the misalignment angle between the orientations in the plane of the sky is zero \citep{ro+ko98}. This is in agreement with what observed in A1689, where the centroid and the orientation of the surface brightness map \citep{ser+al12} coincide with those of the projected mass distribution as inferred from lensing \citep{se+um11}. 

The axial ratios of the gas distribution, $q^\mathrm{ICM}_i$, can be expressed in terms of the corresponding axial ratios of the matter distributions as
\beq
q^\mathrm{ICM}_i=\sqrt{1-(e^\mathrm{ICM}/e^\mathrm{Mat})^2(1-q_i^2)}.
\eeq
Being ${\cal T}^\mathrm{Mat}={\cal T}^\mathrm{ICM}$, then $e^\mathrm{ICM}_1/e^\mathrm{Mat}_1=e^\mathrm{ICM}_2/e^\mathrm{Mat}_2=e^\mathrm{ICM}/e^\mathrm{Mat}$. The above assumptions limit the number of free axial ratios to three: $q_1$ and $q_2$ for the matter and $q_1^\mathrm{ICM}$ for the gas. $q_2^\mathrm{ICM}$ is determined by ${\cal T}(q_1,q_2)$ and $q_1^\mathrm{ICM}$,
\beq
q^\mathrm{ICM}_2=\sqrt{1-\frac{1-[q^\mathrm{ICM}_1]^2}{[{\cal T}^\mathrm{Mat}]^2}}.
\eeq 
In a triaxial analysis, results strongly depend on the minor to major axial ratio whereas the inference about the intermediate axis is more  affected by priors. Then, using $q_1$, $q_2$ and $q_1^\mathrm{ICM}$ as free parameters is not really a limitation with respect to considering all four axial ratios as free. Finally, as expected from both theory and observations, we required that the gas distribution is rounder than the matter one, $q^\mathrm{Mat}_1 \le q^\mathrm{ICM}_1$.

Under these hypotheses, total matter and gas have different ellipticities and elongations but share the same orientation $\theta_\epsilon$. Since we model the gas isodensities as ellipsoids, they project into ellipses in the plane of the sky. The same degeneracy plaguing the inference of shape and orientation of the matter distribution affects the gas too.

\section{X-ray and SZe}
\label{sec_x_sz}

\begin{figure}
       \resizebox{\hsize}{!}{\includegraphics{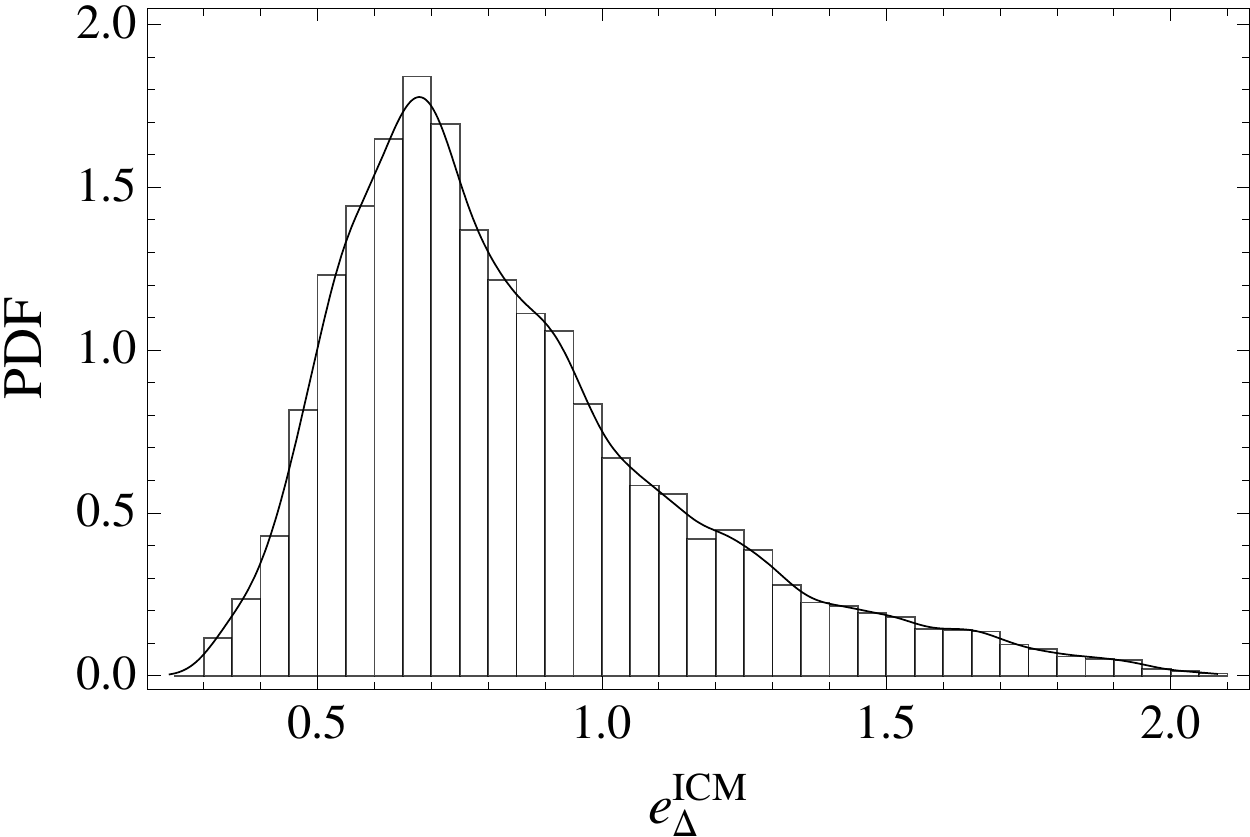}}
       \caption{Probability distribution of the elongation of the gas distribution as inferred from the X-ray plus SZe analysis.}
	\label{fig_pdf_e_Delta_fit}
\end{figure}

The shape and orientation of the gas distribution can be inferred with a combined analysis of X-ray and SZe data \citep{ser+al12}. The first set of constraints comes from the analysis of the projected maps. The ellipticity and the orientation of the isophotes, which could be measured with either X-ray or SZe alone, are related to the size of the gas distribution in the plane of the sky. The inference of the elongation along the line of sight requires a combined analysis. This can be done thanks to the different dependence of X-ray and SZe observables on the gas density.

The analysis of X-ray and SZe data of A1689 was performed in \citet{ser+al12}. Here, we summarize the main results and what is needed in the present paper. We refer to \citet{ser+al12} for details. Images in the 0.7--2 keV band taken by {\it Chandra} are well approximated by a series of concentric, coaligned ellipses with $\epsilon^\mathrm{X}=0.15\pm0.03$ and orientation angle $\theta_\epsilon^\mathrm{X}=12\pm3\deg$ (measured North over East). The 3D electronic density and temperature were modelled with the following parametric profiles \citep{vik+al06,ett+al09},
\beq
\label{n_prof}
n_\mathrm{e}=n_0  \left[ 1+\left( \frac{\zeta}{r_\mathrm{c}} \right)^2 \right]^{-3\beta/2} \left[ 1+\left( \frac{\zeta}{r_\mathrm{t}} \right)^3 \right]^{-\frac{\gamma}{3}},
\eeq
where $n_\mathrm{0}$ is the central electron density, $r_{\rm c}$ is the core radius, $r_\mathrm{t}(>r_\mathrm{c})$ is the tidal radius, $\beta$ is the slope in the intermediate regions, and  $\gamma$ is the outer slope.  For the temperature profile, we used
\beq
\label{T_prof}
T= \frac{ T_0}{  [1+ (\zeta/r_\mathrm{T})^{2}]^{0.45}  },
\eeq
where $T_0$ is the central temperature and the radius $r_\mathrm{T}$ describes a decrement at large radii. The parametric forms in Eqs.~(\ref{n_prof},~\ref{T_prof}) were motivated by the absence of a cool core.

Profiles based on Eqs.~(\ref{n_prof},~\ref{T_prof}) were then compared to observations of X-ray surface brightness (SB), spectroscopic temperature and SZ decrement.
The  surface brightness observed by Chandra was collected in 68 elliptical annuli up to $\xi \ls 1200~\mathrm{kpc}$. The temperature profile from the {\it XMM} satellite was binned in 5 elliptical annuli up to $\xi \ls 900~\mathrm{kpc}$. We considered the integrated Compton parameters $Y$ within the circle of radius $r_{2500} \sim 600~\mathrm{kpc}$ measured from 7 observatories (BIMA, OVRO, AMIBA,  SuZIE, WMAPS, SZA, SCUBA). We re-did the analysis as in \citet{ser+al12}, but we used an improved version of the code, with some better numerical algorithms for numerical integration. 

To asses realistic probability distributions for the parameters we performed a statistical Bayesian analysis. The Bayes theorem states that
\beq
\label{baye}
p(\bfP | \bfd) \propto {\cal L}( \bfP|\bfd) p(\bfP),
\eeq
where $p(\bfP | \bfd)$ is the posterior probability of the parameters $\bfP$ given the data $\bfd$, ${\cal L}( \bfP|\bfd)$ is the likelihood of the data given the model parameters and $p(\bfP)$ is the prior probability distribution for the model parameters. 

The posterior probability distribution for the elongation of the gas $e_\Delta^\mathrm{ICM}$ is plotted in Fig.~\ref{fig_pdf_e_Delta_fit}. The elongation together with the projected ellipticity and orientation enables us to constrain the intrinsic shape of the gas. Whereas the projected ellipticity and the orientation angle of the projected isophotes of the gas can be precisely determined with X-ray data alone, the measurement of the elongation requires a combined X-ray plus SZe analysis.

\section{Weak lensing}
\label{sec_wl}

\begin{table*}
\centering
\begin{tabular}[c]{lr@{$\,\pm\,$}lr@{$\,\pm\,$}lr@{$\,\pm\,$}lr@{$\,\pm\,$}lr@{$\,\pm\,$}lr@{$\,\pm\,$}l}
        \hline
        \noalign{\smallskip}
        NFW & \multicolumn{2}{c}{$\kappa_\mathrm{s}$} & \multicolumn{2}{c}{$\theta_{1,0}$} & \multicolumn{2}{c}{$\theta_{2,0}$}& \multicolumn{2}{c}{$\epsilon$}&\multicolumn{2}{c}{$\theta_\epsilon$} & \multicolumn{2}{c}{$r_\mathrm{sP}$}    \\
        \noalign{\smallskip}
        & \multicolumn{2}{c}{} & \multicolumn{2}{c}{[$\arcsec$]} &\multicolumn{2}{c}{[$\arcsec$]} & \multicolumn{2}{c}{} &
\multicolumn{2}{c}{[$\deg$] }&\multicolumn{2}{c}{[$\mathrm{kpc}$]}  \\
        \noalign{\smallskip}
        \hline
         \multicolumn{12}{c}{Weak Lensing} \\
        & $0.62$&$0.18$& $-0.8$&$1.2$ & $-3.1$&$1.1$ & $0.20$&$0.10$ & $11.6$&$2.9$ & $220$&$50$   \\
        \hline
         \multicolumn{12}{c}{Strong Lensing} \\
        & $0.27$&$0.02$& $-3.0$&$0.7$ & $-4.8$&$2.7$ & $0.16$&$0.03$ & $13.6$&$5.6$ & $910$&$120$   \\
	\hline
         \multicolumn{12}{c}{Weak plus Strong Lensing} \\
        & $0.35$&$0.01$& $-3.8$&$0.5$ & $-0.1$&$1.0$ & $0.11$&$0.02$ & $12.7$&$2.4$ & $410$&$20$   \\
        \hline
       \end{tabular}
\caption{
Projected NFW parameters inferred with either the weak or the strong or the combined lensing analysis. The final WL distributions for the centroid position and the orientation angle mirror the X-ray derived priors. The orientation angle $\theta_\epsilon$ is measured north over east and $r_\mathrm{sP}$ is the projected length scale. Convergence is normalised to a reference source redshift ($z_\mathrm{s}=2.0$). Central location and dispersion of the marginalized posterior density functions (PDFs) are computed as mean and standard deviation. 
}
\label{tab_fit_nfw_GL}
\end{table*}


We fitted the surface mass density of A1689 with a projected ellipsoidal NFW halo. The weak lensing convergence map, described as ``2D MEM" in \citet{ume+bro08}, was obtained from {\it Subaru} data and covers a field of $\sim 30\arcmin \times 24\arcmin$ ($21\times 17$ grid pixels with pixel size of $1.4\arcmin$) \citep{ume+bro08,ume+al09}. We closely followed the method employed in \citet{se+um11}, which we refer to for more details. The likelihood can be written in terms of a $\chi^2_\mathrm{WL}$ \citep{ogu+al05,ogu+al10}, with
\beq
\label{like_wl}
\chi^2_\mathrm{WL}=\sum_{i,j}\left[ \kappa_\mathrm{obs}(\bfr_i)- \kappa(\bfr_i) \right] \left( V^{-1}\right)_{i,j} \left[ \kappa_\mathrm{obs}(\bfr_j)- \kappa(\bfr_j) \right] 
\eeq
where $\kappa_\mathrm{obs}$ is the measured convergence map and $\bfV^{-1}$ is the inverse of the pixel-pixel covariance matrix. 

One major difference with \citet{se+um11} is in the priors used  to determine the NFW projected parameters. Whereas in \citet{se+um11} uniform priors were employed, here we exploit constraints from the X-ray analysis on centroid position and ICM orientation. The position of the centroid of the X-ray surface brightness, $\{\alpha,\delta\}=\{197.87274, -1.3400533\}\deg$ with an accuracy of $\sim 1.2 \arcsec$, is consistent with the centre of the BCG and with the centroid of the total matter distribution as estimated from lensing \citep{se+um11}. The orientation of the X-ray map is consistent too with the orientation of the matter halo as estimated from lensing. The posterior probability is then
\begin{eqnarray}
p & \propto & \exp(-\chi^2_\mathrm{WL}/2)  \times {\cal N}(\theta_\epsilon; \theta_\epsilon^\mathrm{X}, \delta \theta_\epsilon^\mathrm{X}) \\
 & \times & {\cal N}(\theta_{1,0}; \theta_{1}^\mathrm{X}, \delta \theta_{1}^\mathrm{X})\times {\cal N}(\theta_{2,0}; \theta_{2}^\mathrm{X}, \delta \theta_{2}^\mathrm{X}) \nonumber
\end{eqnarray}
where ${\cal N}(P; \mu, \sigma)$ denotes a normal distribution for the parameter $P$ centred in $\mu$ and with variance $\sigma$. $\theta_{1,0}$ and $\theta_{2,0}$ are the Cartesian coordinates of the projected centre of the matter distribution in a reference system centred in the BCG galaxy. $\theta_{1}^\mathrm{X}$ and $\theta_{2}^\mathrm{X}$ are the observed coordinates of the X-ray centroid.

The parameter space was explored with Markov chains. Chain convergence was checked by requiring that the standard var(chain mean)/mean(chain var) indicator was less than 1.2. Results are summarised in Table~\ref{tab_fit_nfw_GL}.

Analyses based on WL alone are not very effective in determining the projected ellipticity. The relative constraints on the intrinsic axial ratios are not sharp and are dominated by the a priori hypothesis employed. Here, since we exploited the X-ray information on the halo orientation in the plane of the sky, the ellipticity is better constrained. With respect to the WL only analysis in \citet{se+um11}, the final distributions on the projected parameters are more peaked and with reduced tails. This is clearly seen in the probability function for $\epsilon$, which would be nearly flat without the X-ray priors.

\section{Strong lensing}
\label{sec_sl}

We performed a strong lensing analysis of the inner regions of A1689 and obtained a pixelated map of the surface mass density by use of the PixeLens software \citep{sa+wi04}. We considered multiple image systems with confirmed spectroscopic redshift, whose detailed description can be found in \citet{lim+al07,coe+al10}. In a second step, we fitted the map with a projected NFW profile. The method is described in detail in \citet{se+zi12}

PixeLens cannot handle all the multiple image systems of A1689 at once. We then divided the SL systems in two groups and analyzed each group separately. The first group includes the systems 1, 2, 7, 10, 11, 18, 22, 24, 40 according to the notation in \citet{lim+al07}, for a total of 31 multiple images. The second group includes the systems 4, 5, 15, 17, 19, 29, 33 and 35 (28 multiple images). For each group, we computed 500 convergence maps within a radius of 94\arcsec around the BCG on a grid of 861 pixels with a pixel size of $12.6~\mathrm{kpc}/h$. 

As explained in \cite{se+zi12}, we excised a central region of $40~\mathrm{kpc}/h$ to minimize the effects of miscentering and baryonic physics \citep{ume+al11}. We also excluded the outer pixels where the logarithmic density slope artificially takes values smaller than -3.

We then modelled the convergence with a projected NFW profile. For each convergence map, we looked for the minimum of
\begin{equation}
\label{eq_fit_1}
\chi^2 = \sum_i \left[ \kappa_\mathrm{obs}(\bfr_i) -\kappa_\mathrm{NFW}(\bfr_i)\right]^2
\end{equation}
where the sum runs over the pixels. From the derived ensemble of maximum likelihood parameters, we obtained the posteriori distribution of projected NFW parameters. We repeated the procedure for the two groups of images, ending up with results statistically not distinguishable.

Results were consistent with previous SL analyses based on the same image systems \citep[ and references therein]{lim+al07,coe+al10,se+um11}. Mass density profiles obtained with PixeLens are usually shallower than reconstructions based on parametric models \citep{gri+al10,ume+al12}. In comparison with the mass profiles in \citet{lim+al07}, who used {\it Lenstool}, or \citet{se+um11}, who used {\it Gravlens}, PixeLens retrieves a profile shallower in the inner $\ls 90~\mathrm{kpc}/h$. On the other hand, the projected masses within the Einstein radius ($\ls 300~\mathrm{kpc}$) are fully consistent. The excision of the inner $40~\mathrm{kpc}/h$ in our analysis makes the impact of the differences between different methods nearly negligible.

Results from SL are marginally consistent with the WL analysis, see Table~\ref{tab_fit_nfw_GL}. WL favors more concentrated profiles but the strong degeneracy between the lensing strength $\kappa_\mathrm{s}$ and the projected length-scale $r_\mathrm{sP}$ makes the two analyses compatible.

The measurement of the orientation angle with the strong lensing analysis of the core region is in very good agreement with the estimate from the X-ray analysis and further supports the working hypothesis that the gas and the total matter distribution share the same triaxiality degree and are then co-aligned in projection.

\section{Deprojection}
\label{sec_depr}

Results from either gravitational lensing or X-ray plus SZe measurements can be combined to deproject the observations and infer the three dimensional structure and orientation of the matter and gas distributions. The combined X-ray plus SZ analysis enabled us to infer the width of the ICM in the plane of the sky (parameterized in terms of the ellipticity $\epsilon^\mathrm{ICM}$) and its size along the line of sight (expressed as the elongation $e_\Delta^\mathrm{ICM}$). 

The lensing analysis describes how the total matter density projects in the plane of the sky. The orientation and the ellipticity of the projected surface density are related to the intrinsic shape and orientation of the total matter halo. How the convergence varies with the radius, i.e., the information contained in the parameters $\kappa_\mathrm{s}$  and $r_\mathrm{sP}$ of the NFW halo, constrains the functional form of the density, i.e, the mass and the concentration. 

The orientation and the shape of the matter halo significantly affect lensing. The more the cluster is elongated along the line of sight, the more the apparent convergence is boosted and the smaller the projected length scale in the plane of the sky.

Information from X-ray, SZe and lensing can be brought together. The likelihood function combining the results from the previous sections can be written as 
\begin{equation}
\label{like_comb}
{\cal L}={\cal L}^\mathrm{GL}\times {\cal L}^\mathrm{ICM},  
\end{equation}
where ${\cal L}^\mathrm{GL}={\cal L}^\mathrm{WL}\times{\cal L}^\mathrm{SL}$ and ${\cal L}^\mathrm{WL,SL}={\cal L}(\kappa_\mathrm{s},r_\mathrm{sP},\epsilon, \theta_\epsilon)$ \citep{se+um11}. For the likelihood of the X-ray plus SZe part \citep{ser+al12},
\begin{eqnarray}
{\cal L}^\mathrm{ICM}& = & \frac{1}{(2\pi)^{1/2} \sigma_\epsilon} \exp \left\{ -\frac{[\epsilon^\mathrm{X}-\epsilon^\mathrm{ICM}]^2}{2\sigma_\epsilon^2}  \right\}  \nonumber \\
& \times &	P(e_\Delta^\mathrm{ICM}-\Delta e_\Delta^\mathrm{sys}) \label{like_icm} \\
& \times	&	\frac{1}{(2\pi)^{1/2} \delta e_\Delta^\mathrm{sys}} \exp \left\{ -\frac{1}{2}\left( \frac{\Delta e_\Delta^\mathrm{sys}}{\delta e_\Delta^\mathrm{sys}} \right)^2\right\}, \nonumber 
\end{eqnarray}
where $P(e_\Delta^\mathrm{ICM})$ is the marginalized posterior probability distribution for the elongation parameter of the gas obtained in Sec.~\ref{sec_x_sz}. The parameter $\Delta e_\Delta^\mathrm{sys}$ quantifies the additional unknown statistical and systematic uncertainty on the elongation. We let it follow a normal distribution \citep{dag03}, centred on 0 and with dispersion $\delta e_\Delta^\mathrm{sys} \simeq 0.13$ \citep{ser+al12}.

The overall ellipsoidal cluster model describing both ICM and dark matter has 9 free parameters. For the total matter halo: 2 parameters determining the profile, i.e, the mass $M_{200}$ and the concentration $c_{200}$; 2 axial ratios ($q_1$ and $q_2$) and 3 orientation angles ($\vartheta$, $\varphi$ and $\psi$). 

For the gas distribution: the axial ratio $q^\mathrm{ICM}_1$, determining the shape of the ICM; 1 parameter, $\Delta e_\Delta^\mathrm{sys}$, quantifying the systematic uncertainty on the elongation of the gas distribution. Mass and gas share the same orientation angles whereas the second axial ratio of the ICM is fixed by the remaining parameters thanks to the priors.

The distributions of matter and gas project into ellipses into the plane of the sky. Lensing constraints reduce to four parameters. The analysis of the ICM adds two more measured quantities. The intrinsic parameters have to be then determined from a smaller number of observational constraints. The problem is under-constrained \citep{ser07}.

From the lensing analysis, we can measure the 4 projected parameters of the NFW halo (the lensing normalization $\kappa_\mathrm{s}$, the projected length-scale $r_\mathrm{sP}$, the ellipticity $\epsilon$ and the orientation $\theta_\epsilon$). From the analysis of the ICM, we can estimate the size of the gas in the plane of the sky ($\epsilon^\mathrm{ICM}$) and along the line of sight ($e_\Delta^\mathrm{ICM}$). The information on the orientation of the X-ray isophotes in the plane of the sky, $\theta_\epsilon^\mathrm{ICM}$, was exploited as a prior in the WL analysis to better constrain $\theta_\epsilon$. The estimate for the systematic error $\delta e_\Delta^\mathrm{sys}$ provides another constraint.

Let us summarize the dependences of each observed quantity. The ellipticity $\epsilon^\mathrm{ICM}$ and the elongation $e_\Delta^\mathrm{ICM}$ of the gas distributions are functions of $q^\mathrm{ICM}_1$, $q^\mathrm{ICM}_2$, and of the orientation angles $\vartheta$ and $\varphi$. Due to the imposed equality of triaxiality between the mass and gas distributions, we can express $q^\mathrm{ICM}_2$ in terms of $q^\mathrm{ICM}_1$, $q_1$ and $q_2$. The systematic shift $\Delta e_\Delta^\mathrm{sys}$ is an additional parameter which will be marginalized over. 

The ellipticity $\epsilon$ of the total matter distribution is function of $q_1$, $q_2$, and of the orientation angles $\vartheta$, and $\varphi$. The orientation angle $\theta_\epsilon$ depends on the third Euler's angle, $\psi$, too. The mass $M_{200}$ and the concentration $c_{200}$ of the total matter halo determine the lensing strength $\kappa_\mathrm{s}$ and the projected length-scale $r_\mathrm{sP}$. $\kappa_\mathrm{s}$ and $r_\mathrm{sP}$ depend also on  the elongation, and, in turn, on $q_1$, $q_2$, $\vartheta$ and $\varphi$. 

Some a priori hypotheses on the cluster shape are needed to disentangle the intrinsic degeneracies. We applied some Bayesian methods already employed in either gravitational lensing investigations \citep{ogu+al05,cor+al09,ser+al10,se+um11} or X-ray plus SZe analyses \citep{ser+al12} and extended them for our GL plus X-ray plus SZe analysis.

We considered two kind of priors for the axial ratio of the matter distributions. The distribution of $q_1$ obtained in high resolution $N$-body simulations can be approximated as \citep{ji+su02},
\beq
\label{nbod3}
p(q_1) \propto \exp \left[ -\frac{(q_1-q_\mu/r_{q_1})^2}{2\sigma_\mathrm{s}^2}\right]
\eeq
where $q_\mu=0.54$, $\sigma_\mathrm{s}=0.113$ and
\beq
r_{q_1} = (M_\mathrm{vir}/M_*)^{0.07 \Omega_\mathrm{M}(z)^{0.7}},
\eeq
with $M_*$ the characteristic nonlinear mass at redshift $z$ and $M_\mathrm{vir}$ the virial mass. The conditional probability for $q_2$ goes as
\beq
\label{nbod4}
p(q_1/q_2|q_1)=\frac{3}{2(1-r_\mathrm{min})}\left[ 1-\frac{2q_1/q_2-1-r_\mathrm{min}}{1-r_\mathrm{min}}\right]
\eeq
for $q_1/q_2 \geq r_\mathrm{min} \equiv \max[q_1,0.5]$, whereas is null otherwise.

We considered also a uniform distribution for the axial ratios in the range $q_\mathrm{min}<q_1 \le 1$ and $q_1 \le q_2 \le 1$. Probabilities are defined such that the marginalized probability $P(q_1)$ and the conditional probability $P(q_2|q_1)$ are constant. The probabilities can then be expressed as
\beq
\label{flat1}
p(q_1) =1/(1-q_\mathrm{min})
\eeq 
for the full range $q_\mathrm{min}<q_1 \le 1$ and
\beq
\label{flat2}
p(q_2|q_1) = (1-q_1)^{-1}
\eeq 
for $q_2 \ge q_1$ and zero otherwise. A flat distribution is also compatible with very triaxial clusters ($q_1 \ls q_2 \ll1$), which are preferentially excluded by $N$-body simulations. We fixed $q_\mathrm{min}=0.1$.

For the axial ratio of the gas we considered a uniform distribution in the interval $q_1 \le q^\mathrm{ICM}_1 \le 1$. The prior on $q^\mathrm{ICM}_1$ is then similar to that of $q_2$ in the case of the uniform distribution for the matter axial ratios.

For the orientation, we considered a population of randomly oriented clusters with
\beq
\label{flat3}
p(\cos \vartheta) = 1
\eeq 
for $0 \le \cos \vartheta \le 1$ and
\beq
p(\varphi)=\frac{1}{\pi}
\eeq
for $-\pi/2 \le \varphi \le \pi/2$.

For the remaining parameters, we employed uniform distributions.

\section{Results}
\label{sec_resu}

\begin{table*}
\begin{tabular}{c r@{$\,\pm\,$}lr@{$\,\pm\,$}lr@{$\,\pm\,$}lr@{$\,\pm\,$}lr@{$\,\pm\,$}l}
        \hline
        \noalign{\smallskip}
	Priors & \multicolumn{2}{c}{$M_{200}$}	& \multicolumn{2}{c}{$c_{200}$}	& \multicolumn{2}{c}{$q_1$}	& \multicolumn{2}{c}{$q_2$}	& \multicolumn{2}{c}{$\cos\vartheta$}	 \\
        \noalign{\smallskip}
        $q_i$	&	\multicolumn{2}{c}{$[10^{15}M_\odot]$} &  \multicolumn{8}{c}{} \\
        \hline
         \noalign{\smallskip}
         flat		&$1.33$	&$0.17$	&$7.8$	&$0.7$	&$0.72$	&$0.11$	&$0.86$	&$0.11$	&$0.57$	&$0.29$	 \\	
         $N$-body	&$1.34$	&$0.18$	&$7.6$	&$1.0$	&$0.51$	&$0.07$	&$0.69$	&$0.12$	&$0.76$	&$0.21$	 \\
\hline
\end{tabular}
\caption{Intrinsic parameters of the matter distribution inferred assuming different priors on the axial ratios. Reported values are the mean and the variance of the posterior PDF.}
\label{tab_pdf_par}
\end{table*}

\begin{figure*}
\begin{center}
$
\begin{tabular}{c@{\hspace{.1cm}}c@{\hspace{.1cm}}c@{\hspace{.1cm}}c@{\hspace{.1cm}}c}
\multicolumn{5}{c}{Flat $q$-distribution}	\\
\noalign{\smallskip}
\includegraphics[width=3.4cm]{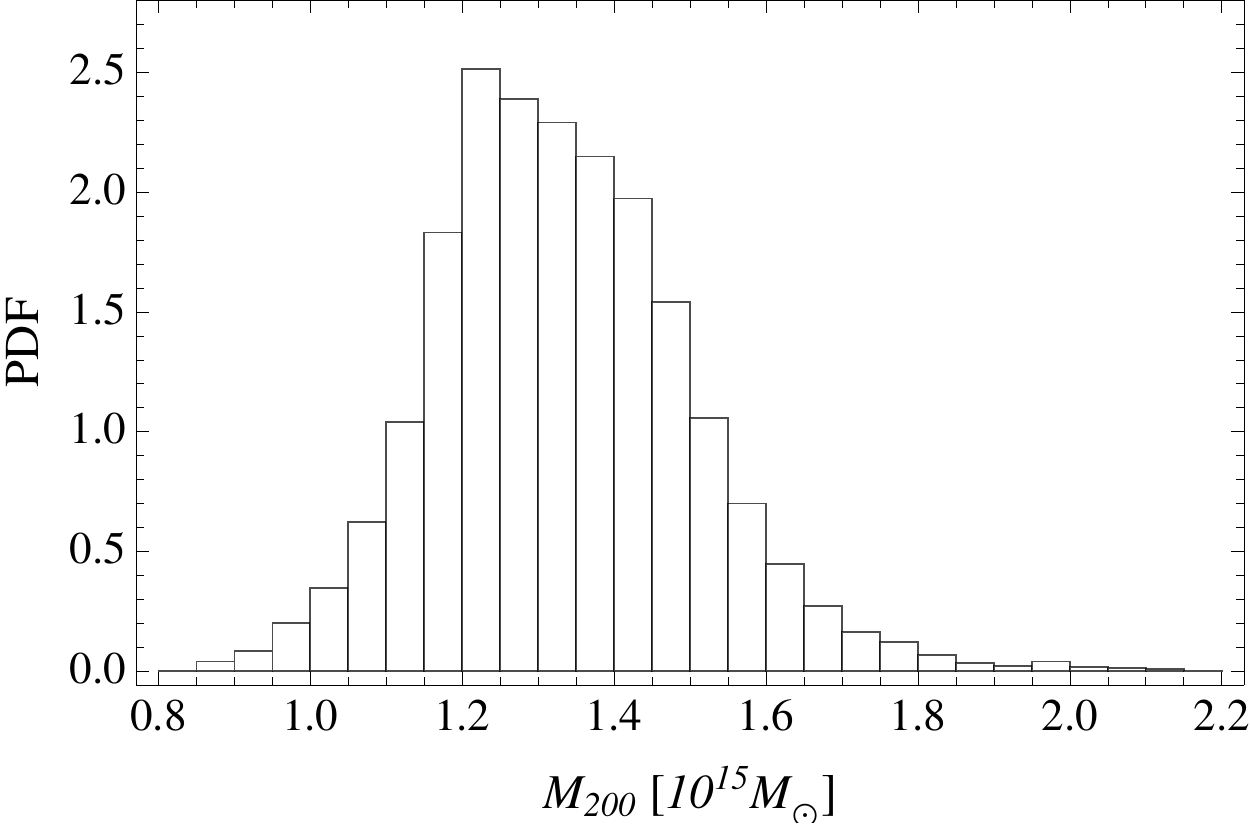} &
\includegraphics[width=3.4cm]{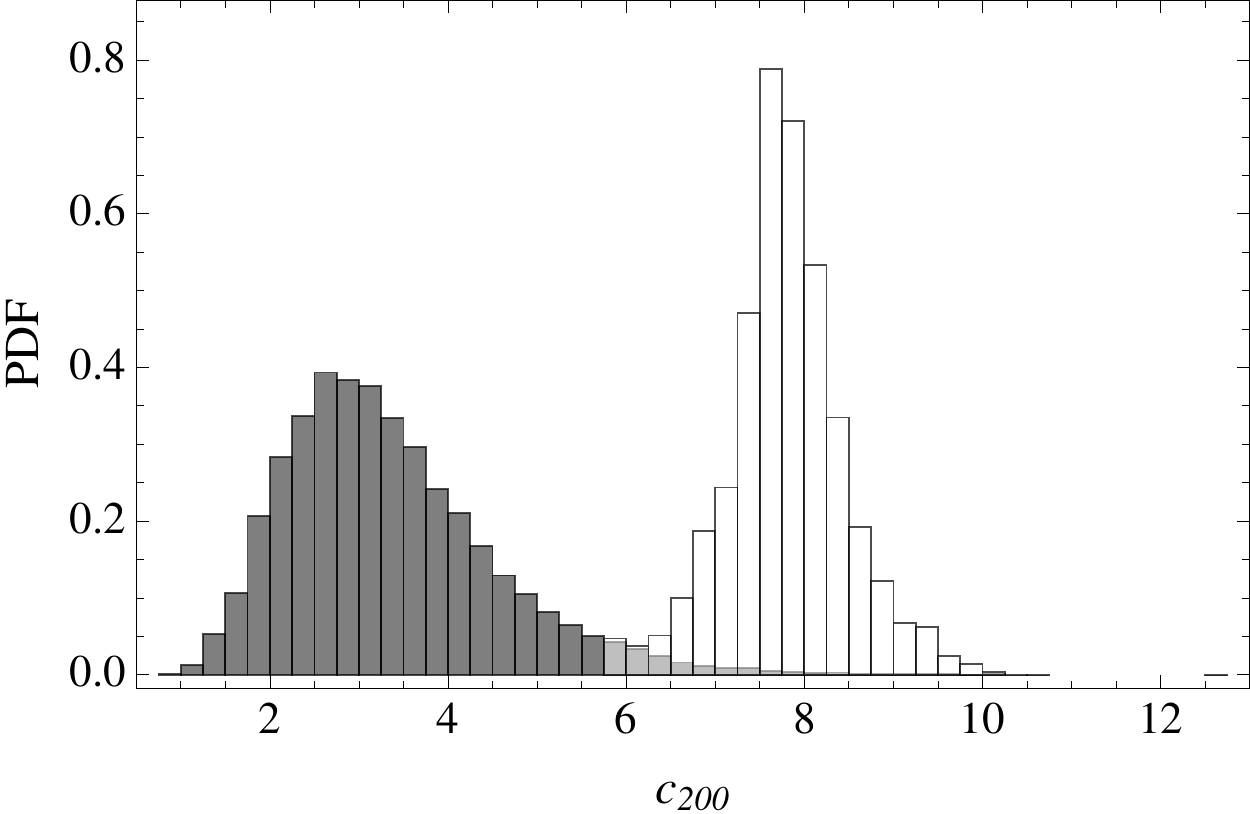} &
\includegraphics[width=3.4cm]{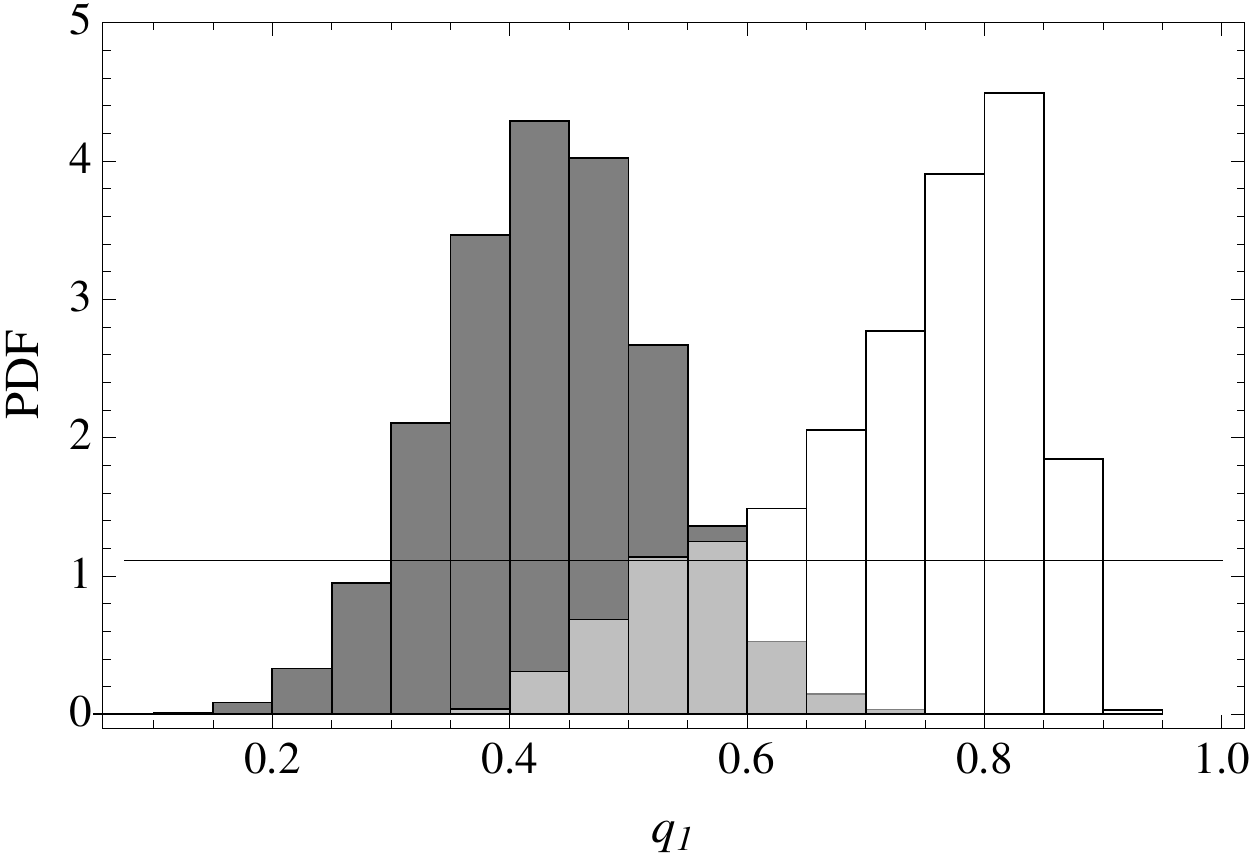} &
\includegraphics[width=3.4cm]{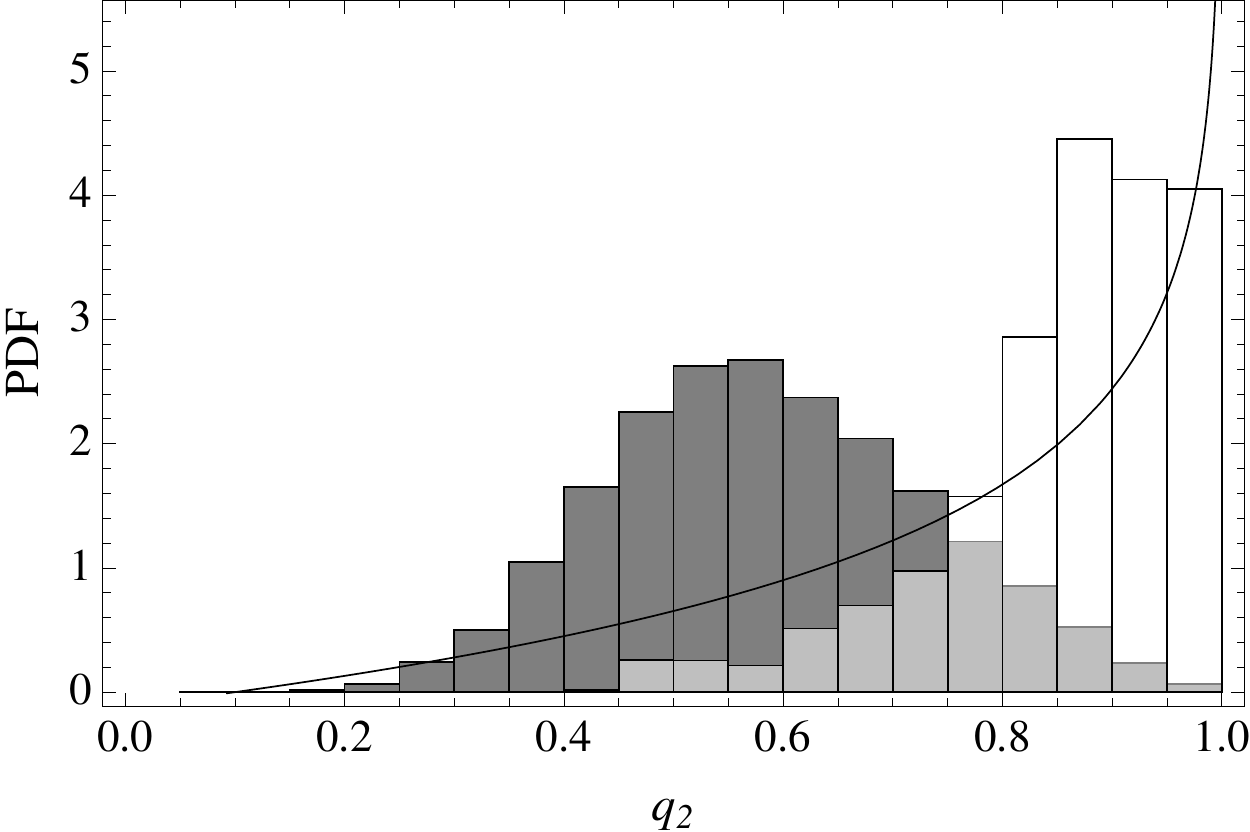} &
\includegraphics[width=3.4cm]{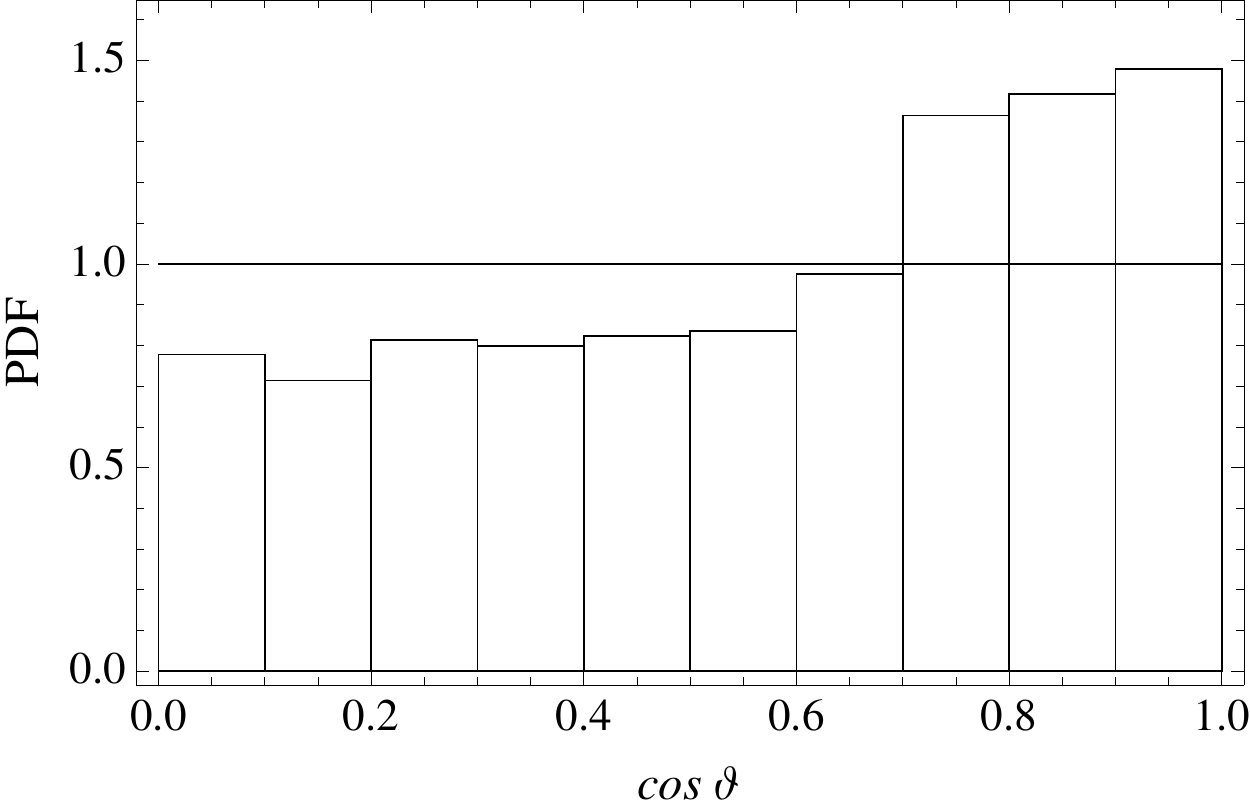}  \\	
\noalign{\smallskip}
\multicolumn{5}{c}{$N$-body $q$-distribution} \\
\noalign{\smallskip}
\includegraphics[width=3.4cm]{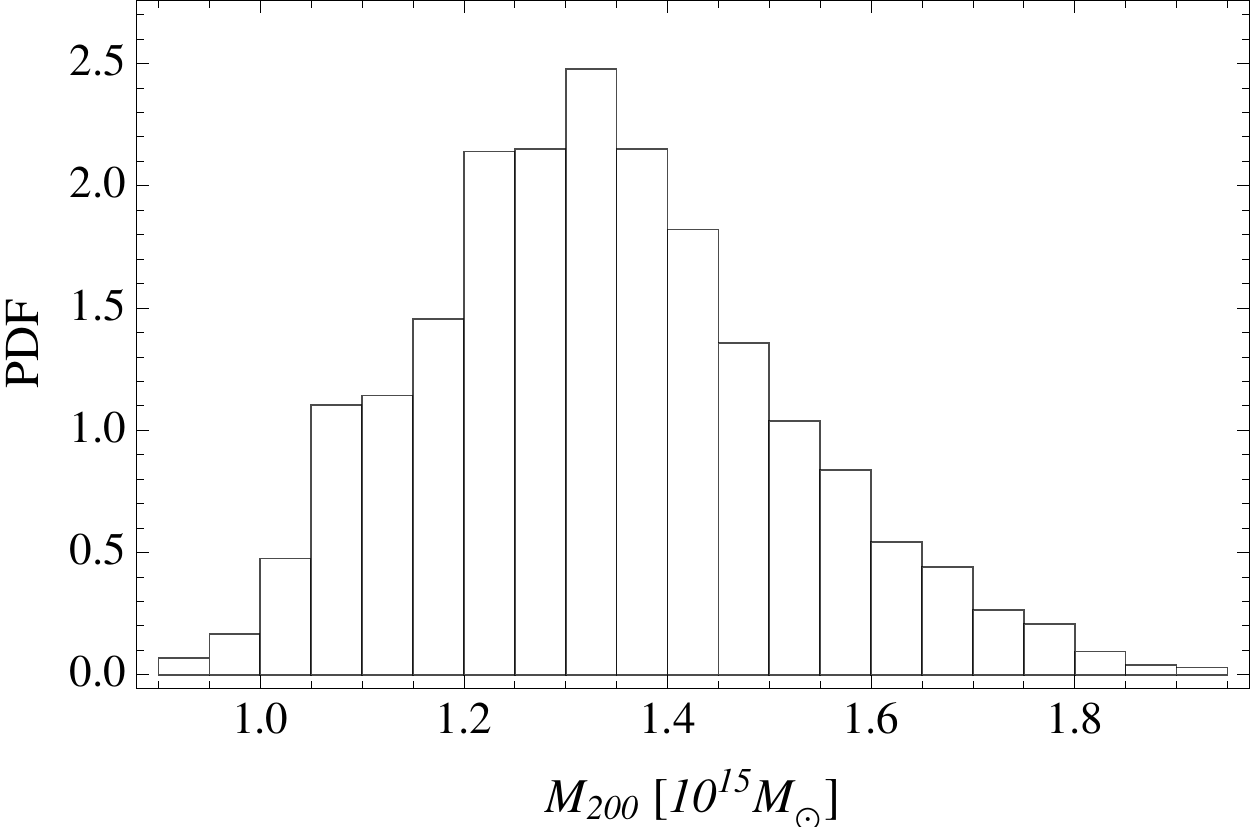} &
\includegraphics[width=3.4cm]{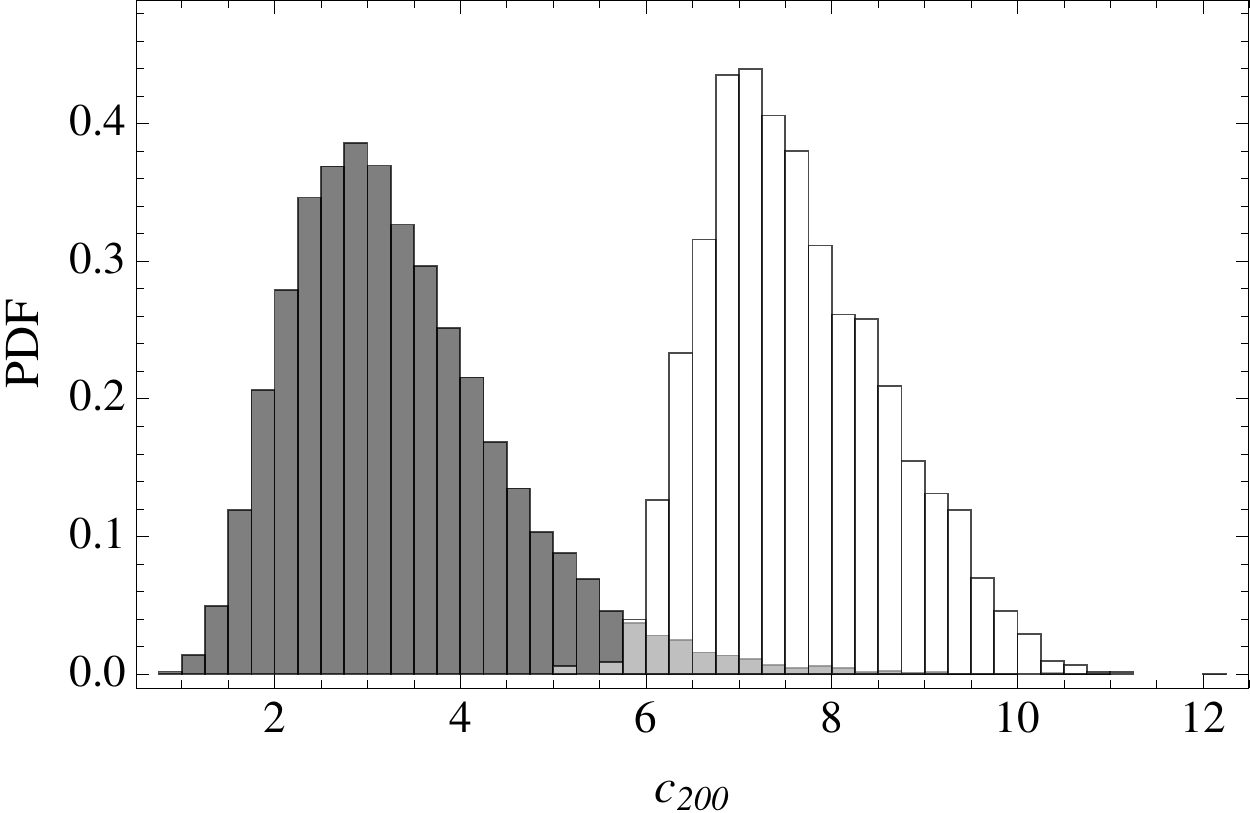} &
\includegraphics[width=3.4cm]{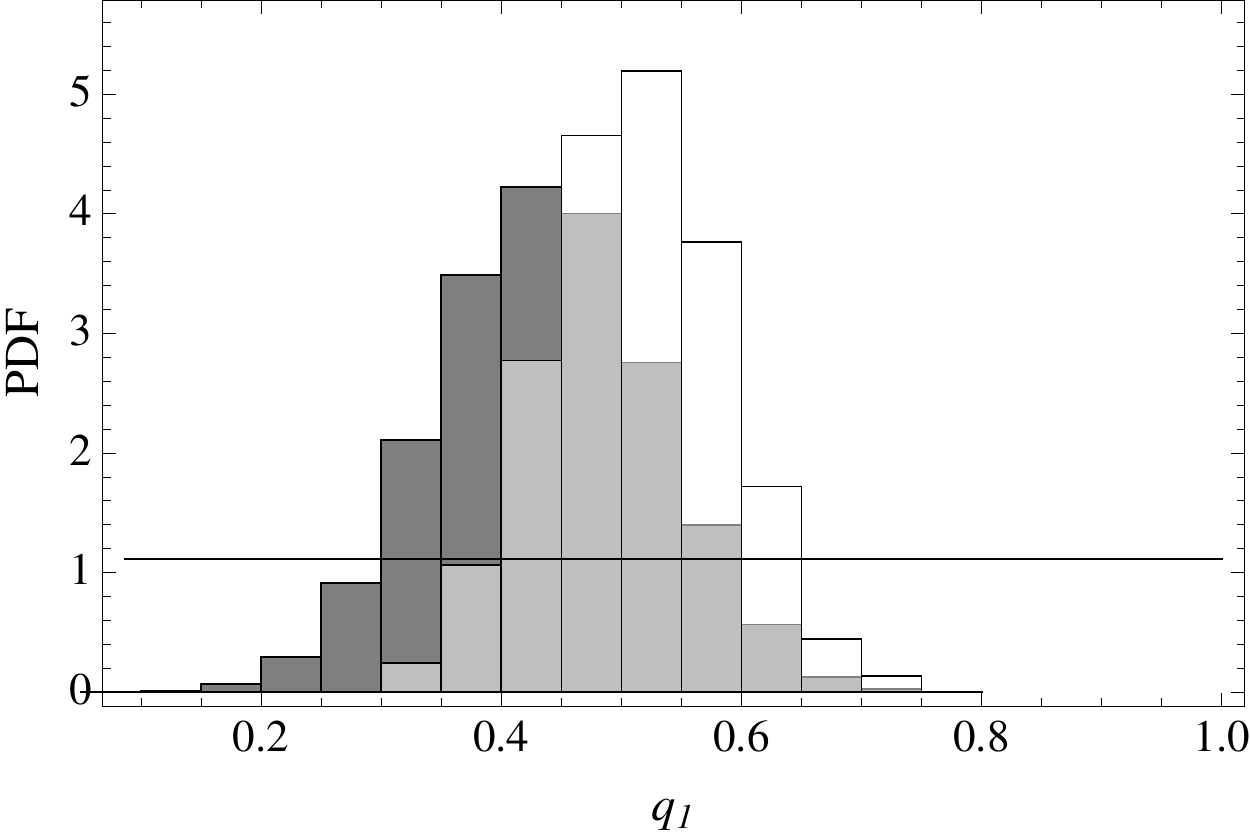} &
\includegraphics[width=3.4cm]{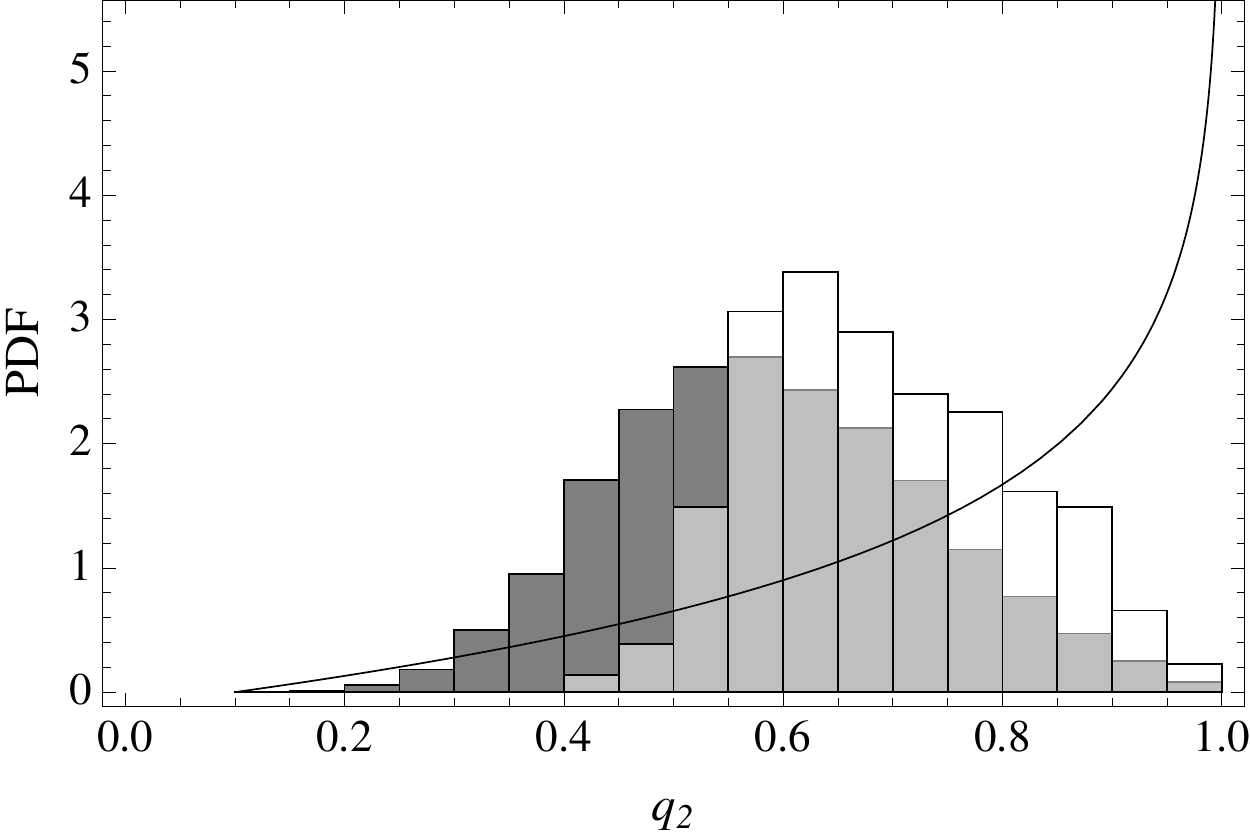} &
\includegraphics[width=3.4cm]{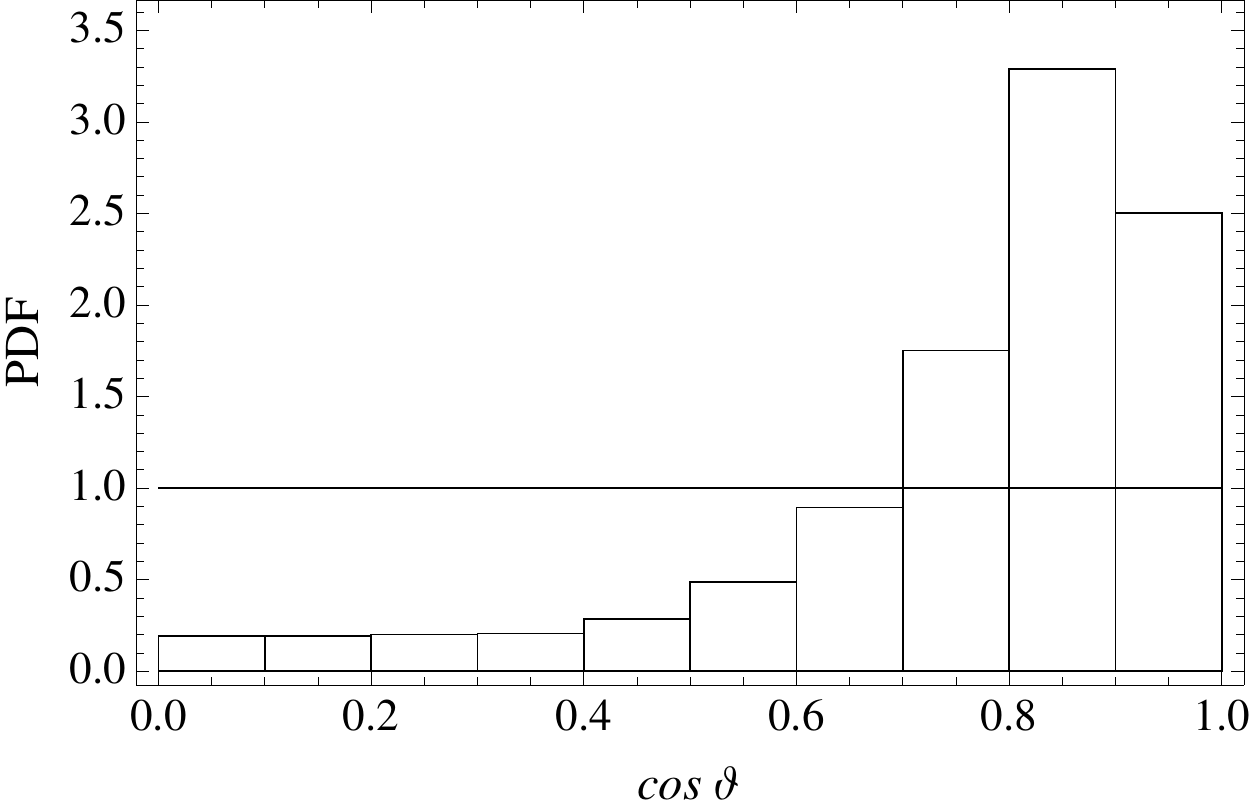} \\
\end{tabular}
$
\end{center}
\caption{PDFs for the matter halo parameters. Panels from the left to the right are for $M_{200}$, $c_{200}$, $q_1$, $q_2$ and $\cos \vartheta$, respectively. The white binned histograms are the posterior PDFs. The dark grey histograms are the theoretical expectations given the measured mass in the left panel. For the concentration, we exploited the $c(M)$ relation from \citet{duf+al08}. For the axial ratios, we considered predictions from $N$-body simulations \citep{ji+su02}. Intermediate grey bins are shared by the two plotted distributions. Thin lines in the $q$-panels denote a flat $q$-distribution. The thin line in the $\cos \vartheta$ panel stands for random orientation. In the top row, we plot the inversion results under the prior hypotheses of flat $q$-distribution and random orientation angles. In the bottom row, prior assumptions are $N$-body like $q$-distribution and random orientation angles. Masses are in units of $10^{15}M_\odot$.}
\label{fig_pdf_All_1D}
\end{figure*}

\begin{figure*}
\begin{center}
$
\begin{tabular}{cccc}
\noalign{\smallskip}
\includegraphics[width=4cm]{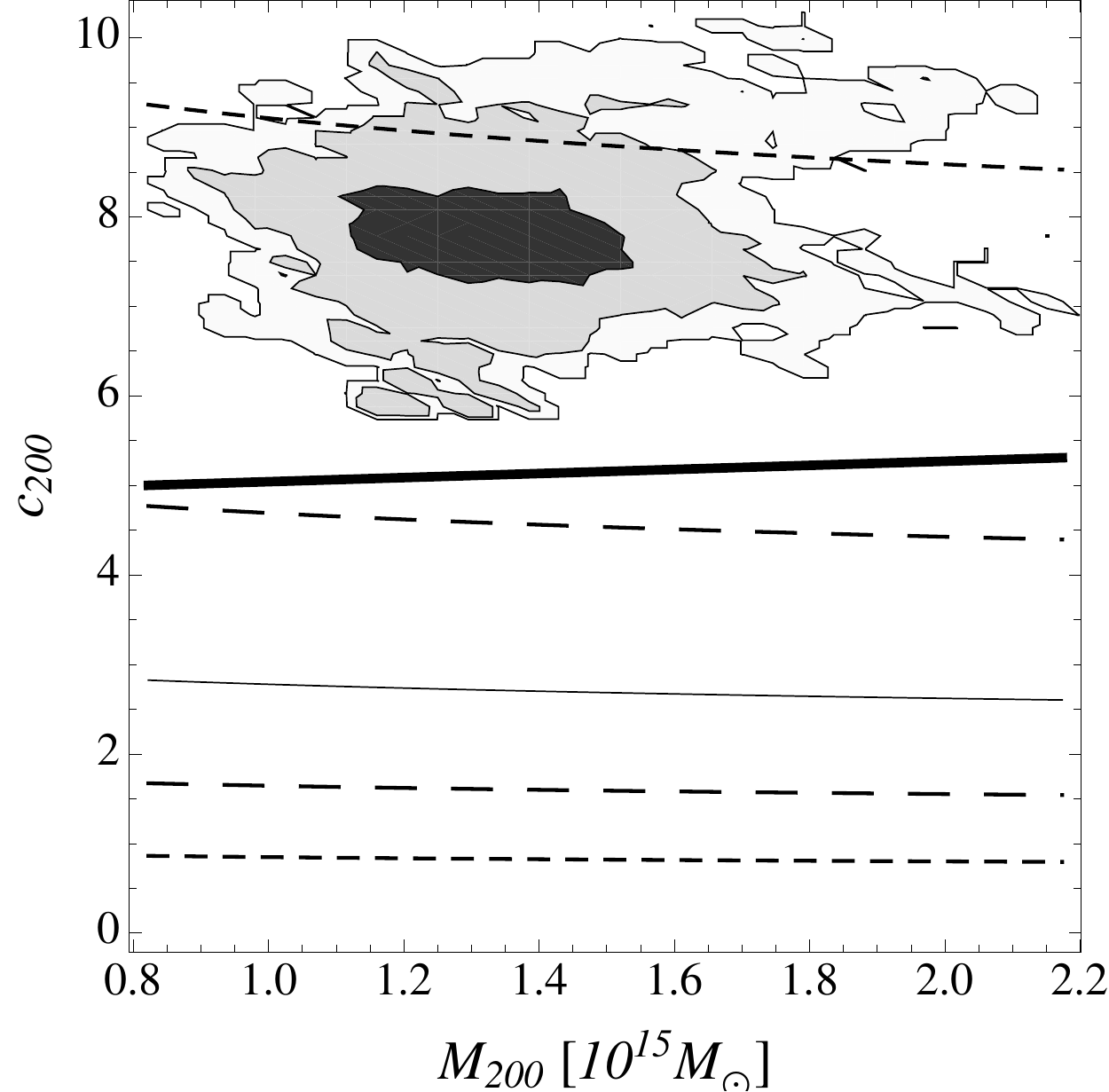} & & & \\
\includegraphics[width=4cm]{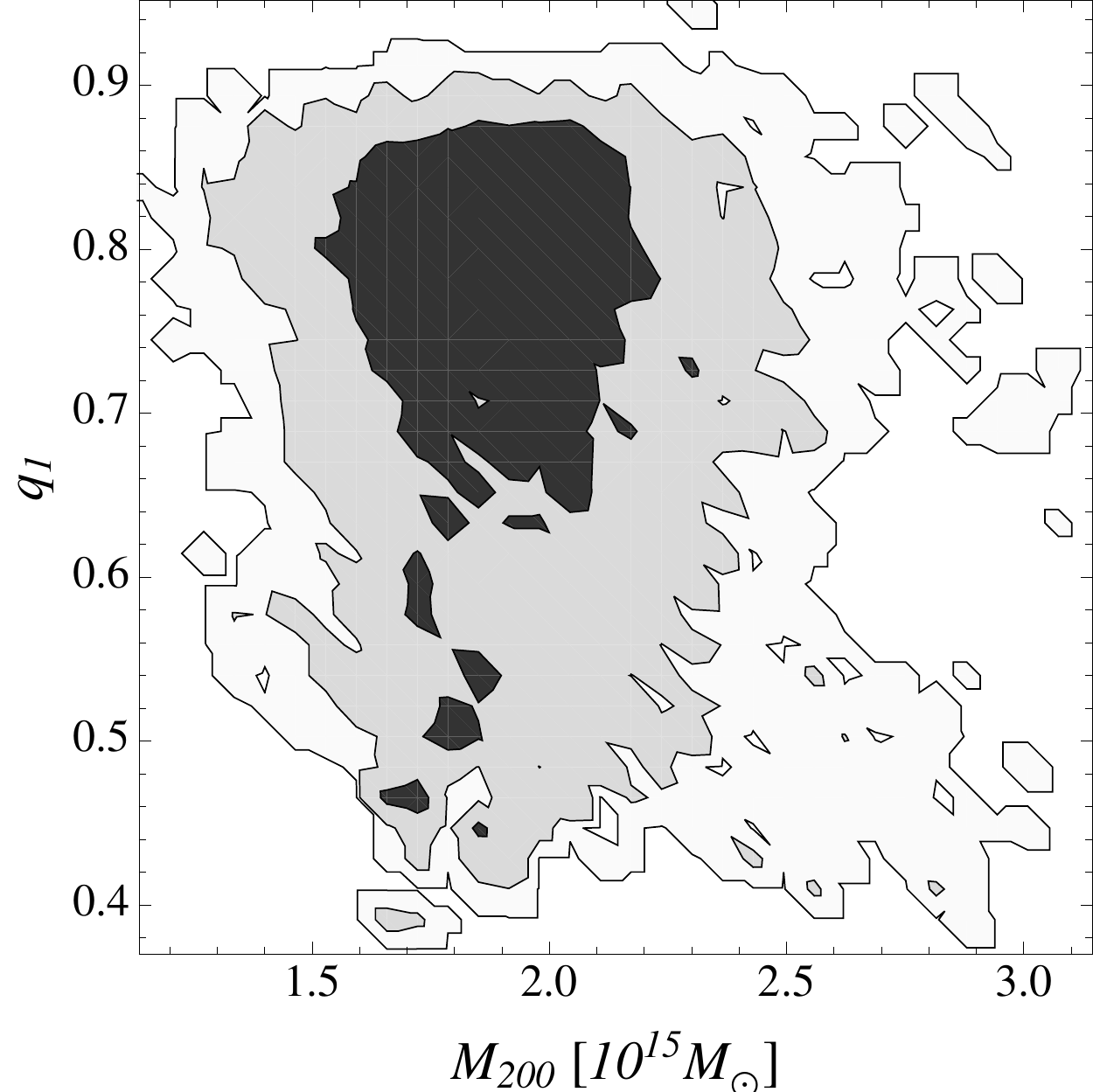} &
\includegraphics[width=4cm]{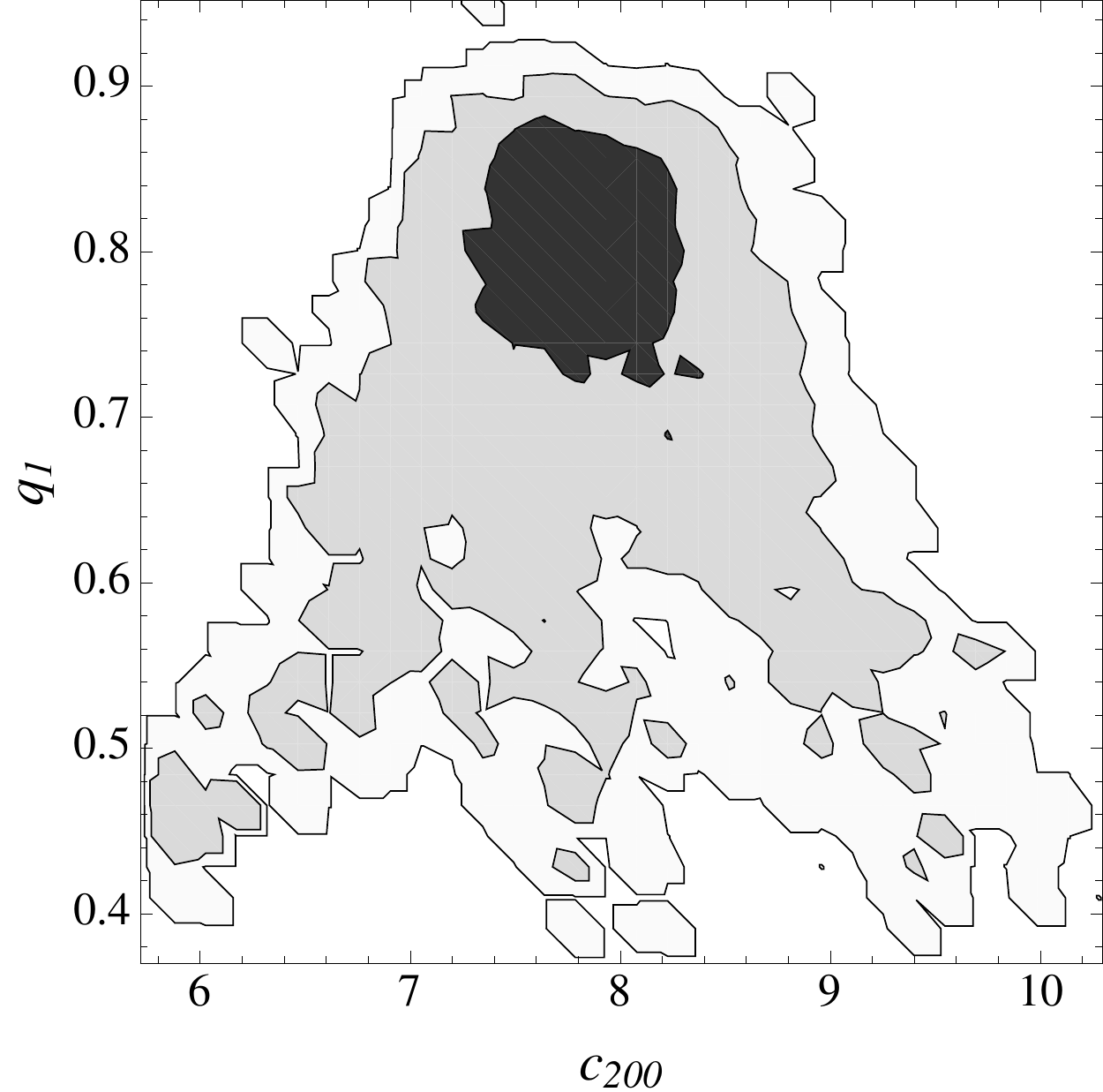} & & \\
\includegraphics[width=4cm]{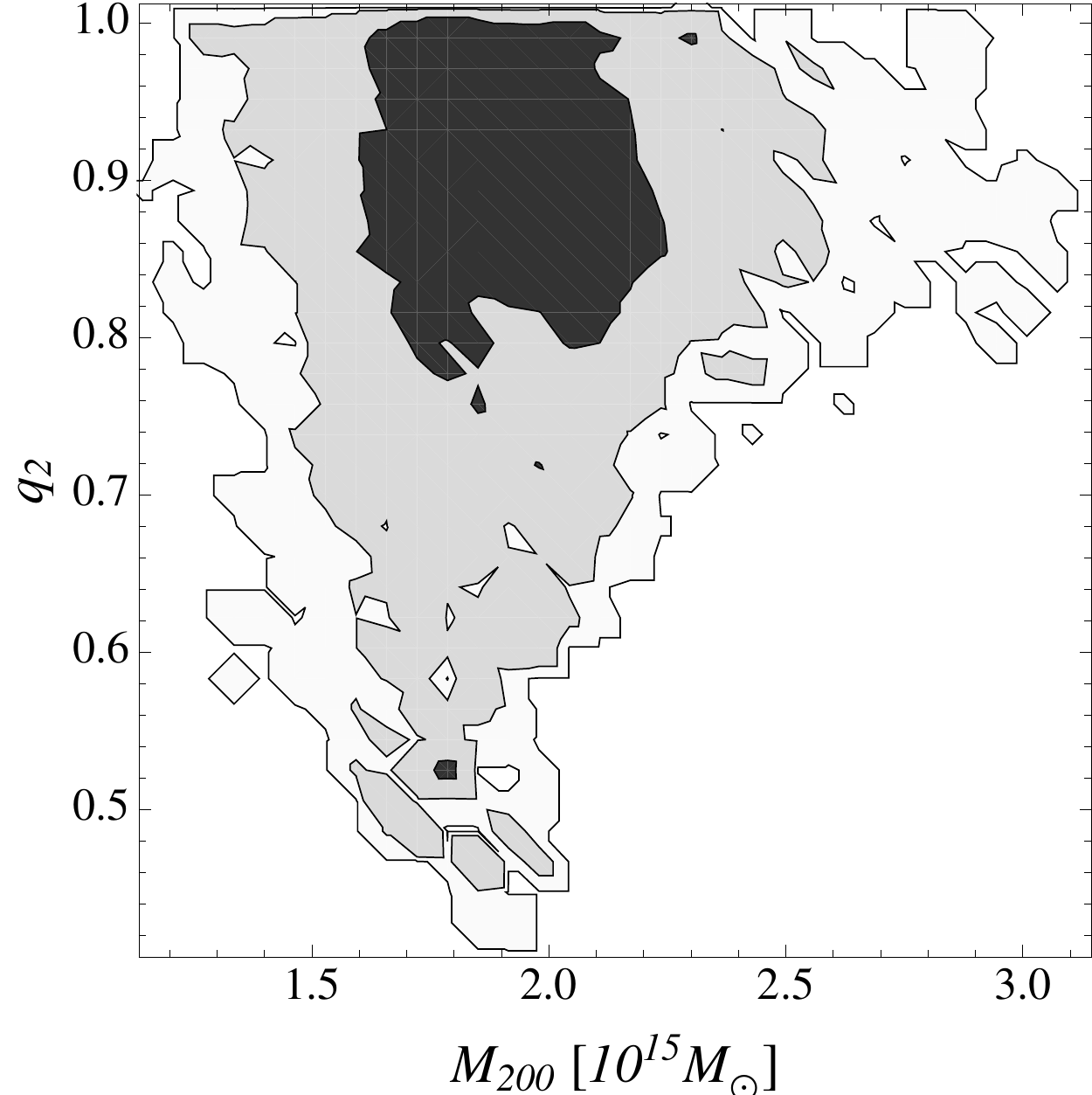} &
\includegraphics[width=4cm]{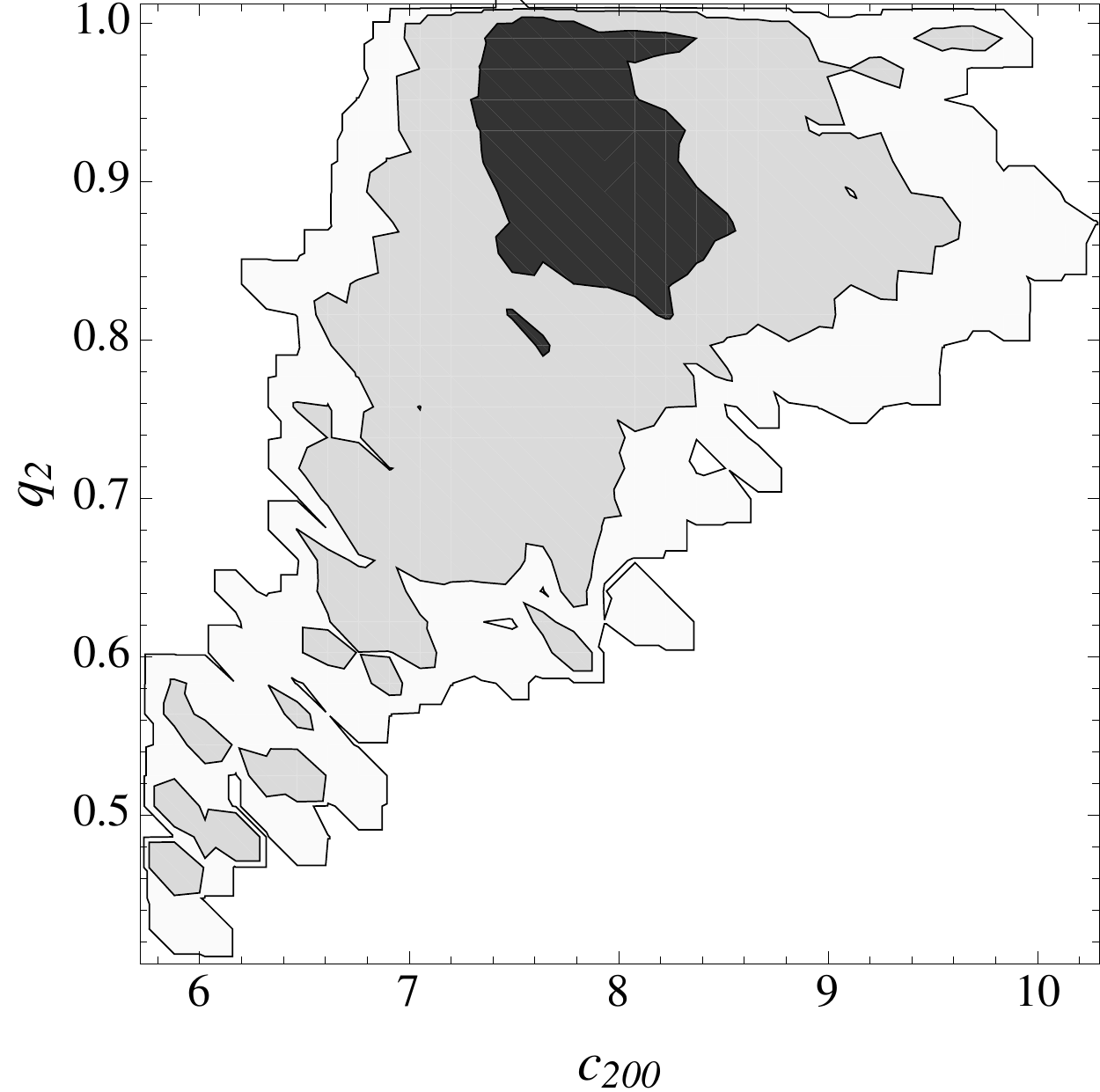} &
\includegraphics[width=4cm]{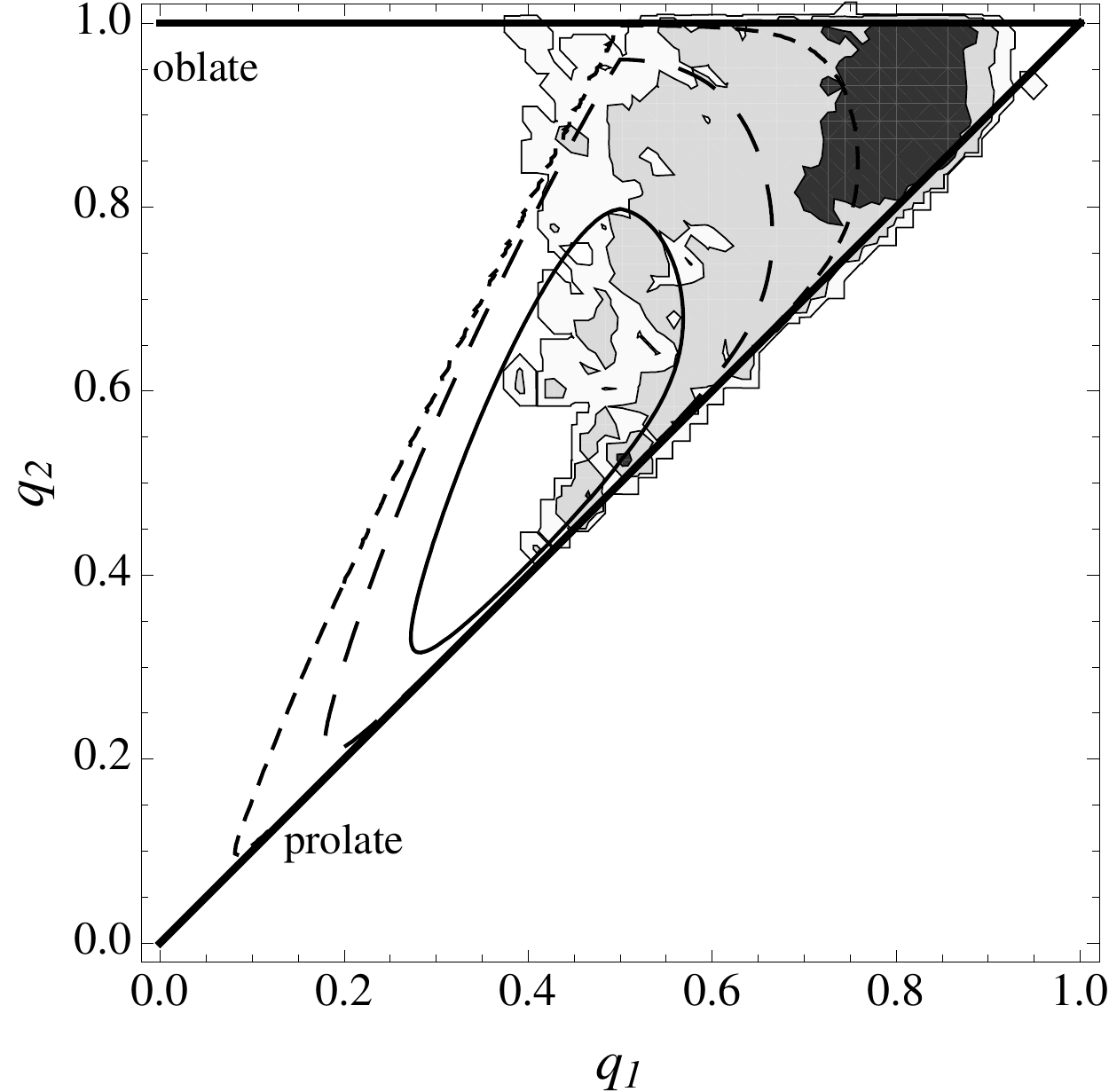} &  \\
\includegraphics[width=4cm]{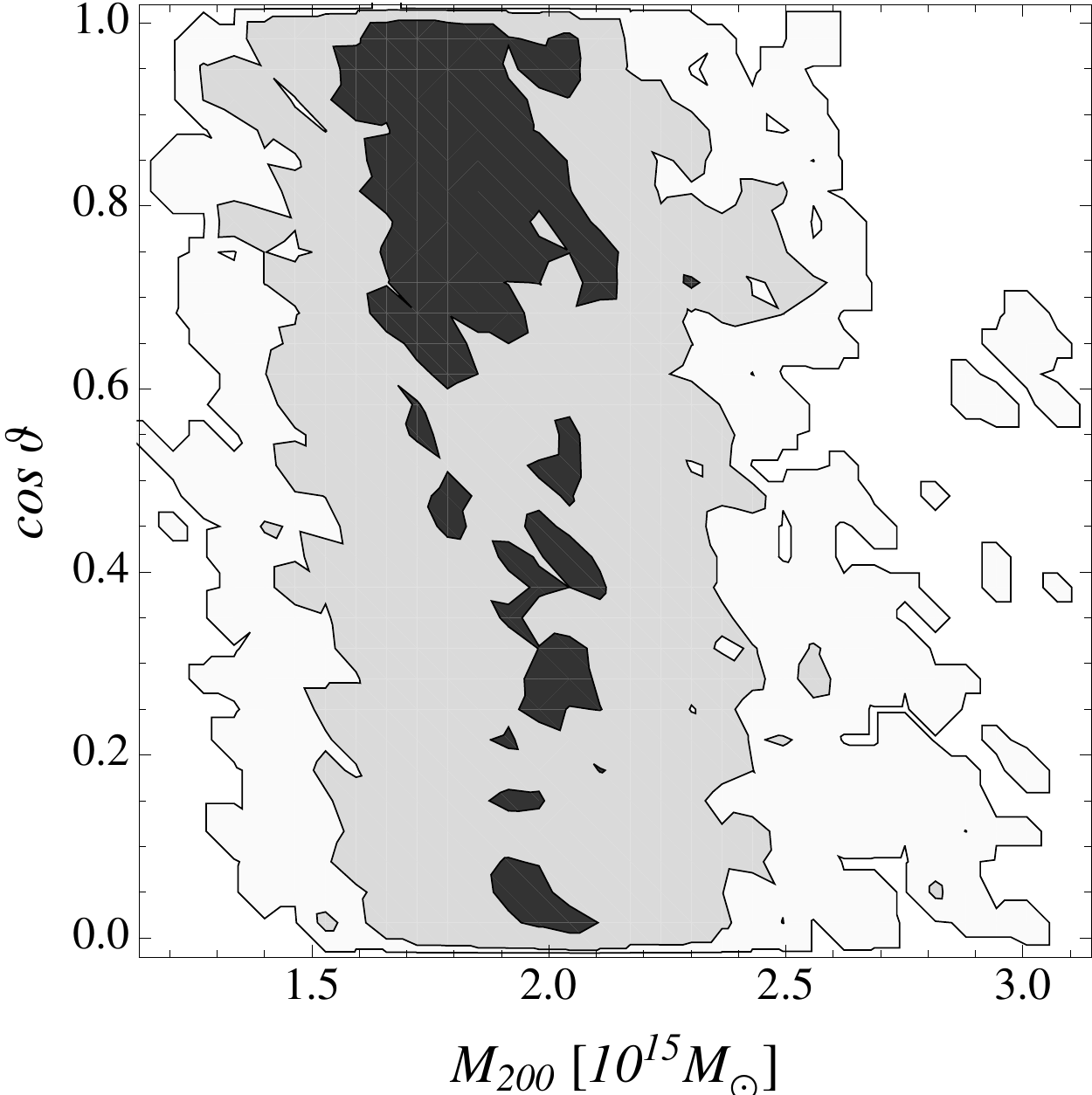} &
\includegraphics[width=4cm]{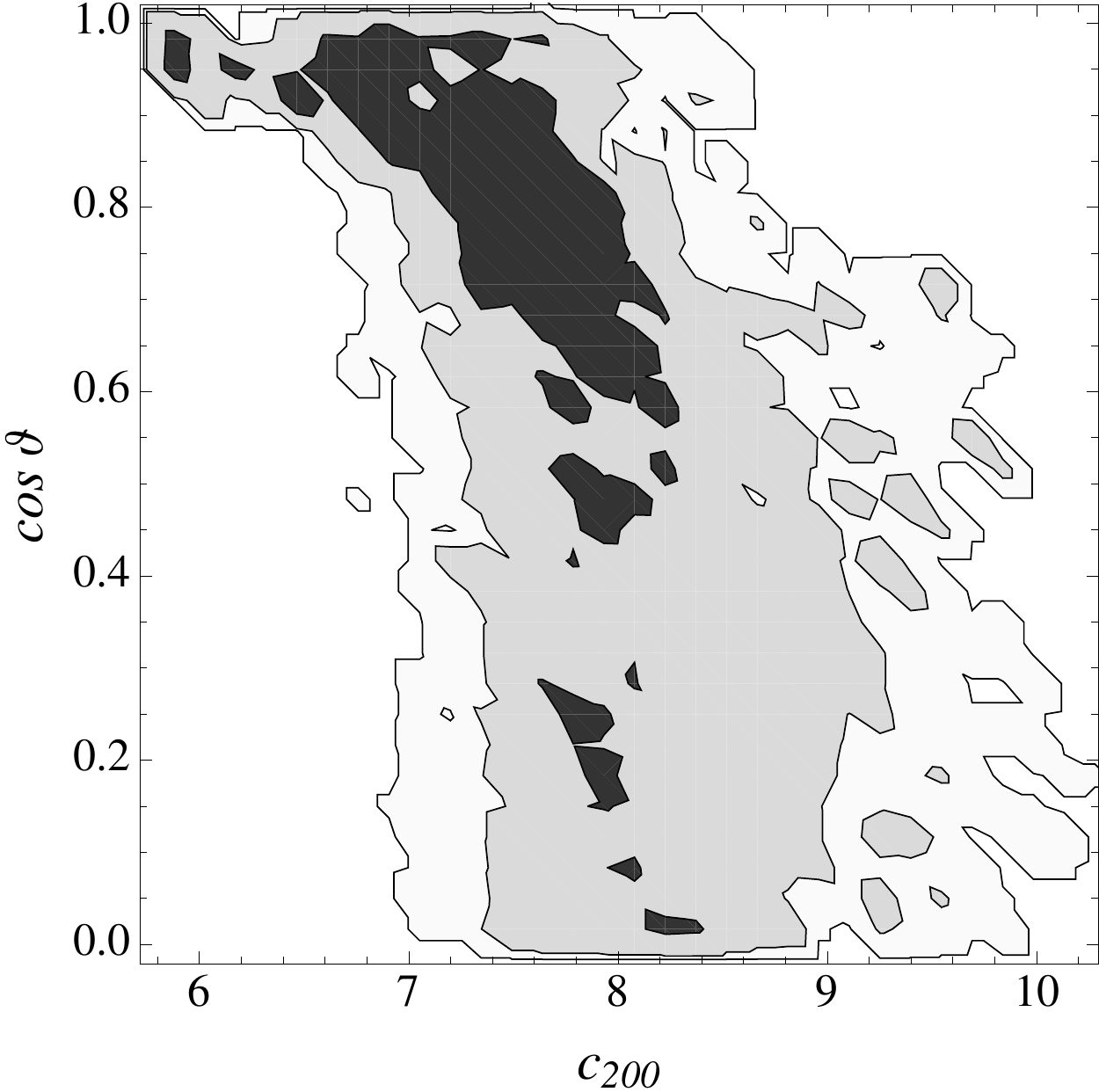} &
\includegraphics[width=4cm]{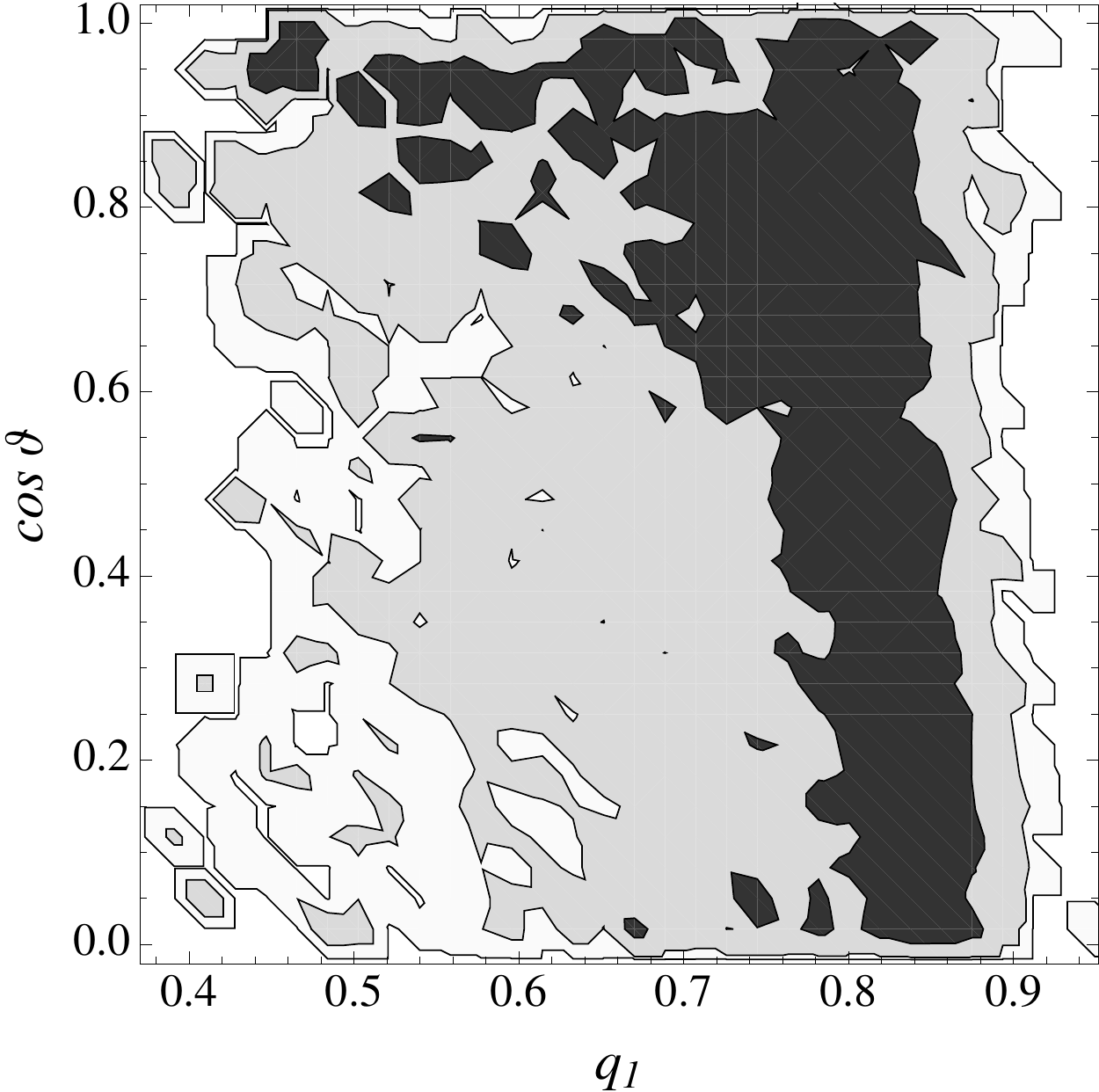} &
\includegraphics[width=4cm]{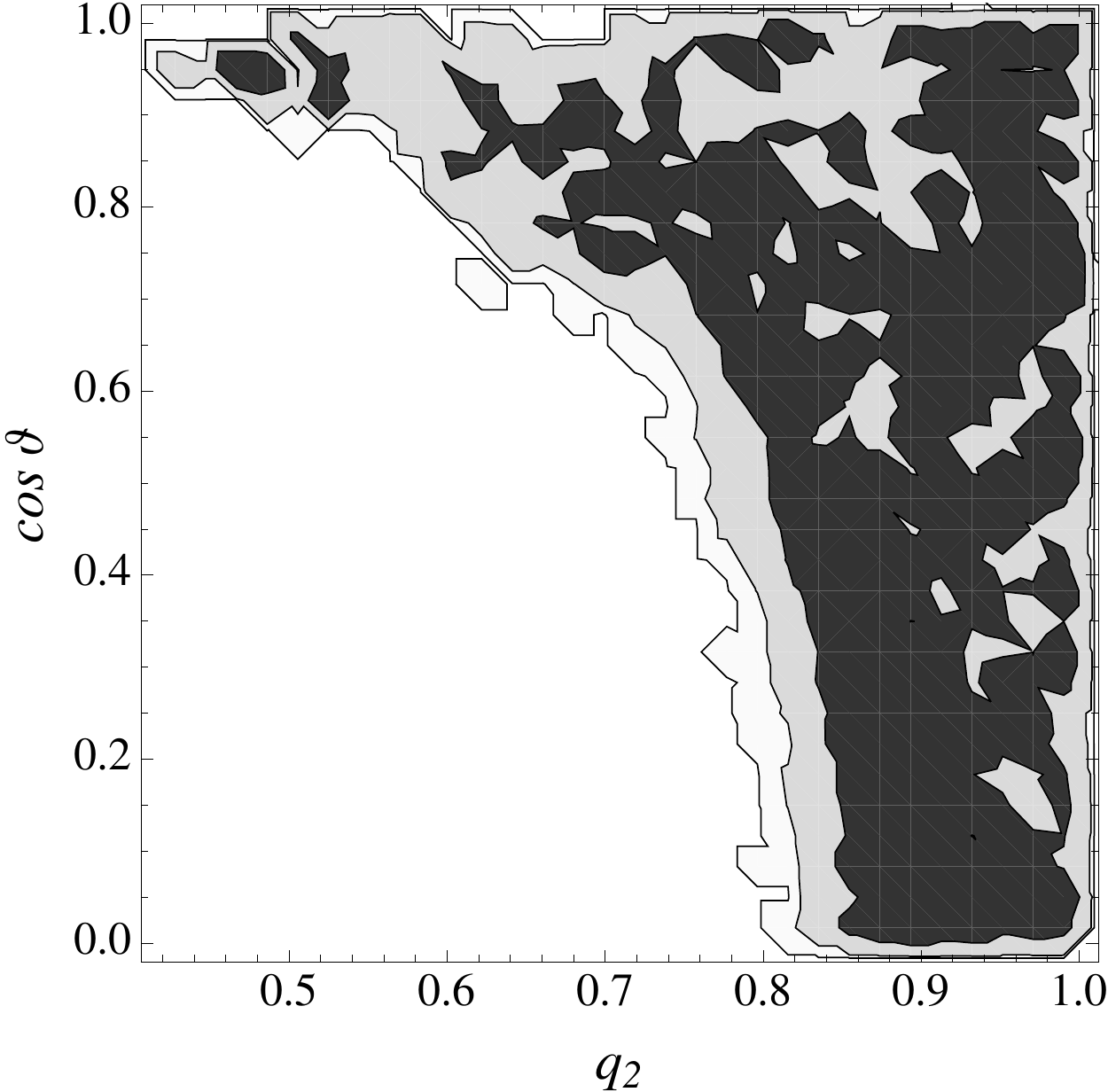} \\
\end{tabular}
$
\end{center}
\caption{Contour plots of the bi-dimensional marginalised PDFs derived under the prior assumptions of uniform $q$-distribution and random orientation angles. Contours are plotted at fraction values $\exp (-2.3/2)$, $\exp(-6.17/2)$, and $\exp(-11.8/2)$ of the maximum, which denote confidence limit regions of 1, 2 and $3\sigma$ in a maximum likelihood investigation, respectively. In the $M_{200}-c_{200}$ plane (top right panel), the thin and thick full line are the predictions from \citet{duf+al08} or \citet{pra+al11}, respectively. Dashed and long-dashed lines enclose the 1 and $3\sigma$ regions for the predicted conditional probability $c(M)$ relation of \citet{duf+al08}, respectively. In the $q_1-q_2$ plane, the thick full, long dashed and dashed lines limit the 1, 2 and $3\sigma$ confidence regions for the $N$-body like expected distributions.
}
\label{fig_pdf_All_flat_random_2D}
\end{figure*}

\begin{figure*}
\begin{center}
$
\begin{tabular}{cccc}
\noalign{\smallskip}
\includegraphics[width=4cm]{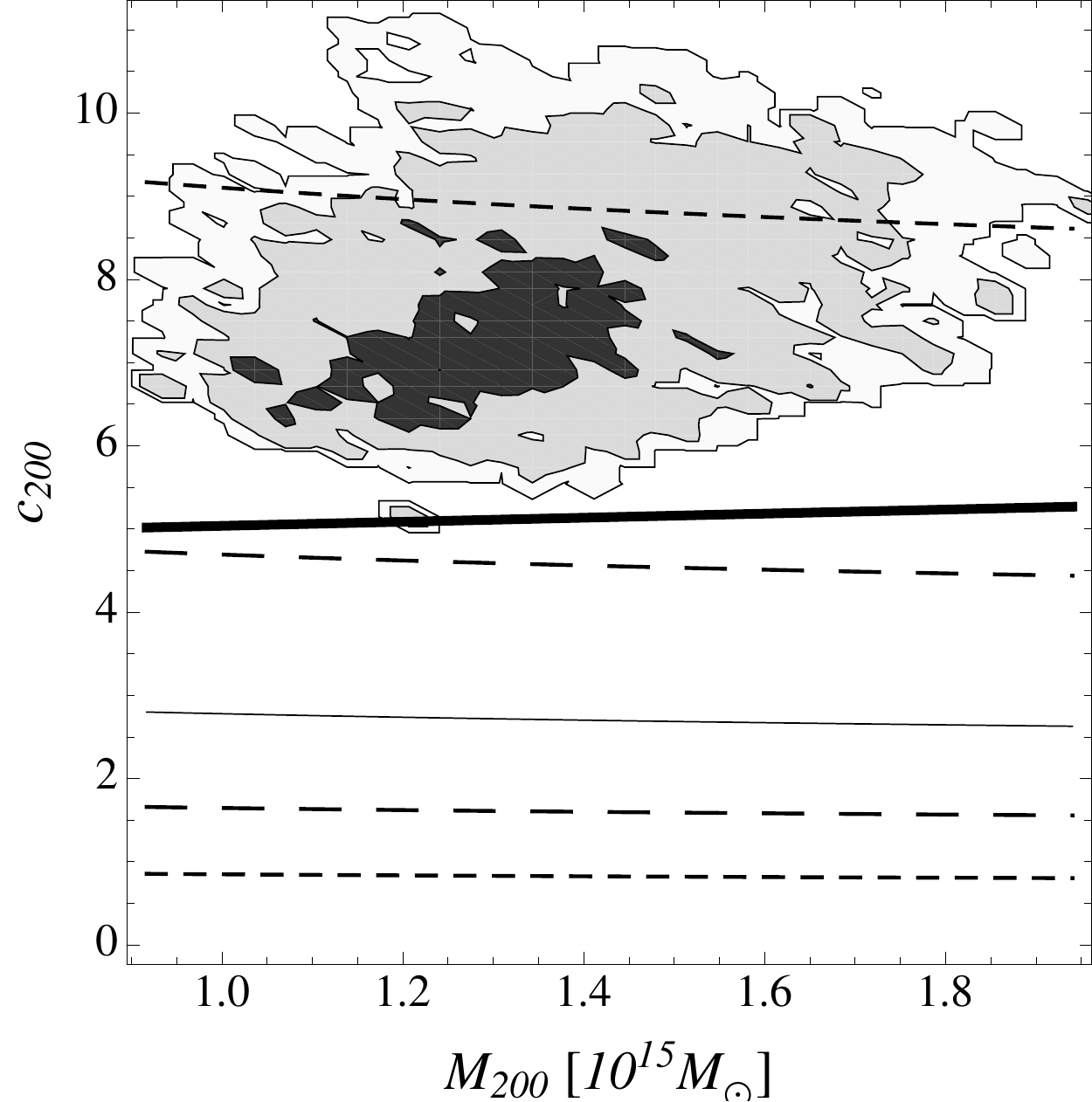} & & & \\
\includegraphics[width=4cm]{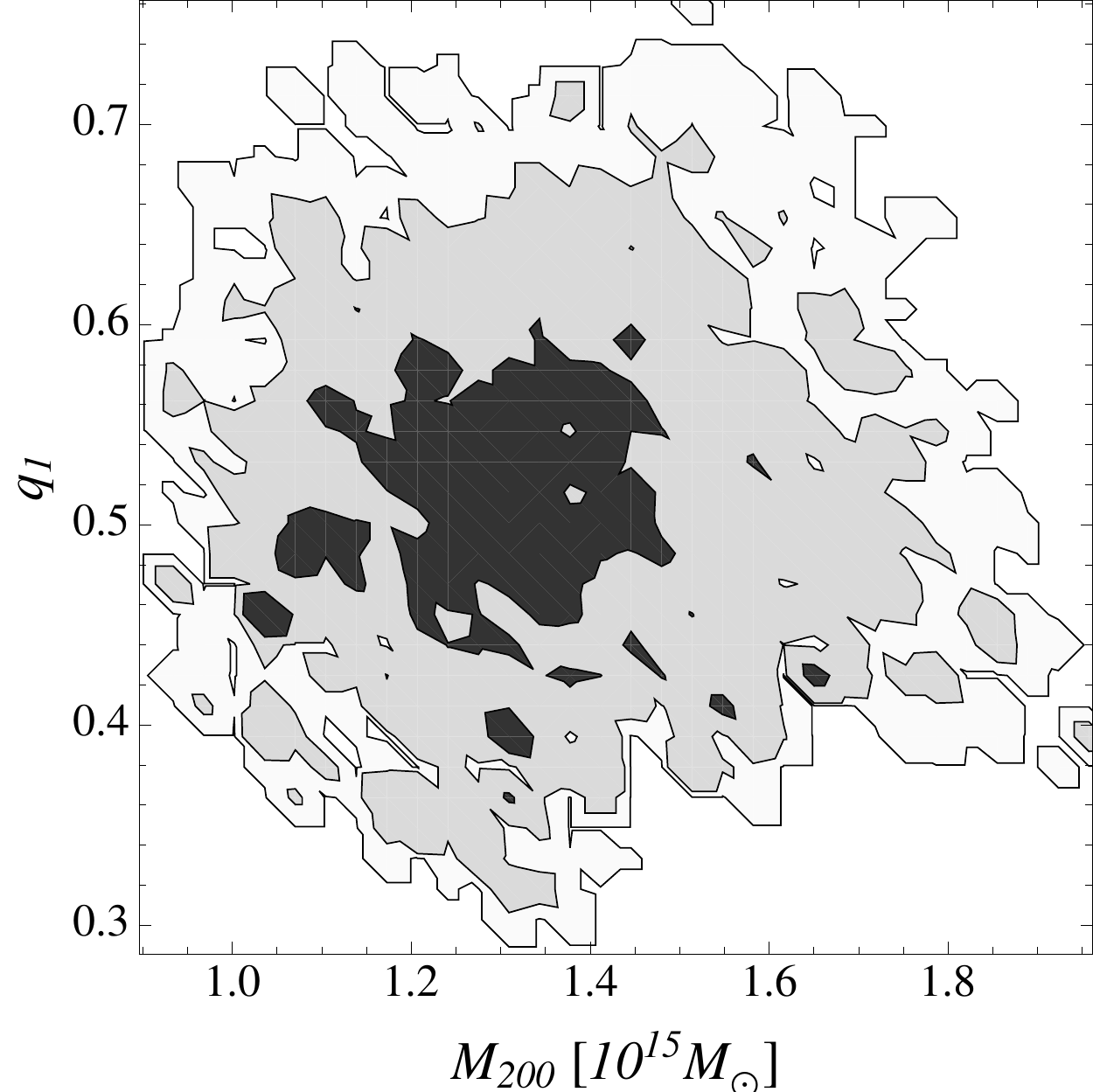} &
\includegraphics[width=4cm]{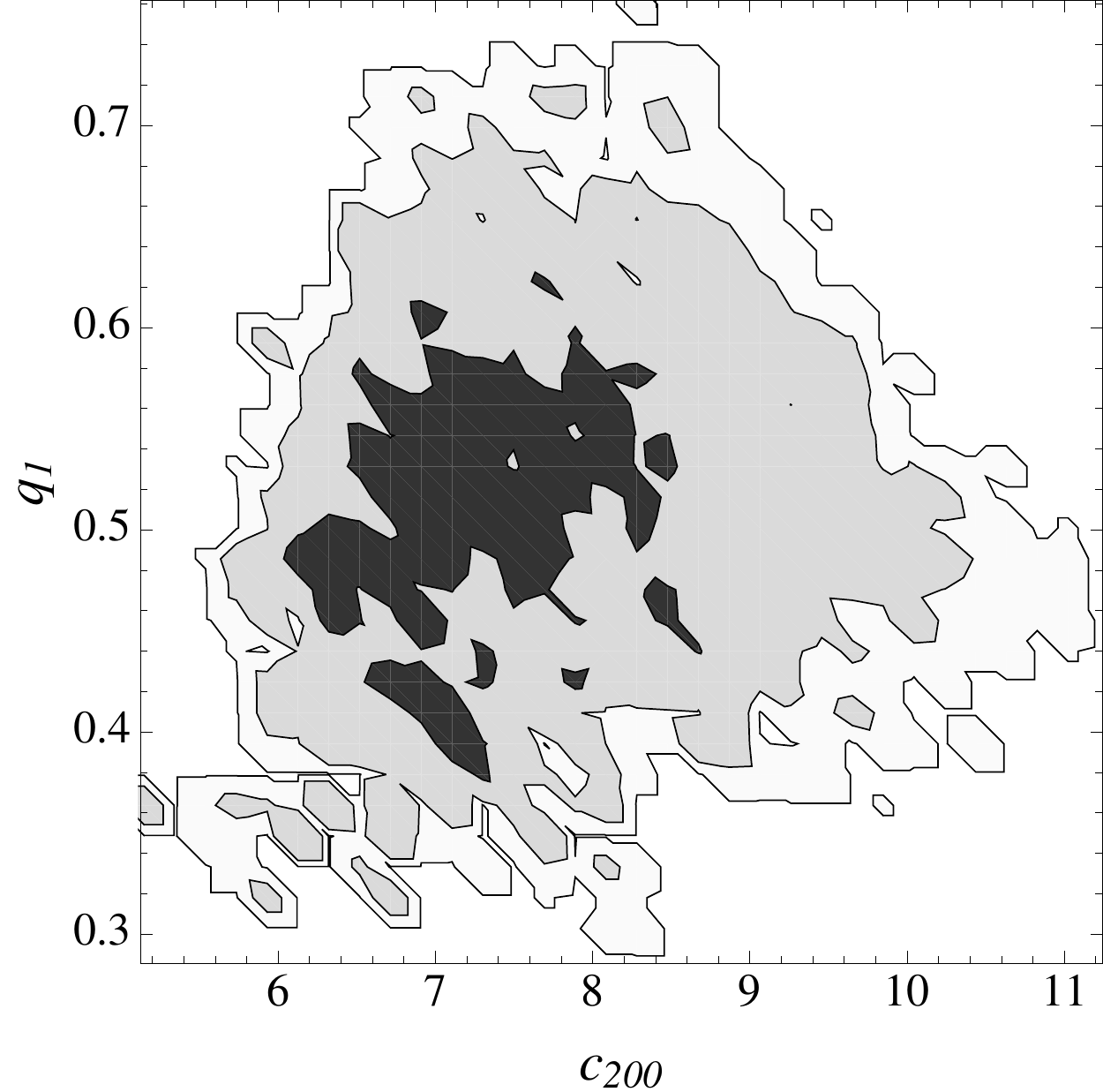} & & \\
\includegraphics[width=4cm]{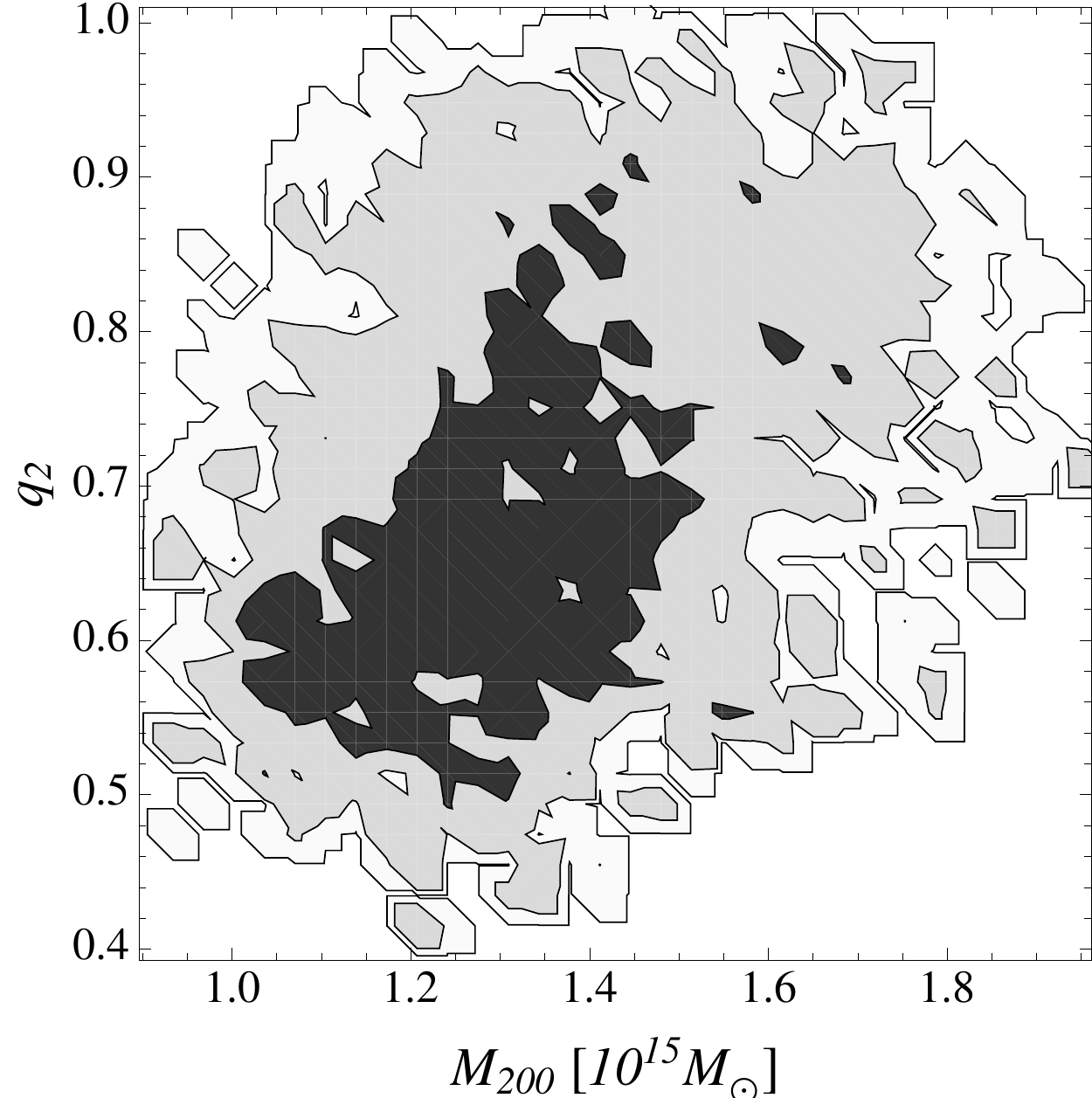} &
\includegraphics[width=4cm]{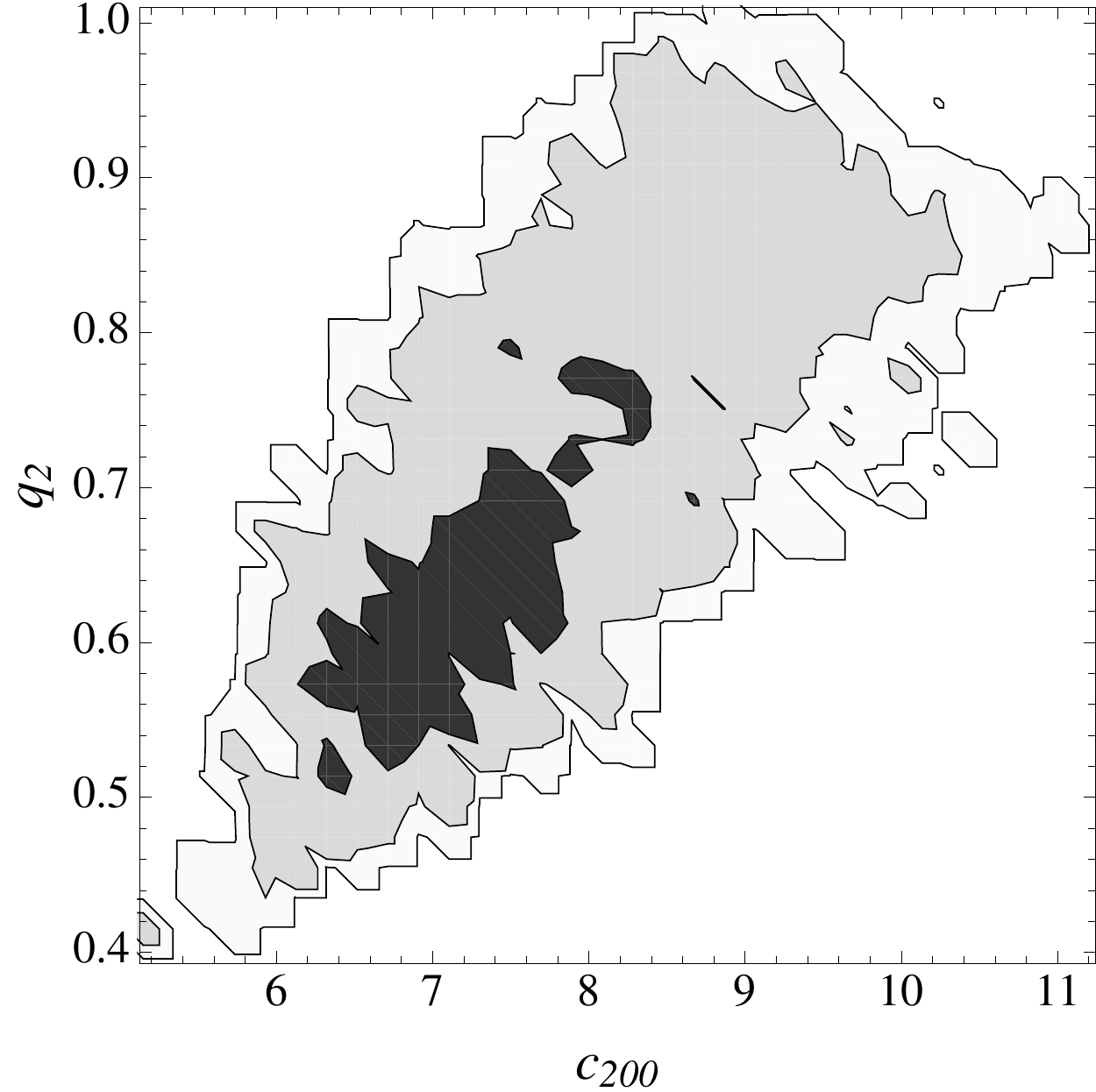} &
\includegraphics[width=4cm]{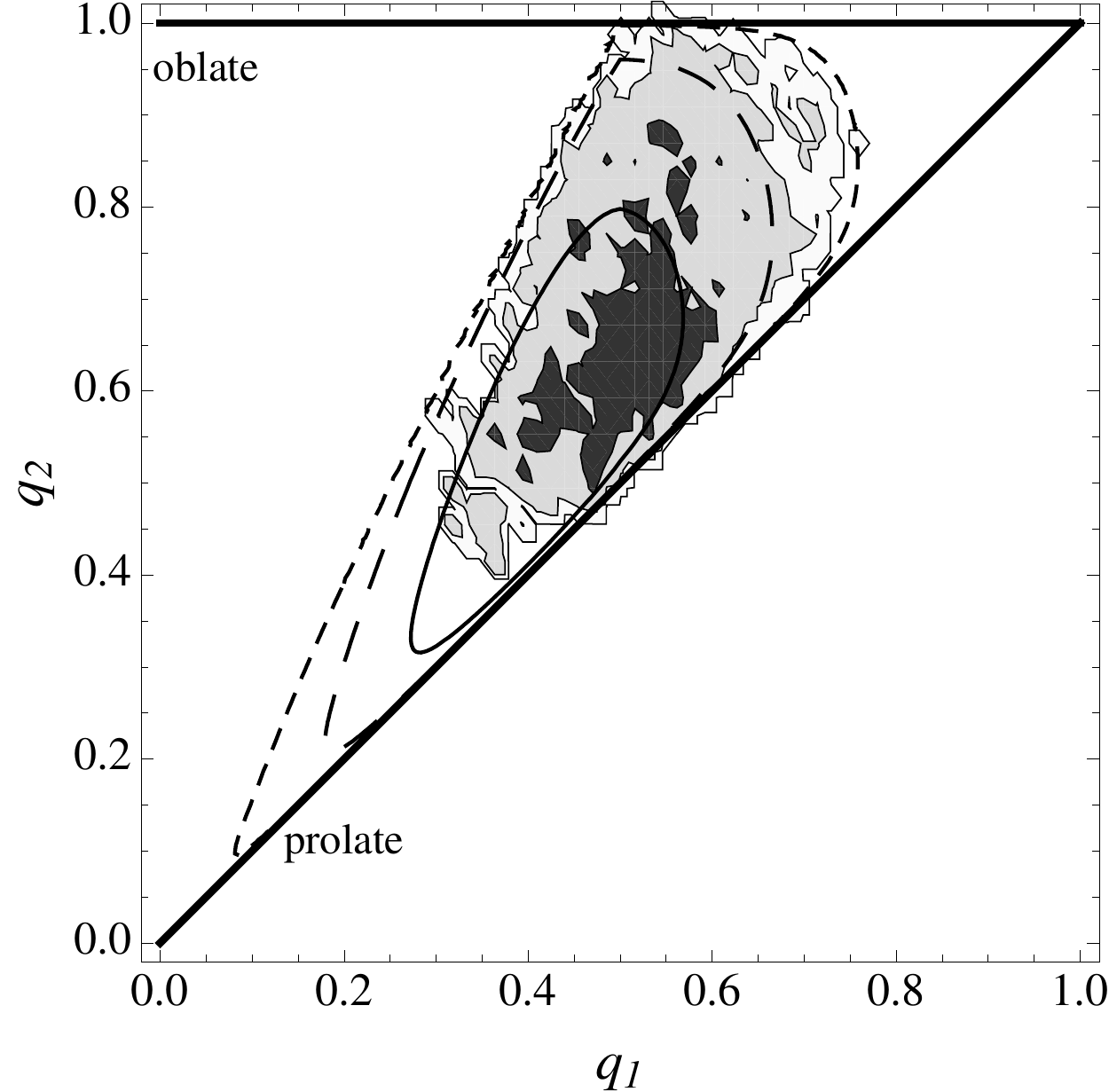} &  \\
\includegraphics[width=4cm]{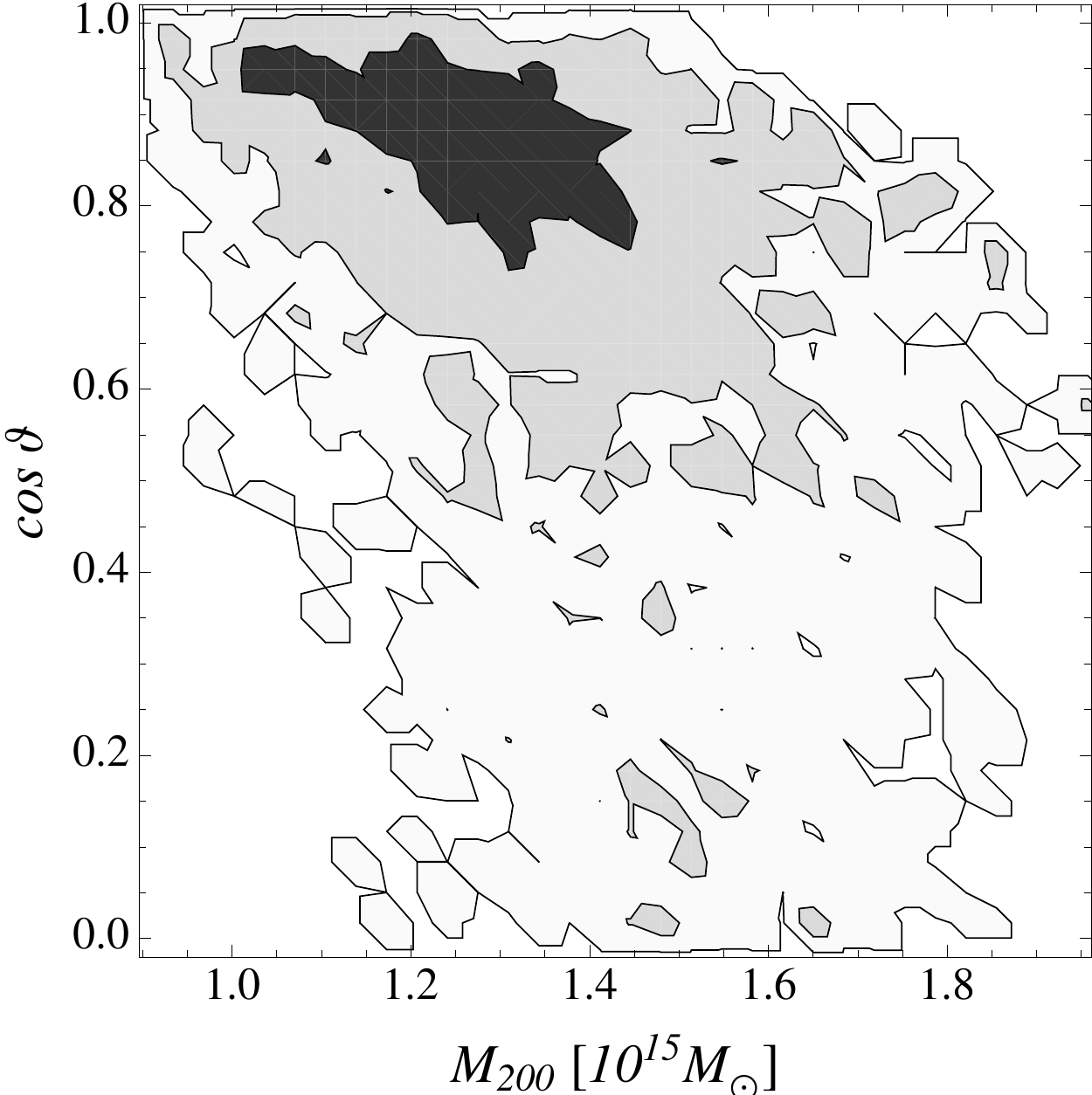} &
\includegraphics[width=4cm]{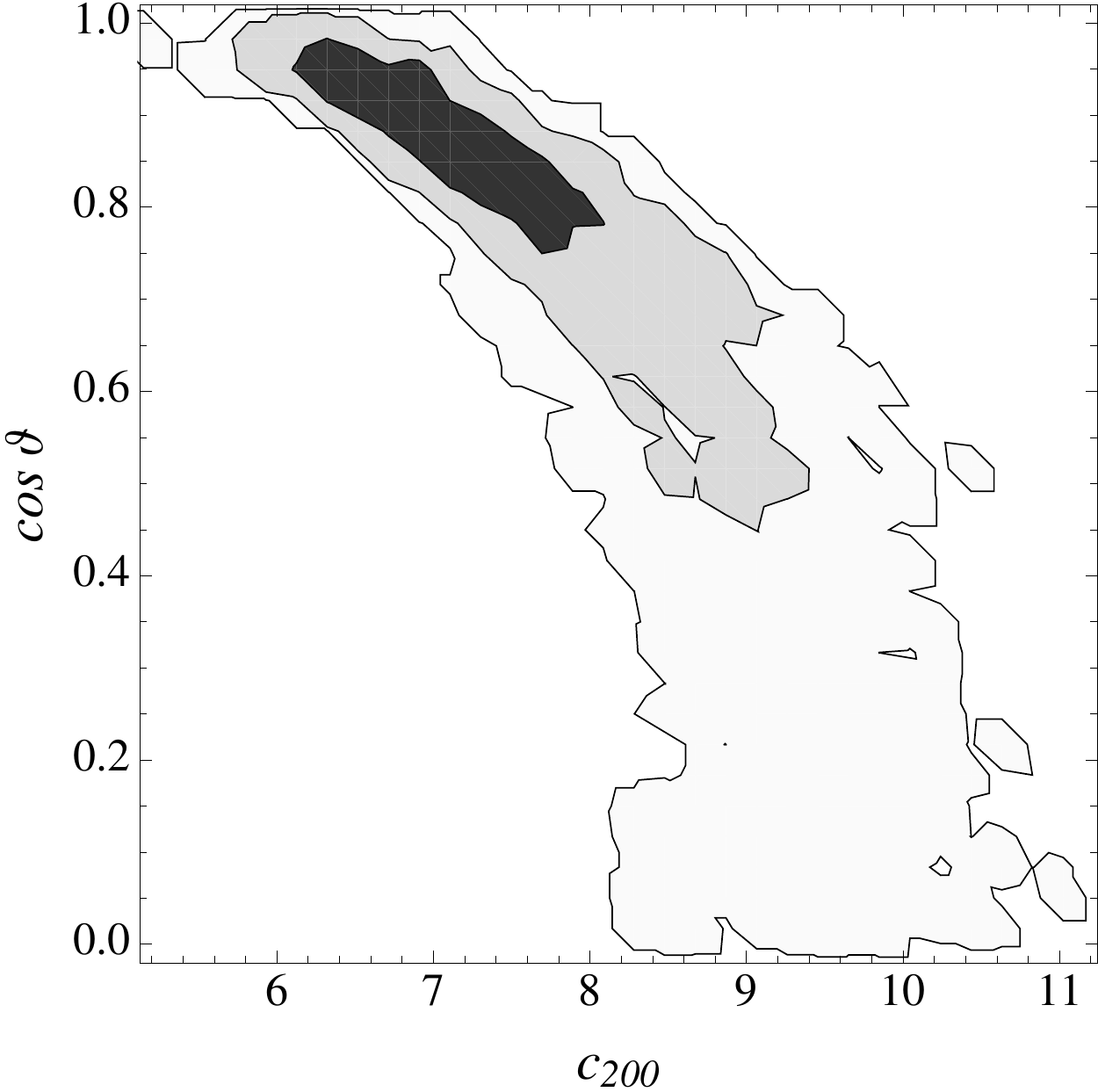} &
\includegraphics[width=4cm]{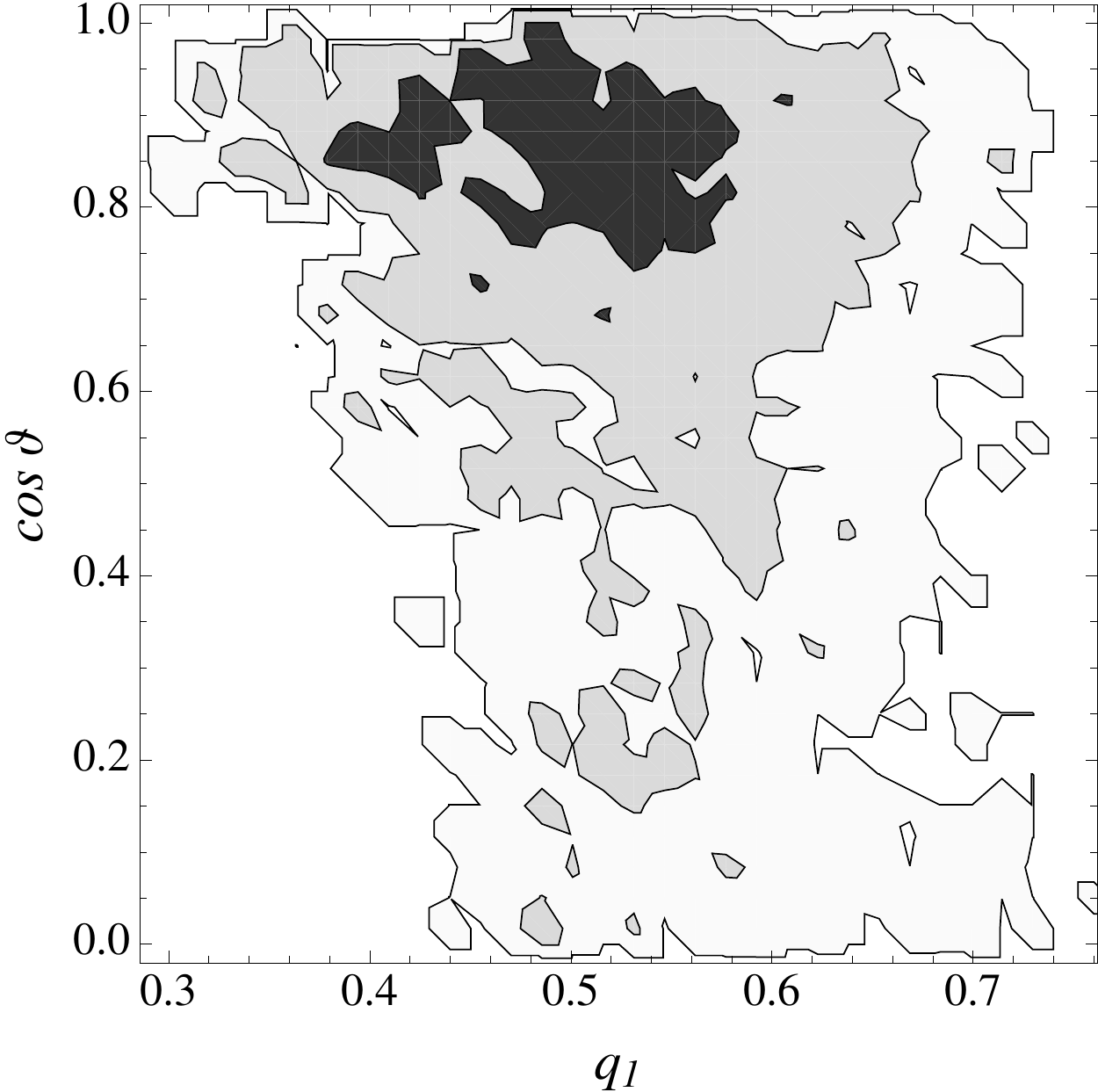} &
\includegraphics[width=4cm]{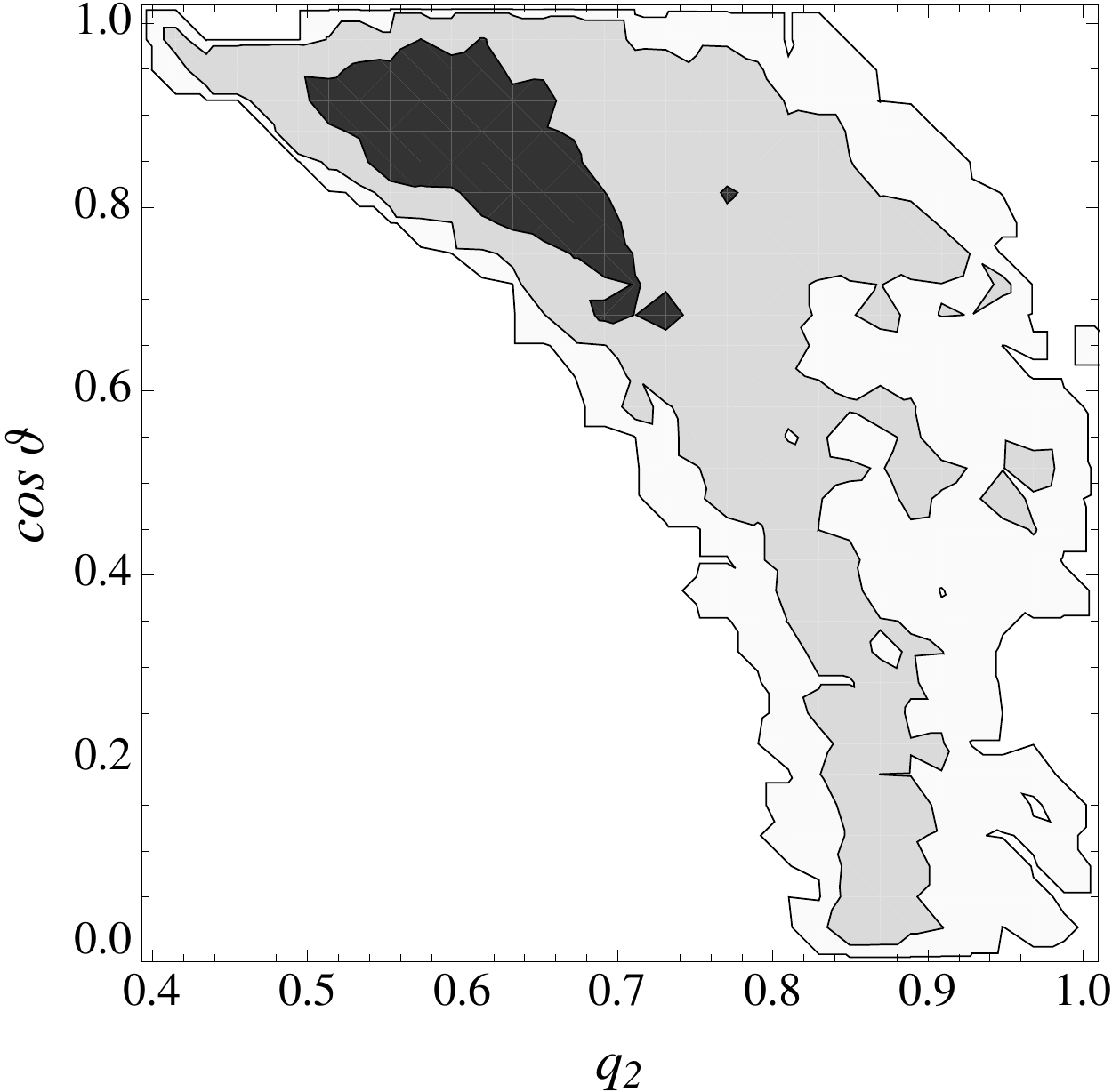} \\
\end{tabular}
$
\end{center}
\caption{Contour plots of the bi-dimensional marginalised PDFs derived under the prior assumptions of $N$-body like axial ratios and random orientation angles. Contours and lines are as in Fig.~\ref{fig_pdf_All_flat_random_2D}.
}
\label{fig_pdf_All_nbody_random_2D}
\end{figure*}

Trends and correlations retrieved in a triaxial analysis are predictable based on simple considerations. The efficiency of a halo as a lens is larger the larger either the mass or the concentration, all the rest being equal. The strength of a lens also increases with triaxiality and elongation along the line of sight, which boost convergence and cross section for lensing. A spherical analysis can not account for orientation and shape effects so that mass and concentration estimates are systematically biased higher if the halo is oriented along the line of sight. On the other hand a full triaxial analysis is not affected by such bias, and the search for the best values of mass and concentration is not artificially constrained in a biased niche of the parameter space.

The lensing analysis provides a reliable estimate of the projected total mass. The projected ellipticity can be estimated either for the total mass from lensing or the ICM from X-ray maps. Methods combining X-ray plus SZe constrain the elongation of the gas. Mass, concentration and shape and orientation parameters have then to be chosen such that they reproduce the total lensing efficiency measured by GL within the geometrical limits which come from the knowledge of the size of the ICM along the line of sight and in the plane of the sky, as well as the size of the total matter in the plane of the sky. 

The resulting counterbalancing effects can be seen in Figs.~\ref{fig_pdf_All_flat_random_2D} and~\ref{fig_pdf_All_nbody_random_2D}. Lesser values of mass and concentration are compatible with halos elongated along the line of sight ($\cos \vartheta \ls 1$), i.e., elongation supplies the lensing strength lost due to the decline in mass or concentration. Being the size in the plane of the sky fixed by observations, such elongation has to be fueled by a larger degree of triaxiality (lower values of $q_1$). To account for the X-ray plus SZe constraints, the gas has to be more triaxial too (larger values of $e^\mathrm{ICM}/e^\mathrm{Mat}$). 

The full triaxial analysis is required to quantify these trends and correlations. The results of our analysis are summarized in Table~\ref{tab_pdf_par}. They agree with our previous lensing \citep{se+um11} or X-ray plus SZe analysis \citep{ser+al12}. This is expected since we used the same data sets and similar methodology. There are two main improvements: {\it i}) intrinsic halo parameters are derived with a better accuracy; {\it ii}) we rely on a considerably smaller number of hypotheses. In particular, the X-ray plus SZe data give directly the information on the elongation of the halo and we do not have to assume any orientation bias to reconcile results with theoretical predictions. Likelihood functions combining X-ray plus SZe with lensing are more peaked than the corresponding functions exploiting only one data-set and final results are less sensitive to the a priori hypotheses. Results assuming either a flat distribution of axial ratios or inputs from $N$-body simulations are in very good agreement. Thanks to the information from X-ray plus SZe, which positively constrains the orientation, the alignment of the cluster is clearly seen whatever the priors.

The posterior distributions were investigated by running four Markov chains and checking for convergence. The marginalized 1D posterior probabilities are plotted in Fig.~\ref{fig_pdf_All_1D}, whereas the bidimensional probabilities are represented in Figs.~\ref{fig_pdf_All_flat_random_2D} and~\ref{fig_pdf_All_nbody_random_2D}. Results for mass and concentration are sensitive to the assumed priors to a very small extent. The multi-probe approach is dominated by the data, and in turn by the likelihood, whereas the priors play a negligible role.

As already touched in Sec.~\ref{sec_sl}, our SL and WL constraints on the projected NFW parameters are marginally consistent and it makes sense to combine them. We used lensing data-sets already exploited in previous works \citep{lim+al07,ume+bro08,coe+al10,ume+al11} and we found consistent results. In particular, we refer to \citet{se+um11} for a detailed discussion of the compatibility of WL and SL results for A1689 in a triaxial context. For completeness, we quote here the results for our multi probe approach exploiting just one lensing data-set. Exploiting only the WL, we found $M_{200}=(1.39\pm0.32) \times10^{15}M_{\odot}$ and $c_{200}=13.7\pm3.9$. SL alone favors less concentrated haloes, $c_{200}=4.8\pm0.8$, whereas the mass is poorly constrained.

\subsection{Mass and concentration}

The final results on mass and concentration are independent of the assumed priors. A1689 is a very massive cluster with high concentration. Comparisons with theoretical predictions point to an over-concentrated cluster, see Figs.~\ref{fig_pdf_All_1D},~\ref{fig_pdf_All_flat_random_2D} and \ref{fig_pdf_All_nbody_random_2D}. We considered the $c(M)$ relation from the analysis of a full sample of clusters in \citet{duf+al08}. The estimated concentration is in agreement with the tail at large values of the population of clusters of that given mass. Recently, \citet{pra+al11} claimed that the $c(M)$ relation features a flattening and upturn with increasing mass with substantially larger estimated concentrations for galaxy clusters. In that case, the agreement with theoretical predictions further improves.

The distribution of the ellipsoidal radius $r_{200}$ can be derived too. We found $r_{200}=3060 \pm 290~\mathrm{kpc}$ ($2530 \pm 280~\mathrm{kpc}$) for a priori $N$-body (uniform) axial ratios. The spherical  $r_{200}^\mathrm{Sph}$, which comprises the same over-density and mass of the ellipsoidal $r_{200}$ and can be more useful for comparison with works based on spherical analyses, is $2135 \pm 90~\mathrm{kpc}$ in both cases.

Concentration and orientation are strongly correlated, see Figs.~\ref{fig_pdf_All_flat_random_2D} and~\ref{fig_pdf_All_nbody_random_2D}. For $N$-body axial ratios, the concentration corresponding to orientation angles $40\deg \ls \vartheta \ls 45\deg$ is $c_{200}=8.2\pm0.4$. For $0\deg \ls \vartheta \ls 5\deg$, $c_{200} =6.2\pm0.3$. Even if the a posteriori probability distribution in the full parameter space peaks at lower concentrations and more pronounced alignments with the line of sight, the tail corresponding to larger values of the orientation angle shifts the peak of the marginalized distribution towards larger concentration values.

\subsection{Shape and orientation of the matter distribution}

We found evidence for a triaxial matter distribution. This claim comes mainly from the observed ellipticity of the total projected matter. The measured value of $\epsilon$ deviates from the spherical case, $\epsilon=0$, by 2- and 5-$\sigma$ in the WL or SL analysis, respectively, see Table~\ref{tab_fit_nfw_GL}. The case of a prolate/oblate ellipsoid with the symmetry axis along the line of sight is ruled out at the same confidence level.

When assuming a priori flat distribution, the posterior probability of $q_1$ peaks at $\sim 0.8$ with a tail in correspondence of more triaxial shapes which brings the mean value at $q_1\sim 0.7$. The large tail at small values makes the results fully consistent with theoretical predictions. When assuming the prior from $N$-body simulations, the posterior probability of $q_1$ follows the prior, but with a marginal shift towards rounder values and a smaller dispersion. 

The intermediate to major axis ratio is less constrained. The prior plays an heavier role for the final distribution of $q_2$, even if values of $q_2 \sim 0.7$--$0.9$ perform well under different assumptions. Prolate configurations ($q_1=q_2$) are slightly preferred over oblate ones ($q_2 =1$), but triaxial shapes give much better fits than axially symmetric haloes.

The halo turns out to be elongated along the line of sight. Biased orientations are favoured even if a priori the orientations were random. The a priori probability of a randomly oriented cluster to have $\vartheta < 45\deg$ is $\sim 29$ per cent. A posteriori such probability is $\sim 75$ (42) per cent assuming $N$-body like (flat) axial ratios. This result could be obtained only combining the lensing analysis with the information from X-ray and SZe. The pure lensing analysis in \citet{se+um11} had to assume a priori a biased orientation to get similar results on mass and concentration. Orientations in the plane of the sky or intermediate inclinations are compatible with less triaxial shapes and more massive and concentrated haloes, but their statistical significance is lower than for alignments along the line of sight.

\subsection{Shape of the gas distribution}

\begin{figure*}
\begin{center}
$
\begin{tabular}{ccc}
\multicolumn{3}{c}{Flat $q$-distribution}	\\
\noalign{\smallskip}
\includegraphics[width=3.4cm]{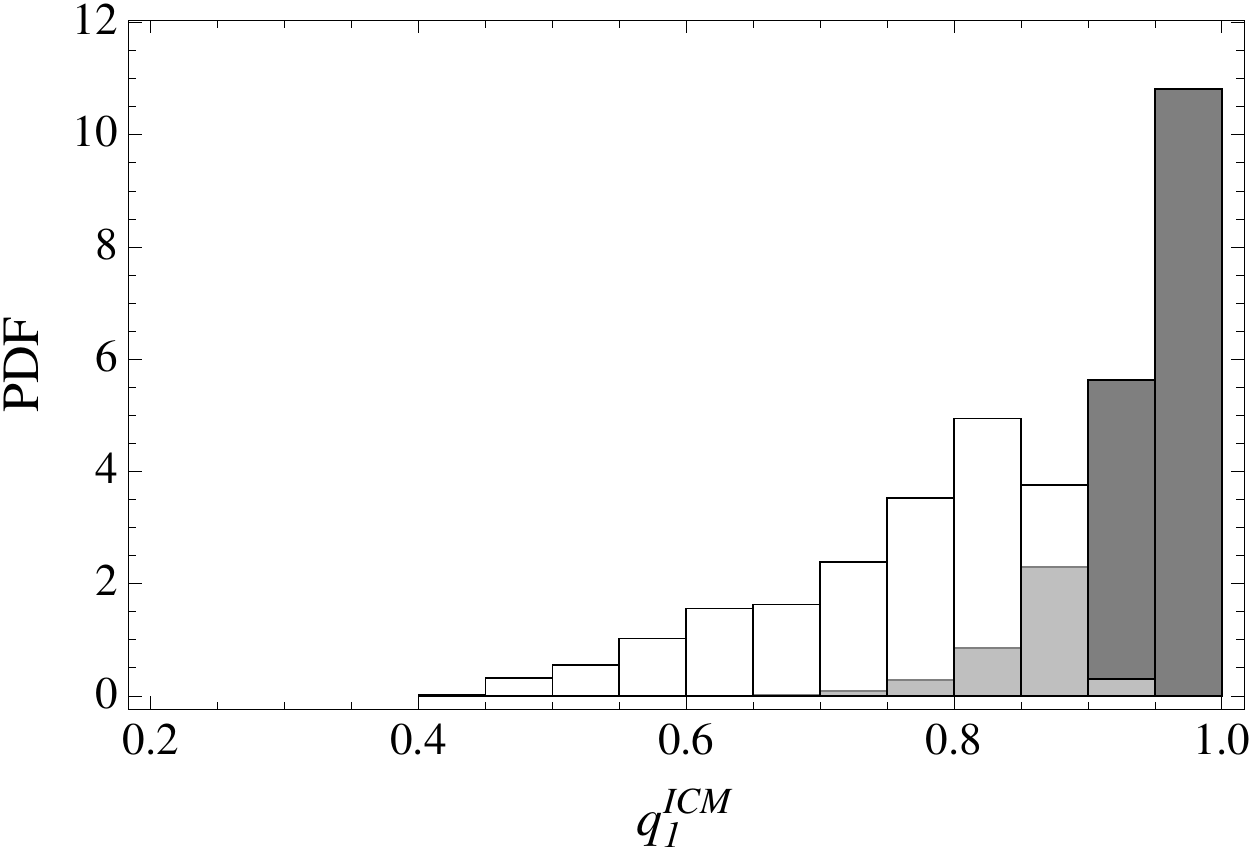} &
\includegraphics[width=3.4cm]{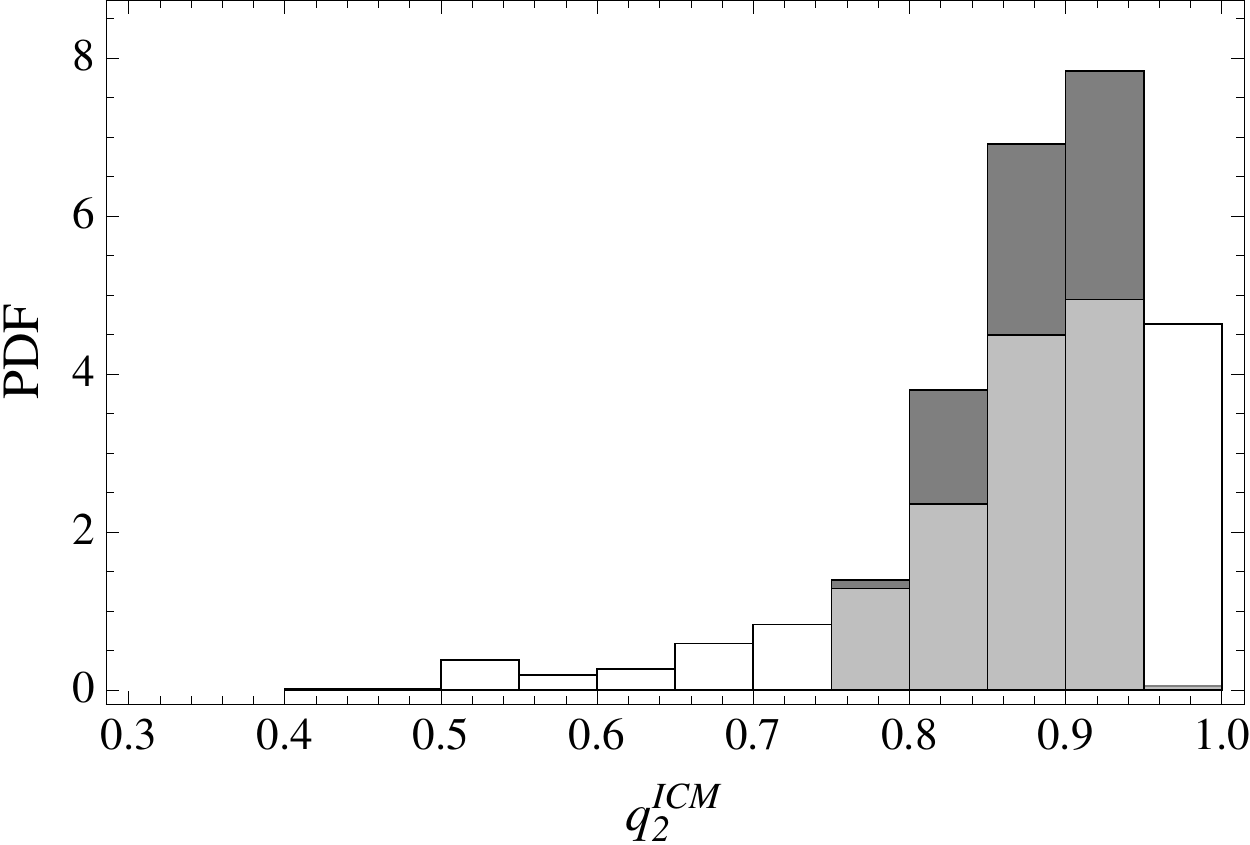} &
\includegraphics[width=3.4cm]{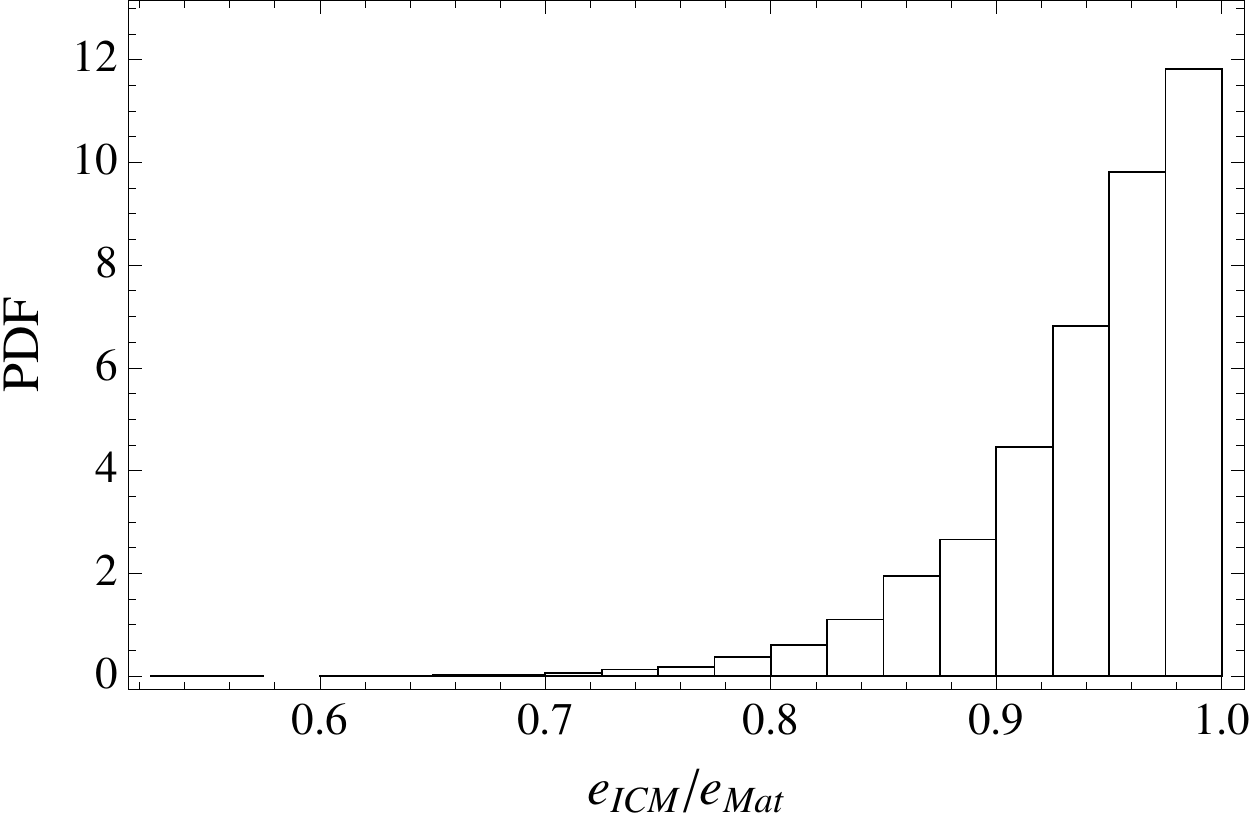}  \\	
\noalign{\smallskip}
\multicolumn{3}{c}{$N$-body $q$-distribution} \\
\noalign{\smallskip}
\includegraphics[width=3.4cm]{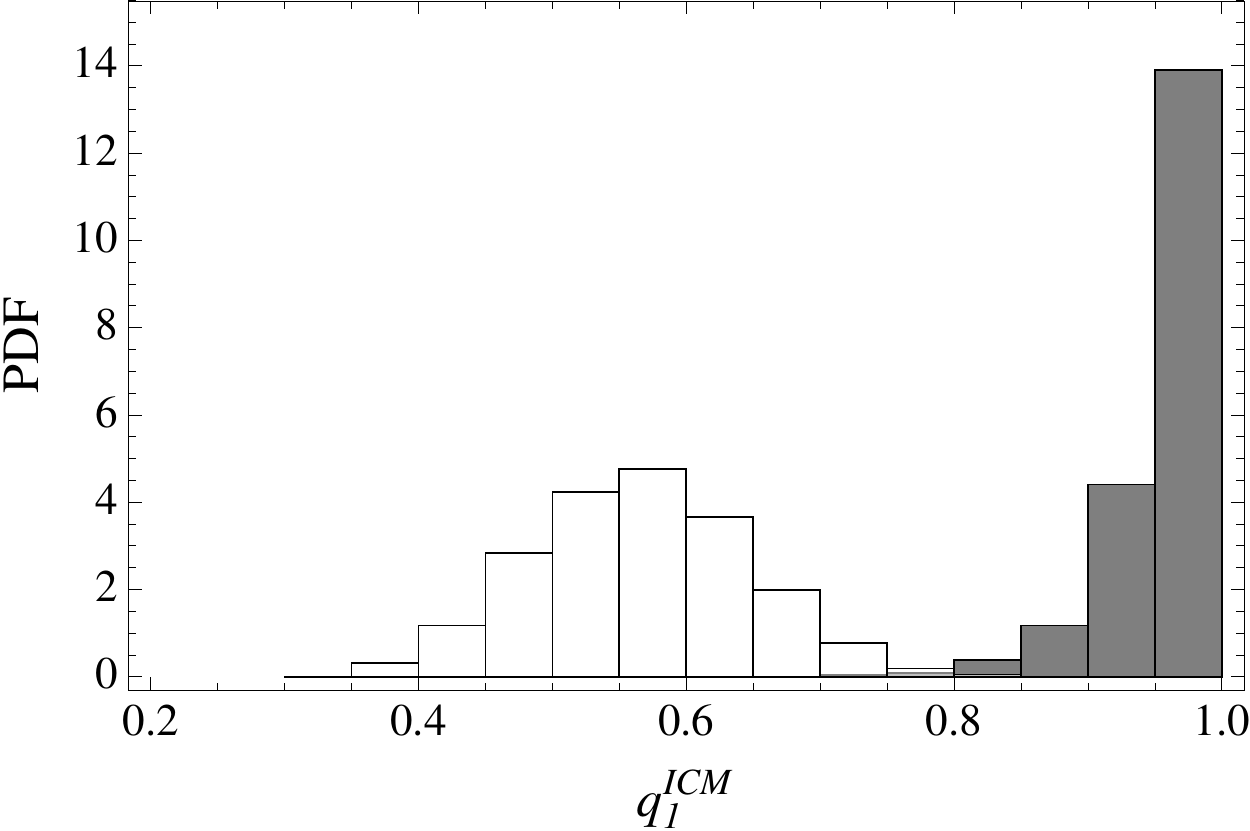} &
\includegraphics[width=3.4cm]{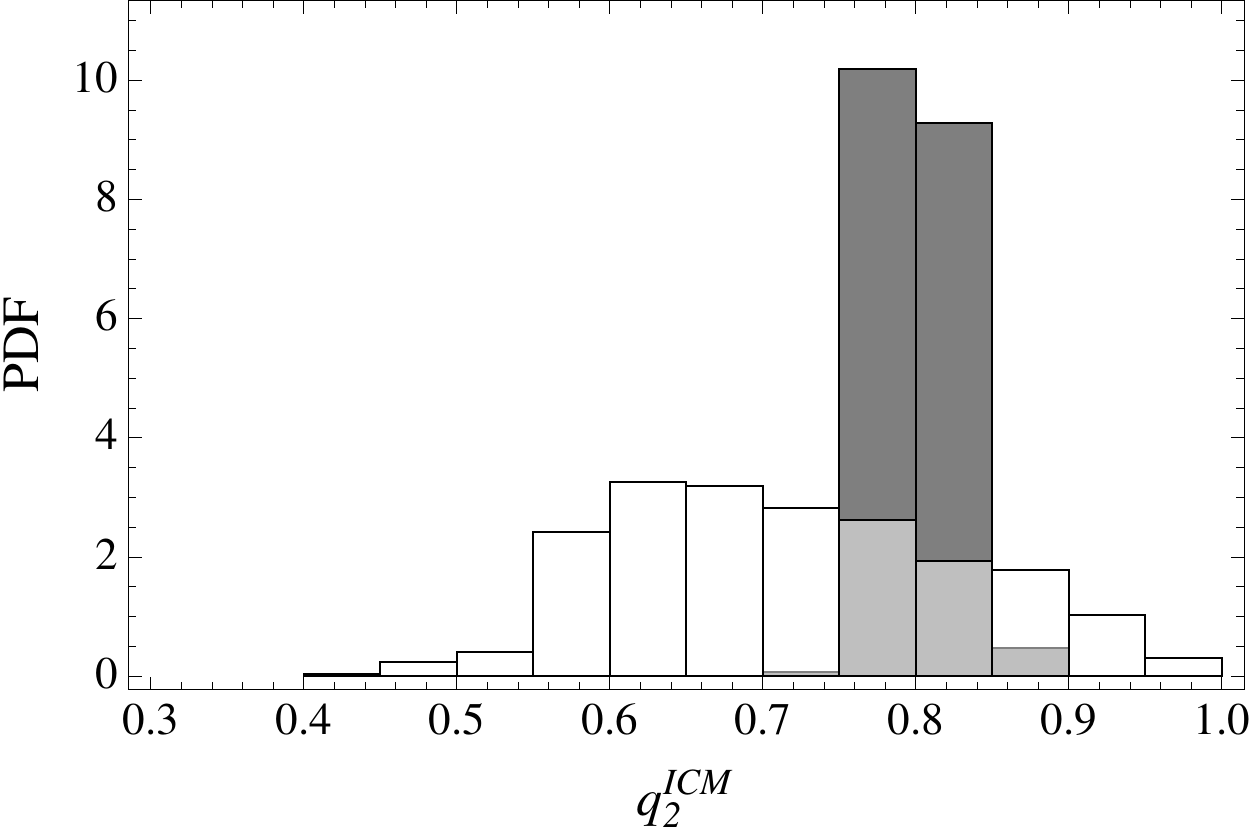} &
\includegraphics[width=3.4cm]{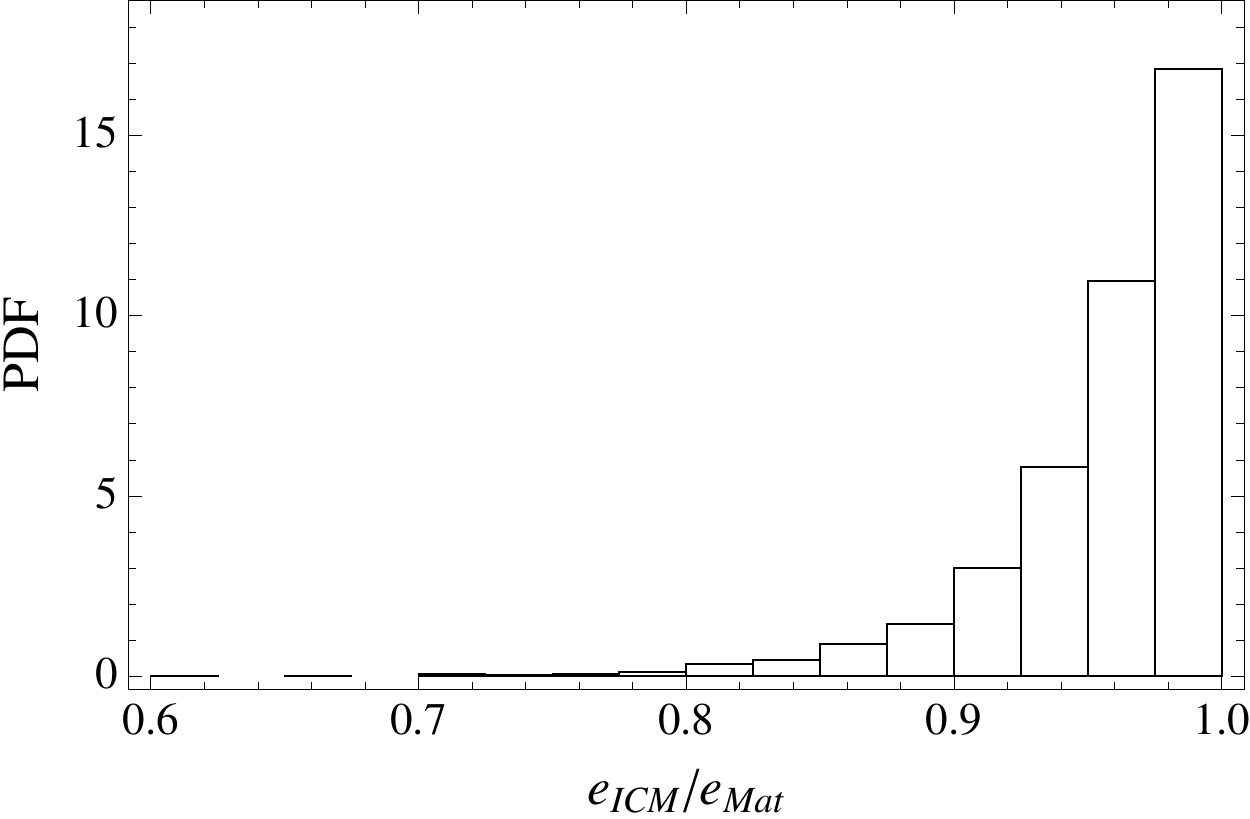}  \\	
\end{tabular}
$
\end{center}
\caption{Marginalized PDFs (plotted as white histograms) for the shape parameters of the gas distribution. Panels from the left to the right are for the axial ratios $q^\mathrm{ICM}_1$, $q^\mathrm{ICM}_2$ and for $e^\mathrm{ICM}/e^\mathrm{Mat}$, respectively. The top (bottom) row refers to uniform ($N$-body like) priors for the matter axial ratios. The grey histograms represent the expected distributions of the axial ratios for a gas in hydrostatic equilibrium under the derived mass density, i.e, by assuming $e^\mathrm{ICM}/e^\mathrm{Mat}=0.7$.}
\label{fig_pdf_All_ICM_1D}
\end{figure*}

\begin{table}
\centering
\begin{tabular}{c r@{$\,\pm\,$}lr@{$\,\pm\,$}lr@{$\,\pm\,$}l}
        \hline
        \noalign{\smallskip}
	Priors &  \multicolumn{2}{c}{$q_1^\mathrm{ICM}$}	& \multicolumn{2}{c}{$q_2^\mathrm{ICM}$}	& \multicolumn{2}{c}{$e^\mathrm{ICM}/e^\mathrm{Mat}$}	 \\
        \noalign{\smallskip}
        \hline
         \noalign{\smallskip}
	flat		&$0.76$	&$0.10$	&$0.87$	&$0.10$	&$0.94$	&$0.05$	 \\
	$N$-body	&$0.56$	&$0.08$	&$0.72$	&$0.11$	&$0.96$	&$0.04$	 \\
\hline
\end{tabular}
\caption{Inferred  intrinsic parameters for the shape of the gas distribution (axis ratios $q_1^\mathrm{ICM}$ and $q_2^\mathrm{ICM}$) and relation with the total matter distribution ($e^\mathrm{ICM}/e^\mathrm{Mat}$). $q_2^\mathrm{ICM}$ and $e^\mathrm{ICM}/e^\mathrm{Mat}$ are derived parameters. Central values and dispersions are the mean and the standard deviation of the PDF, respectively.}
\label{tab_pdf_par_icm}
\end{table}

\begin{figure*}
\begin{center}
$
\begin{tabular}{c@{\hspace{.1cm}}c@{\hspace{.1cm}}c@{\hspace{.1cm}}c@{\hspace{.1cm}}c}
\multicolumn{5}{c}{Flat $q$-distribution}	\\
\noalign{\smallskip}
\includegraphics[width=3.4cm]{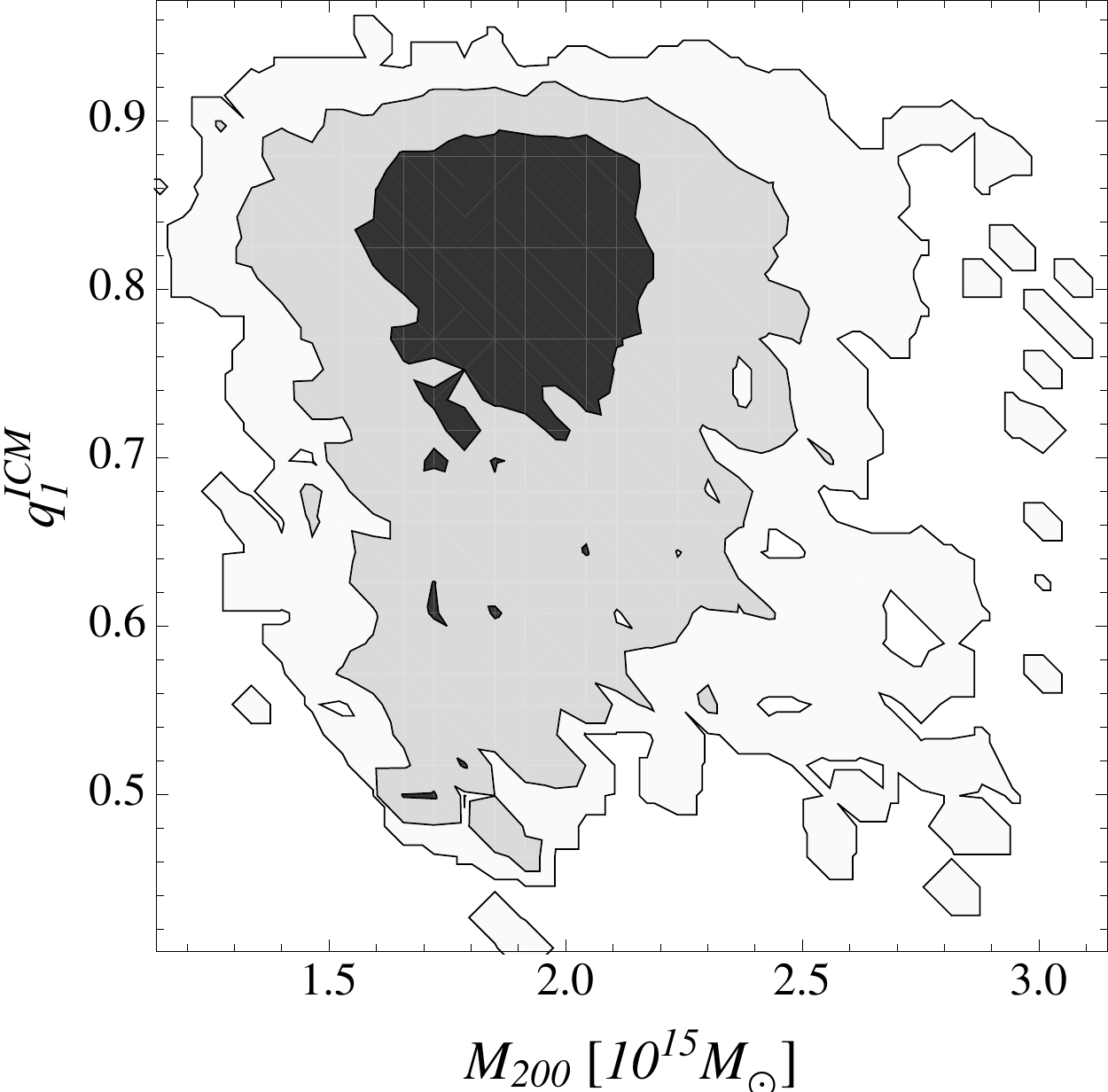} &
\includegraphics[width=3.4cm]{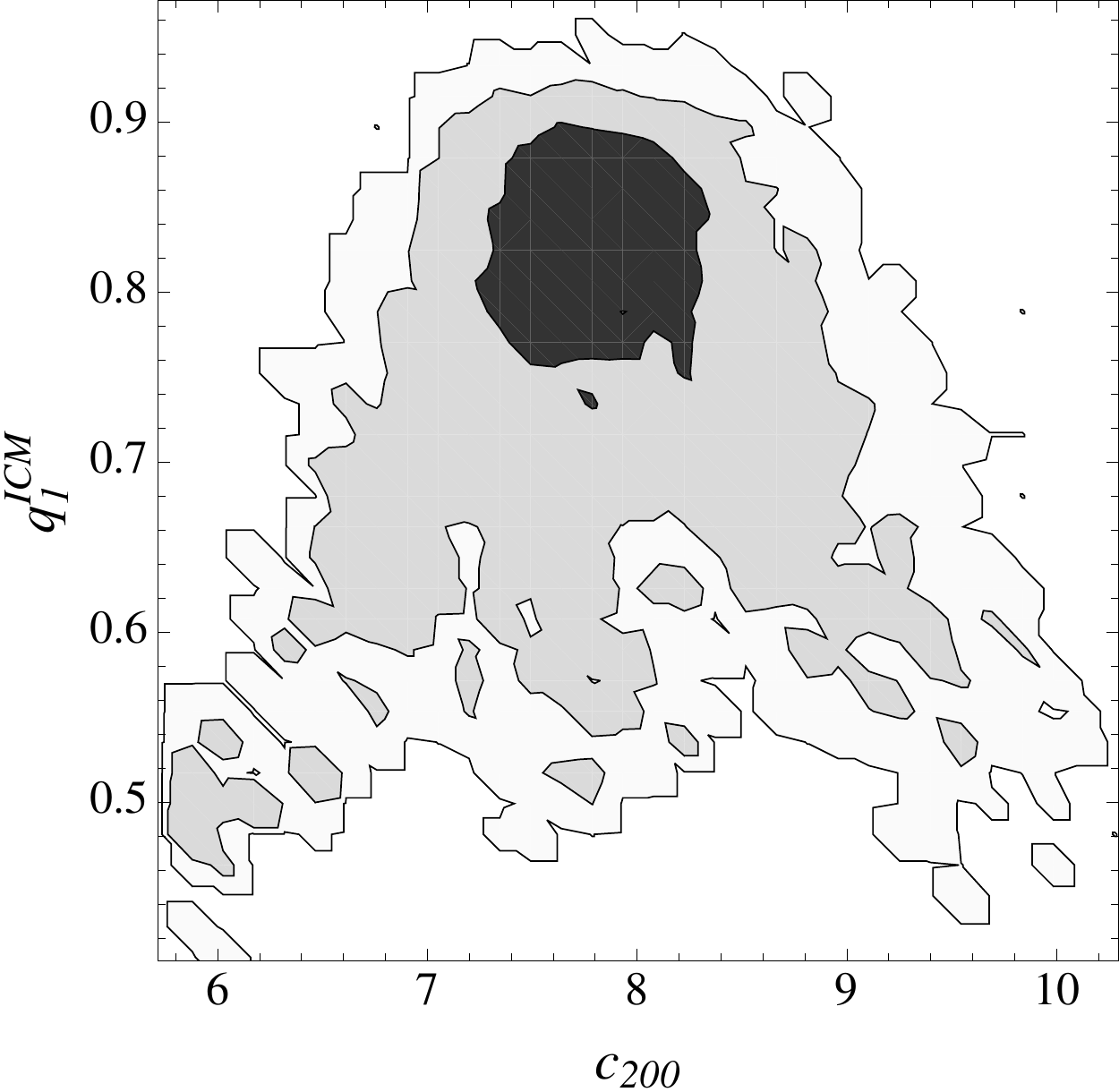} &
\includegraphics[width=3.4cm]{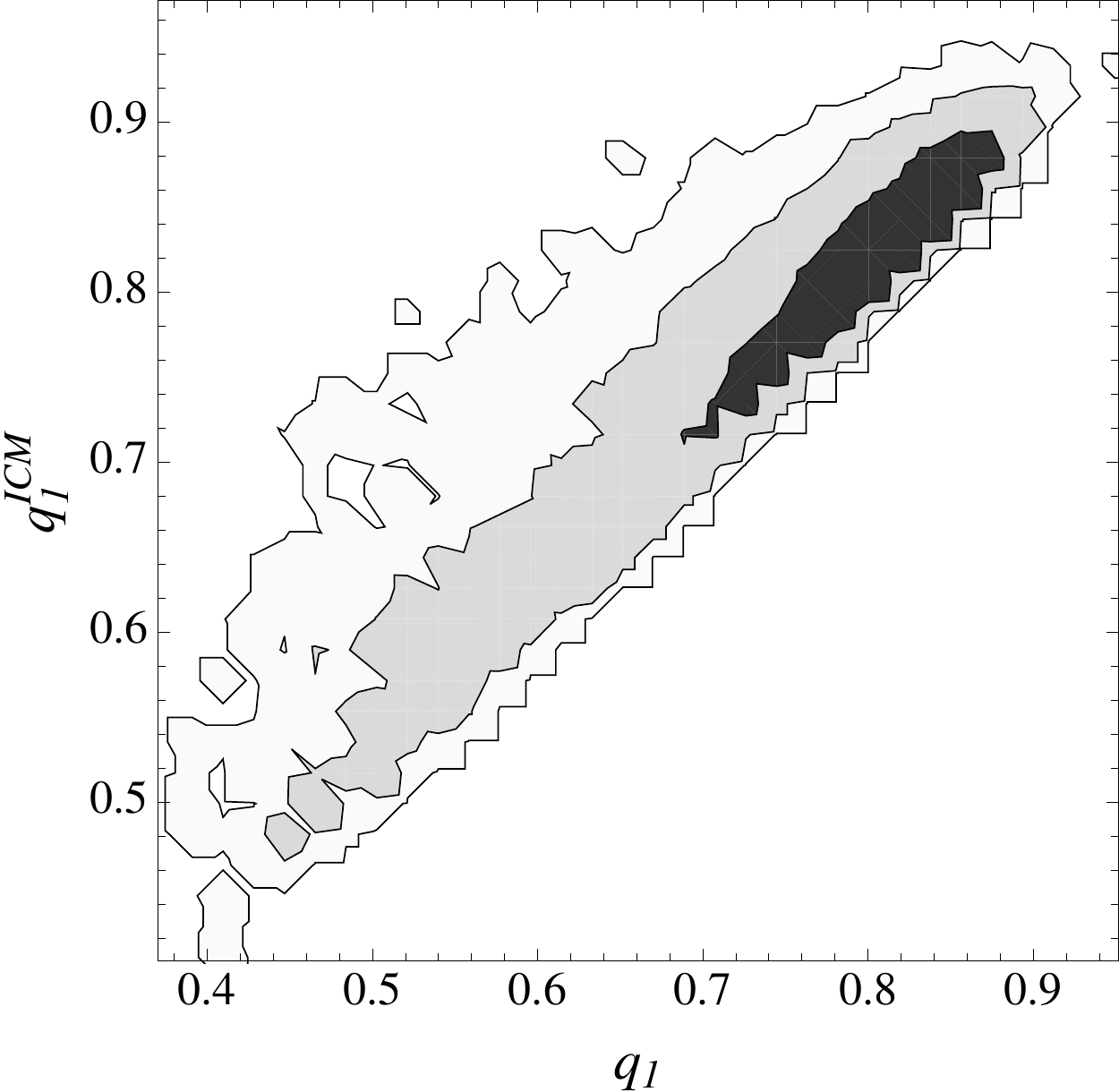} &
\includegraphics[width=3.4cm]{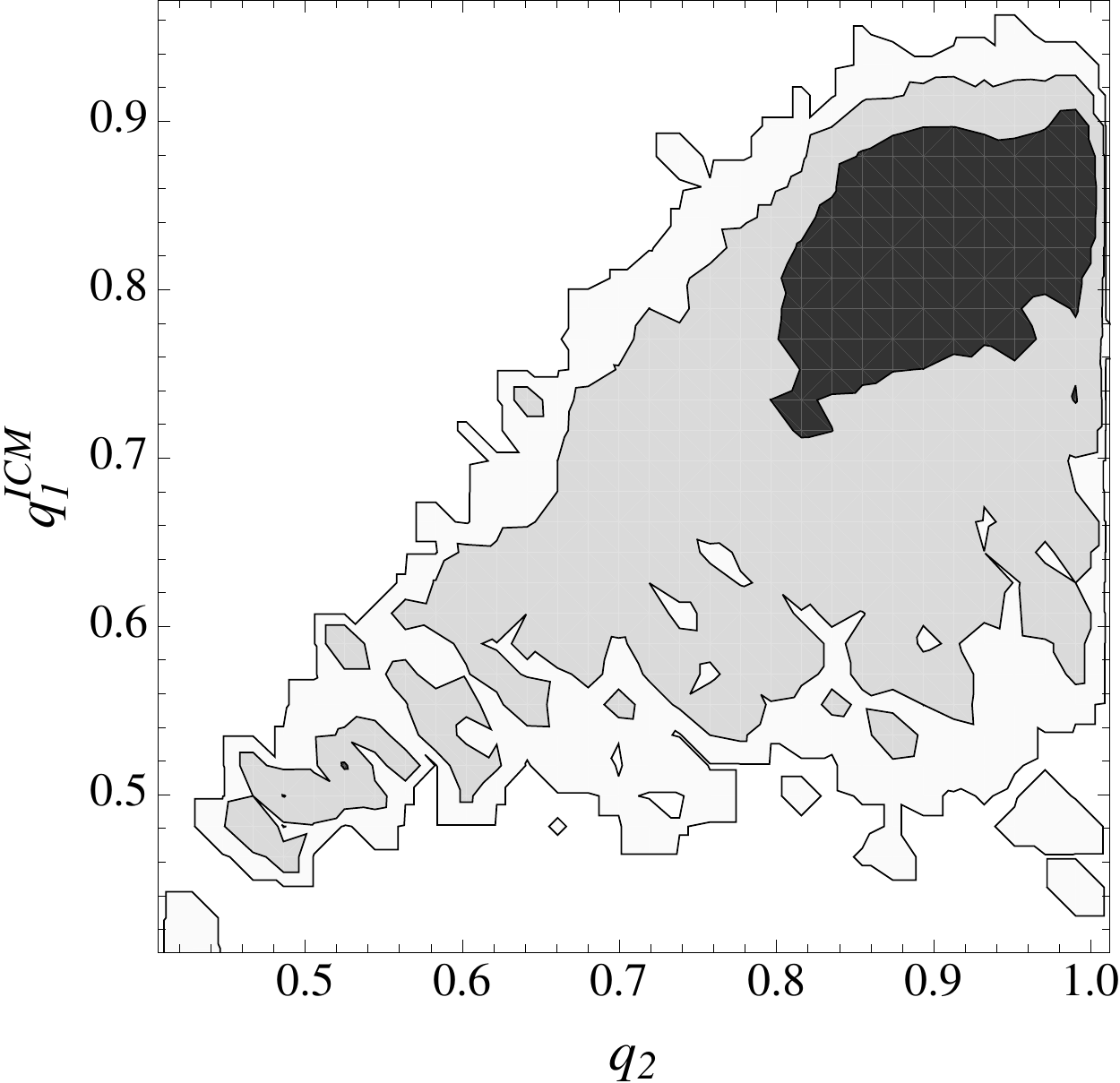} &
\includegraphics[width=3.4cm]{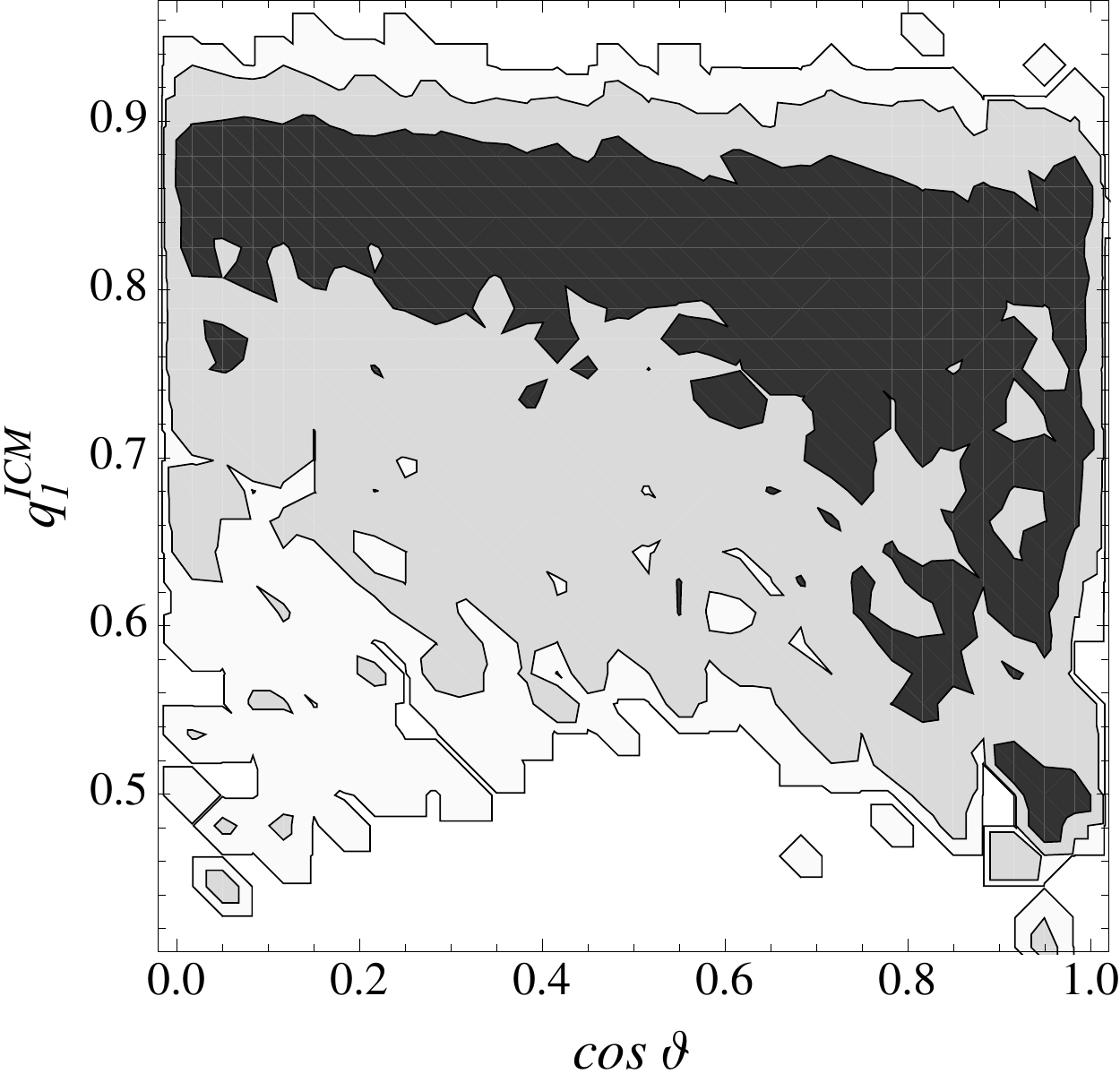}  \\	
\noalign{\smallskip}
\multicolumn{5}{c}{$N$-body $q$-distribution} \\
\noalign{\smallskip}
\includegraphics[width=3.4cm]{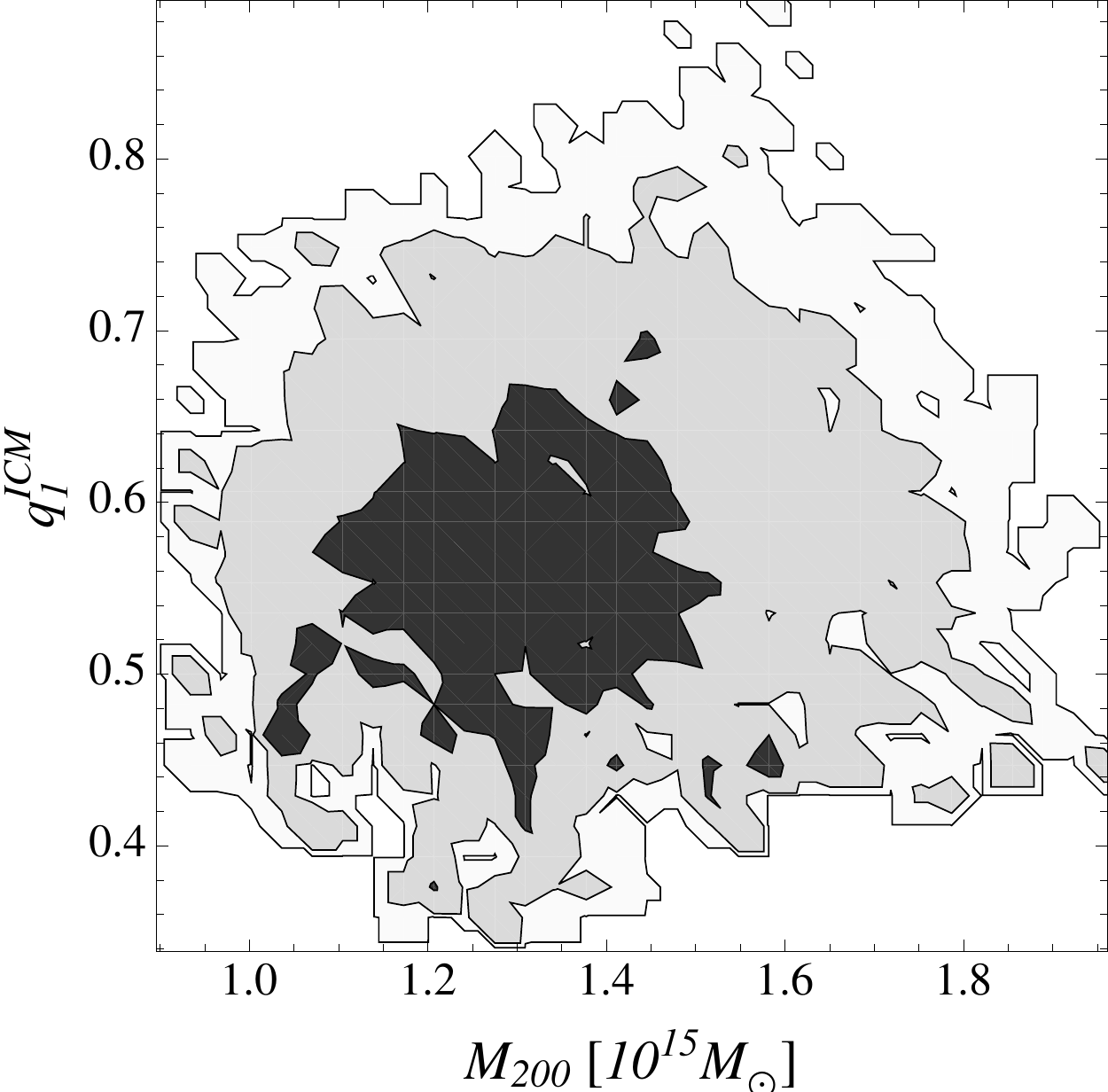} &
\includegraphics[width=3.4cm]{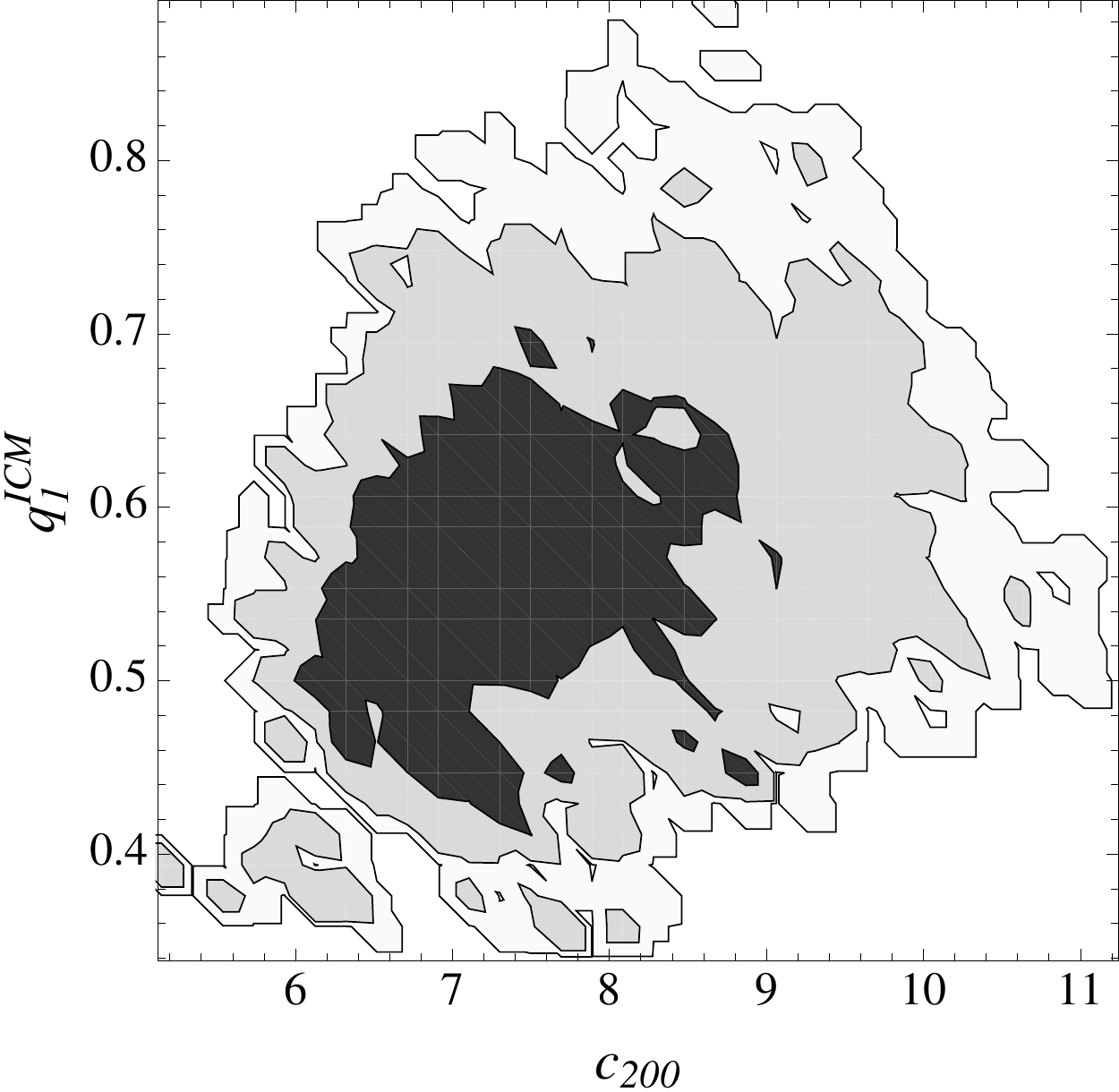} &
\includegraphics[width=3.4cm]{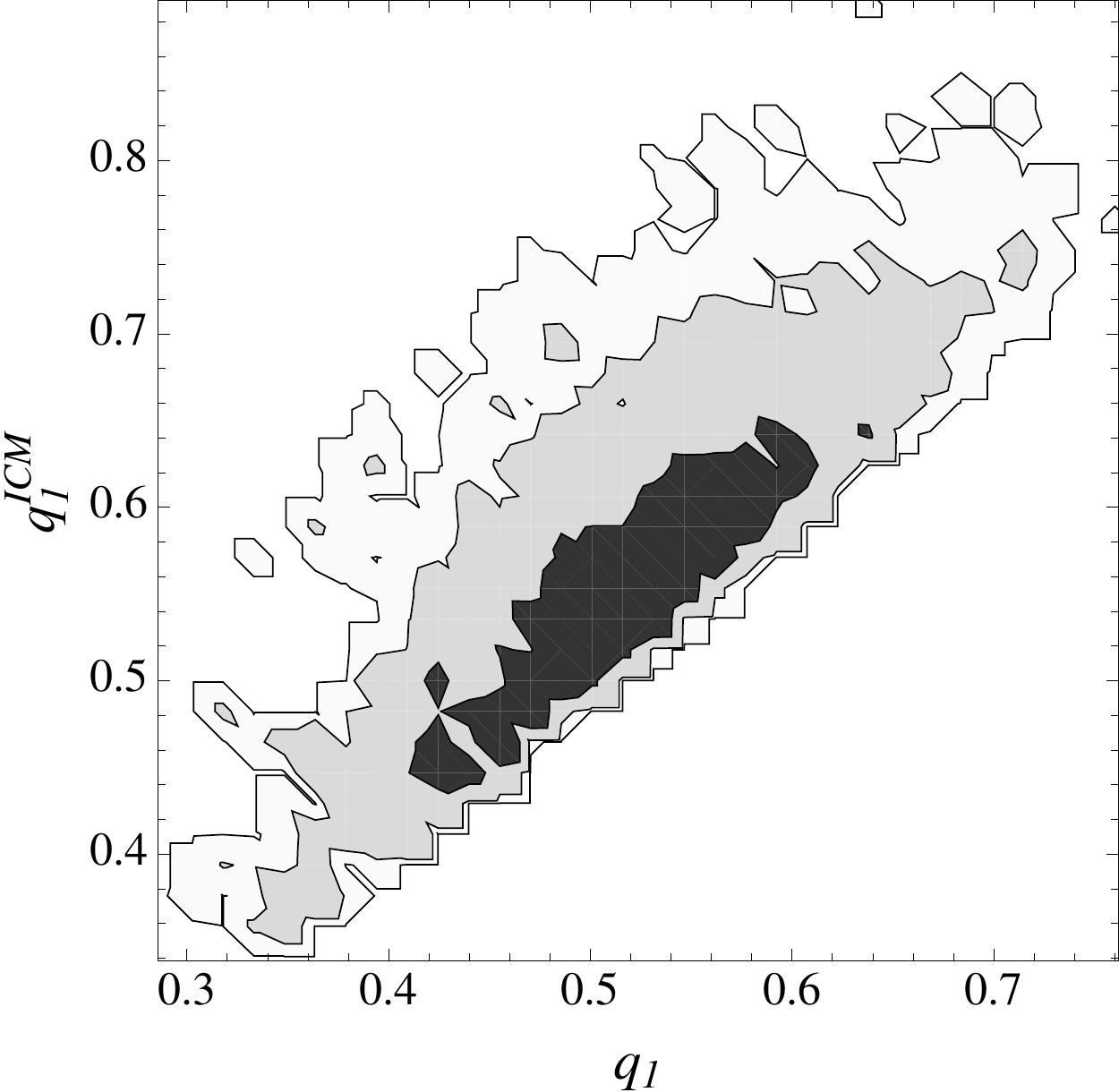} &
\includegraphics[width=3.4cm]{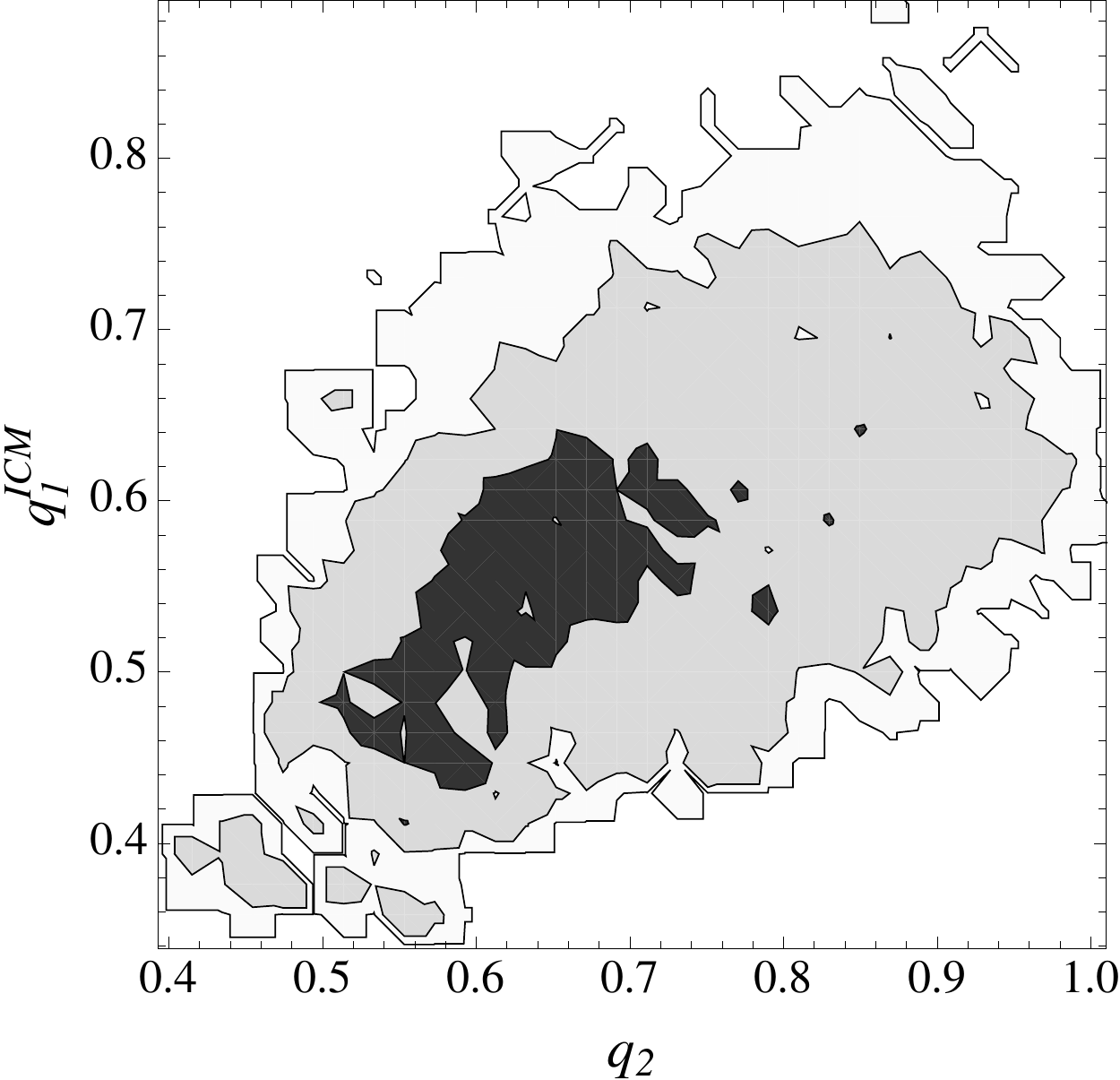} &
\includegraphics[width=3.4cm]{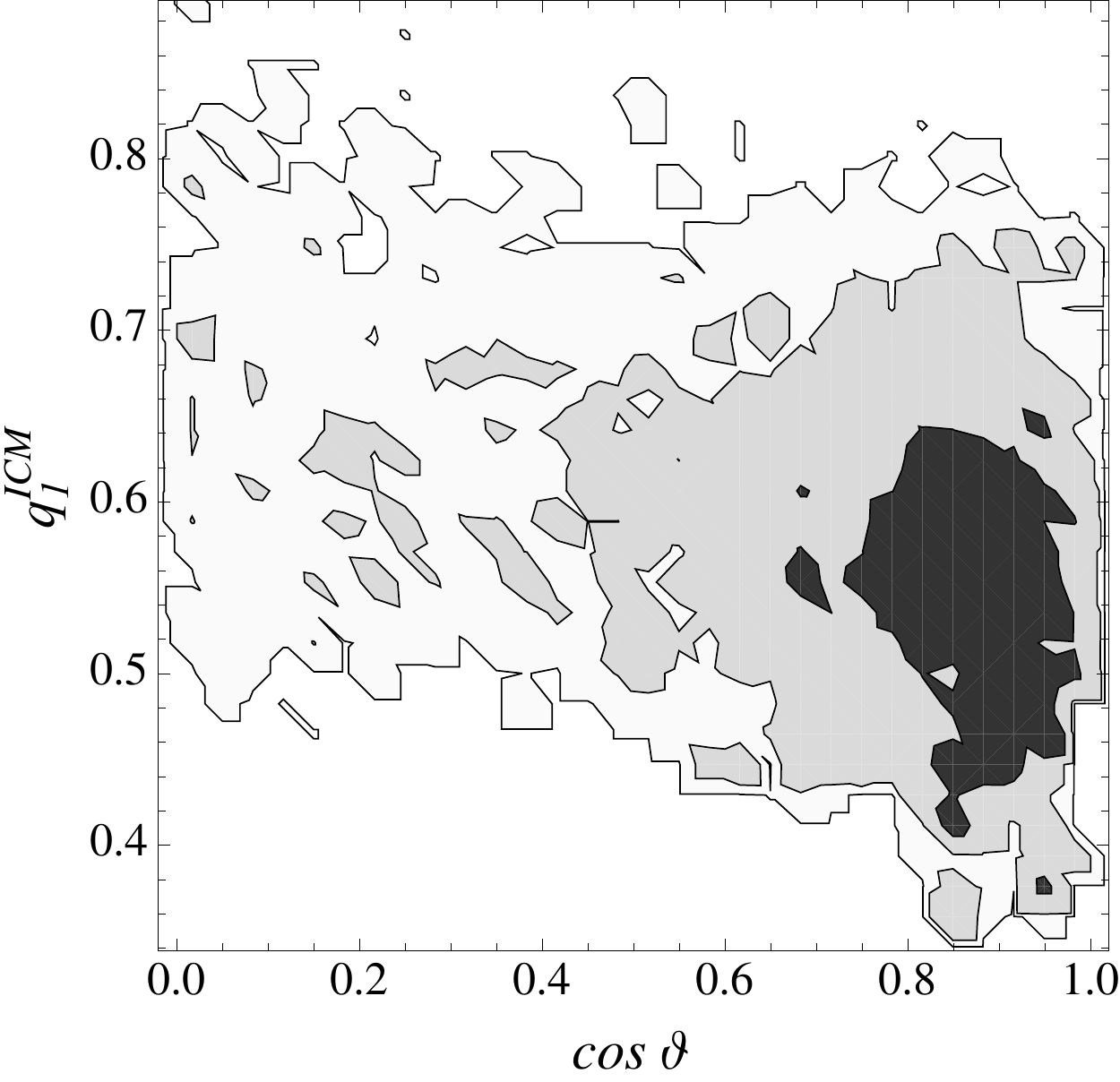} \\
\end{tabular}
$
\end{center}
\caption{Marginalised PDFs in the plane of $q^\mathrm{ICM}_1$ versus matter halo parameters. Panels from the left to the right are for $q^\mathrm{ICM}_1$-$M_{200}$, -$c_{200}$, -$q_1$, -$q_2$ and -$\cos \vartheta$, respectively. Contours are as in Fig.~\ref{fig_pdf_All_flat_random_2D}.  In the top row, we plot the inversion results under the prior hypotheses of flat $q$-distribution and random orientation angles. In the bottom row, prior assumptions are $N$-body like $q$-distribution and random orientation angles. Masses are in units of $10^{15}M_\odot$.}
\label{fig_pdf_All_2D_q1ICM}
\end{figure*}

The X-ray plus SZe part of the inversion method directly constrains the size of the gas distribution along the line of sight and in the plane of the sky. The axial ratios for the ICM, $q^\mathrm{ICM}_1$ and $q^\mathrm{ICM}_2$, are then determined with a better accuracy than their counterparts for the matter distribution, $q_1$ and $q_2$. Results are summarised in Table~\ref{tab_pdf_par_icm}. Posterior PDFs are plotted in Fig.~\ref{fig_pdf_All_ICM_1D}. Final results are nearly independent of the priors. The estimate of the elongation $e_\Delta^\mathrm{ICM}$ for the gas and the exquisite accuracy in the measured ellipticity of the X-ray surface, $\epsilon^\mathrm{X}$, drive the final results on orientation. Even if there is some interplay with GL, ICM shape and cluster orientation are mainly determined by the X-ray plus SZe likelihood. Results are then very similar to those in \citet{ser+al12}.\footnote{This analysis differs in the used priors from \citet{ser+al12}. In \citet{ser+al12}, a flat prior on $q^\mathrm{ICM}_1$ was used. Here, $q^\mathrm{ICM}_1\ge q_1$, so that a priori $p(q^\mathrm{ICM}_1)$ is not flat but peaks at $q^\mathrm{ICM}_1=1$. Being the analysis dominated by the likelihood, this brings about only a small difference in the final results.}  The gas is mildly triaxial, $q^\mathrm{ICM}_1\sim0.6$--$0.8$. A tail of the distribution extends to very low values, $q^\mathrm{ICM}_1 \ls 0.4$, which are associated with very well aligned configurations ($\cos \vartheta \ls 1$).

Some trends between gas shape and matter halo parameters can be seen in Fig.~\ref{fig_pdf_All_2D_q1ICM}. For $e^\mathrm{ICM}/e^\mathrm{Mat} \ls 1$, the total matter would be forced to follow the gas shape and orientations, which are well determined by the X-ray plus SZe part of the analysis. The matter would then be slightly rounder and very well elongated along the line of sight. As a consequence concentration and mass would be lower.

We made the theoretical assumption that the gas is rounder than the matter distribution and data further supports this view. The ellipticity of the ICM, $\epsilon^\mathrm{X}\sim0.15$, is lower than the projected total mass one, see Table~\ref{tab_fit_nfw_GL}, even if not by a large margin. From the three-dimensional analysis, we found $e^\mathrm{ICM}/e^\mathrm{Mat} \sim 0.95$, see Table~\ref{tab_pdf_par_icm}. Very low values of $e^\mathrm{ICM}/e^\mathrm{Mat}$ correspond to a nearly spherical gas distribution, which is excluded. On the other hand, very high values of $e^\mathrm{ICM}/e^\mathrm{Mat}$ would mean that the gas follows the matter distribution rather than the gravitational potential and are not excluded by data.

We found $q^\mathrm{ICM}_1-q_1 = 0.05\pm0.04$ ($= 0.04\pm0.04$), for a $N$-body like (flat) prior. The probability that the difference $q^\mathrm{ICM}_1-q_1$ is larger than 0.1 is $\sim$ 15 (6) per cent for a $N$-body like (flat) prior.

\subsection{Comparison with the spherical approach}

The deprojection method we employed was based on a minimum set of hypotheses. Since we did not assume hydrostatic equilibrium, the analyses of the ICM, based on X-ray and SZe observations, and that of the total mass, relying on lensing, are mostly independent. Their only tie is of geometrical nature, since we required the ICM to share the same orientation of the total mass, both in the space and in the plane of the sky. This can be seen in Eq.~(\ref{like_comb}), where the mass and concentration enter the likelihood only through the lensing part, whereas the X-ray and SZe contribution is only related to the ICM shape, and as a consequence, to the orientation of the matter halo. Since the total lensing strength depends on the shape and orientation, the X-ray and SZe can then play a role in the overall properties of the halo.

By assuming the spherical geometry, ellipticity and elongation of the total and gas mass distribution are fixed ($\epsilon=0$, $e_\Delta=1$). Therefore, the likelihood in Eq.~(\ref{like_icm}) is reduced just to the lensing part, with the ICM that can affect the mass reconstruction from lensing only through the constraints of the halo centroid, since the orientation is not anymore a parameter. Under these assumptions, $M_{200}=(1.26\pm0.12)\times 10^{15}M_\odot$ and $c_{200}=7.8\pm 0.2$. As expected, the central values are compatible with the full triaxial analysis but have associated a statistical error that is small because it does not include the systematic part due to the relaxation of the assumption on the geometrical shape.

Values of $c_{200}\sim 6$, which are fully compatible with the triaxial analysis in the likely case of nearly alignment with the line of sight, are excluded assuming a spherical shape. As discussed in Sec.~\ref{sec_resu}, the SL and WL analyses are marginally consistent assuming a triaxial form. The conflict between the results in the two lensing regimes is further aggravated in the spherical hypothesis.

\section{Hydrostatic equilibrium}
\label{sec_he}

A significant part of X-ray analyses relies on the assumption of hydrostatic equilibrium,
\beq
\label{eq_he_1}
\nabla P_\mathrm{Tot}=-\rho_\mathrm{ICM} \nabla \phi_\mathrm{Mat},
\eeq
where $P_\mathrm{Tot}(=P_\mathrm{Th}+P_\mathrm{nTh})$ is the total pressure, $\rho_\mathrm{ICM}$ the gas density and $\phi_\mathrm{Mat}$ the gravitational potential. When hydrostatic equilibrium holds, the pressure is only thermal, $P_\mathrm{Th}= k_\mathrm{B} T n_\mathrm{ICM}$ for an ideal gas, where $k_\mathrm{B}$ is the Boltzmann constant. Neglecting a non thermal contribution $P_\mathrm{nTh}$, such as bulk and/or turbulent motions, systematically biases low the X-ray mass determination of the cluster \citep{men+al10,ras+al12}.

Assessing the level of hydrostatic equilibrium in a cluster can be problematic and usually relies on either multiple data sets \citep{mor+al11b} or numerical simulations \citep{mol+al10}. Since we determined mass and shape of A1689 without relying on any assumption on the status of the cluster, we can check if and how A1689 departs from hydrostatic equilibrium. Firstly, we know that gas in equilibrium follows the gravitational potential of the halo. In that case, gas and matter eccentricities are related and $e^\mathrm{ICM}/e^\mathrm{Mat}\sim 0.7$ \citep{le+su03}. This prediction can be compared with our results, see Fig.~\ref{fig_pdf_All_ICM_1D} and Table~\ref{tab_pdf_par_icm}. We found that the gas distribution is more triaxial than the shapes expected under the assumption of complete hydrostatic equilibrium. The distribution of $e^\mathrm{ICM}/e^\mathrm{Mat}$ peaks at $\ls 1$, and significance at 0.7 is very low. The chance to have $e^\mathrm{ICM}/e^\mathrm{Mat} \le 0.8$ is of order of $\ls 1$ per cent. We can conclude from this first test that the shape of gas and matter are only marginally compatible with the hypothesis of hydrostatic equilibrium, the gas being decisively more triaxial than expected. Radiative processes can make the gas more triaxial in the central regions \citep{lau+al11}, which would explain the high degree of gas triaxiality found in A1689

\begin{figure}
\resizebox{\hsize}{!}{\includegraphics{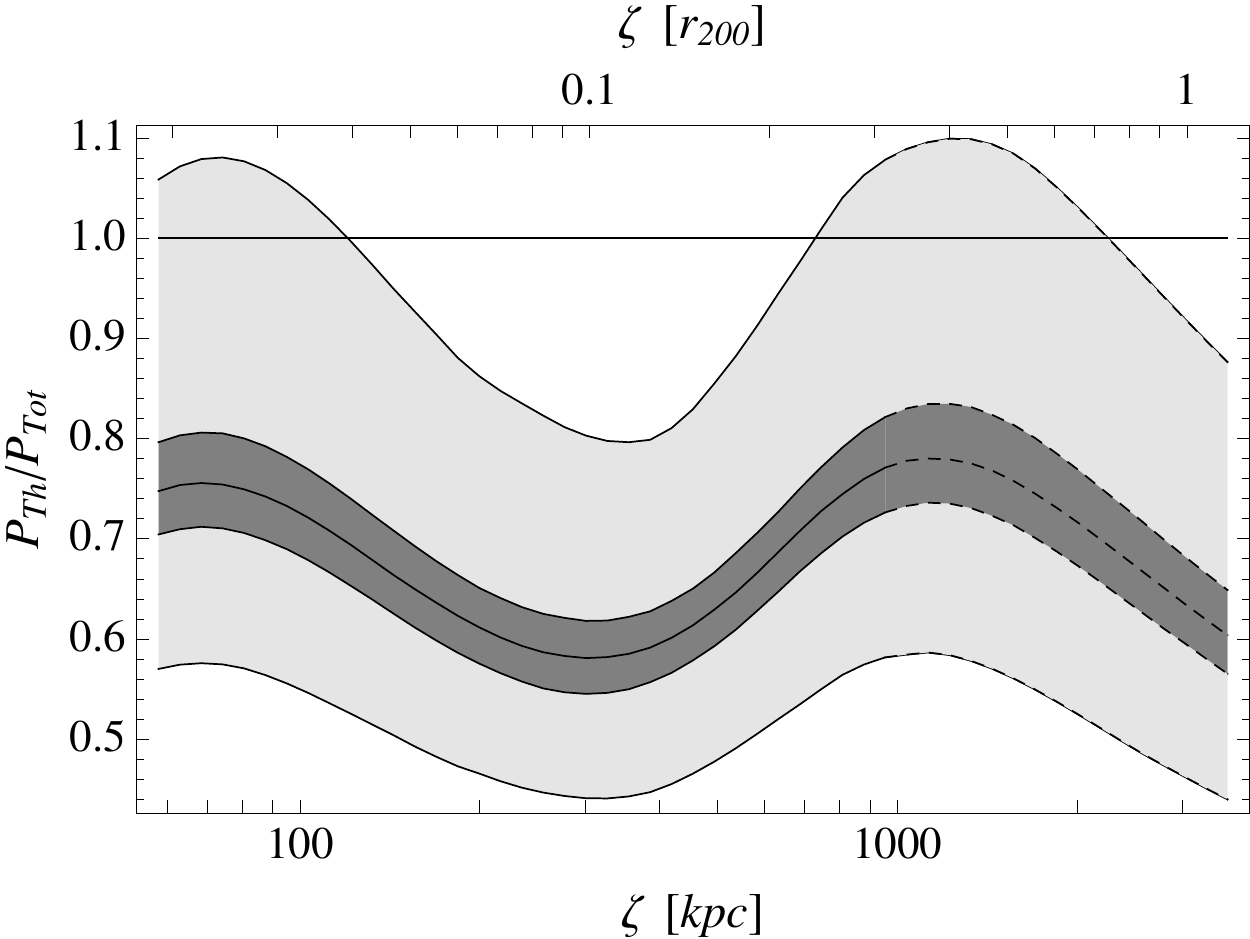}}
\caption{
Fraction of total contributed by the thermal pressure in A1689 as a function of the ellipsoidal radius $\zeta$ of the gravitational potential. The full thick line tracks the median; the dark (light) gray shadowed regions are limited by the 68.3 (99.7) per cent quantiles. The horizontal line at 1 is the value expected for an halo in hydrostatic equilibrium. Lines are dashed  in the radial range where results are extrapolated.
}
\label{fig_hydrostatic_equilibrium}
\end{figure}

As a second test, we checked if the hydrostatic equilibrium condition expressed in Eq.~(\ref{eq_he_1}) is fulfilled. To this aim, we recomputed the posteriori probabilities for the cluster parameters under the sharp prior of $e^\mathrm{ICM}/e^\mathrm{Mat}=0.7$. We exploited $N$-body like priors for the axial ratios. The gravitational potential of the ellipsoidal NFW halo was computed using the approximated formulae in \citet{le+su03}. The final result is summarized in Fig.~\ref{fig_hydrostatic_equilibrium}, which was obtained assuming that the gas follows the potential. Thermal pressure is systematically lower than what needed. Hydrostatic equilibrium is compatible with our results at the 3$\sigma$ confidence level. The non thermal contribution to the total pressure required for equilibrium, i.e., the value of $P_\mathrm{Tot}$ solving Eq.~(\ref{eq_he_1}), is of the order of 20--30 per cent in the outer regions at $\zeta \simeq 1.4~\mathrm{Mpc}$, and even higher towards the center ($\sim$ 20--50 per cent). 

High resolution cosmological simulations showed that there is a significant contribution from non-thermal pressure in the core region of relaxed clusters \citep{lau+al09,mol+al10}. \citet{mol+al10} found in ten simulated massive relaxed clusters that non-thermal pressure support from subsonic random gas motions can contribute up to 40 per cent in the inner regions and up to 20 per cent within one tenth of the virial radius. They also found that the non thermal contribution increases with radius in the very outer regions. \citet{lau+al09} found a similar level of non thermal pressure of the order of 5-15 per cent at about one tenth of the viral radius, also increasing with radius in the outer regions. These trends are retrieved in Fig.~\ref{fig_hydrostatic_equilibrium}, even if we have to caution that {\it XMM} spectroscopic data covers the cluster up to $\ls 900~\mathrm{kpc}$ so that the decrement at large radii is an extrapolation of fitted profiles.


\section{Comparison with previous analyses}
\label{sec_comp}

Our multi-probe analysis of A1689, combining a data-set spanning from X-ray to lensing to SZe data, can be compared to some recent works. \citet{ett+al10} performed an X-ray analysis and recovered the profiles of gas and dark mass in A1689 under the assumption that hydrostatic equilibrium holds between the intracluster medium and the gravitational potential. Even if \citet{ett+al10} used different techniques from ours, they still assumed a NFW functional form for the dark matter distribution and based their analysis on {\it XMM} data. Their results are then comparable to ours and any discrepancy should be interpreted in terms of either triaxiality or non-thermal pressure. They found $c_{200}= 8.31^{+0.64}_{-0.63}$,  and $M_{200} = (0.74\pm0.04)\times10^{15}M_{\odot}$ for the dark matter component and $c_{200}= 7.90^{+0.60}_{-0.53}$ and $M_{200} = (0.91\pm0.05)\times10^{15}M_{\odot}$ for the total matter distribution. The estimated concentration is in excellent agreement with our findings whereas $M_{200}$ is lower by $\sim$30 per cent. This is exactly how much we expect the X-ray mass to be biased low due to the level of non-thermal pressure. Being A1689 nearly aligned with the line of sight, the spherical symmetry hypothesis brings about a bias high of $\sim$5--10 per cent \citep{gav05}, which is secondary with respect to the systematic error connected to the non-thermal pressure.

Previous results from gravitational lensing found a concentration of $c_{200}\sim$ 5--7 for strong lensing analyses or $\sim$ 8--12 for weak lensing analyses. A collection from literature can be found in \citet[ table~4]{cor+al09} and  \citet[ table~2]{coe+al10}. Our results are in line with what already understood. In general, triaxiality and orientation issues reduces the gap between SL and WL analyses by allowing for smaller concentrations and larger dispersions \citep{ogu+al05,cor+al09,se+um11}. We retrieved the same trends already discussed in \citet{se+um11}, whose analysis is here potentiated in two main respects. First, we performed a non-parametric analysis of the SL core regions, which enables us to fully explore the parameter space and to be less sensitive to local minima. Secondly, the additional information from X-ray plus SZe data reduces sensibly the parameter uncertainties and makes results much less sensitive to priors.

\citet{mor+al11} found $c_{200}=5.6\pm0.4$ and $q_1\sim0.5$ from a combined lensing plus X-ray analysis. The level of triaxiality is similar to what found here whereas the concentration is quite smaller. This can be due to their hypothesis of halo perfectly aligned with the line of sight, which bias low the concentration.

The level of hydrostatic equilibrium has been addressed in several works. \citet{mor+al11} found non-thermal pressure of order of $\sim 20$ per cent. They assumed the fraction to be constant with radius and fixed the gas shape to what expected in hydrostatic equilibrium. \citet{lem+al08} derived a temperature profile consistent with hydrostatic equilibrium from combined {\it Chandra} X-ray brightness measurements and joint strong/weak lensing measurements. They found that the resulting equilibrium temperature exceeds the observed temperature by 30 per cent at all radii. The existence of significant cool and dense gas components was proposed as the source of this temperature discrepancy \citep{kaw+al07,lem+al08}.  Assuming spherical symmetry, \citet{pen+al09} derived an hydrostatic mass within the central 200\,kpc$\,h^{-1}$ region of A1689 from {\it Chandra} observations, which is about 30--50 per cent lower than lensing-based mass estimates. They also showed that dense and cool gas clumps alone can not cause such a strong bias in the X-ray temperature determinations and proposed an orientation bias to reconcile mass determinations. \citet{kaw+al10} detected anisotropic temperature and entropy distributions in {\it Suzaku} images of the cluster outskirts correlated with large-scale structure of galaxies, with regions of low gas temperature and entropy deviating from hydrostatic equilibrium. They estimated that the thermal gas pressure within half the virial radius is at most about half of the equilibrium pressure required to balance the gravity predicted by gravitational lensing under the hypothesis of spherical symmetry. 

Our derived ratio of thermal to total pressure as a function of radius agree with \citet{mol+al10}, who assumed spherical symmetry and used {\it Chandra} data for the temperature profile, which are larger by 10--20 per cent than the {\it XMM} measurements considered here and consequently may overestimate the thermal contribution. They found $P_\mathrm{th}/P_\mathrm{Tot}\simeq 0.6$ within the core region.

\section{Conclusions}
\label{sec_conc}

The multi-probe approach to galaxy clusters can tackle in an efficient way one of the classical problems in astronomy, the determination of the mass. An unbiased estimate requires at the same time the knowledge of the shape and the orientation of the halo, of the concentration and of the equilibrium status of the gas. This complete picture can be achieved with an analysis exploiting gravitational lensing, which describes the total mass, and X-ray and SZe observations, which directly constrain the gas. 

We proposed a novel method based on minimal geometrical principles. We only assumed the gas and the total matter distributions to be approximately ellipsoidal and co-aligned. The matter and the ICM were separately modelled with parametric profiles, suitable with comparison with numerical simulations and theoretical predictions. We did not assume any hypothesis on the equilibrium status of the gas or the profile of non-thermal pressure. In this regard, our method is alternative to other recently proposed multi-probe approaches \citep{mor+al11,mor+al11b}.

We obtained pixellated maps of the projected mass density covering a large radial range by combining strong lensing in the inner core and weak lensing in the outer regions. The maps were then fitted with a NFW profile. Photometric and spectroscopic X-ray data, as well as the SZ temperature decrement, were described with a single parameterization for the gas density and temperature profile. The combined X-ray plus SZe analysis enabled us to determine the elongation of the gas along the line of sight, the only geometrical quantity not defined in the plane of the sky that can be directly derived from the combined analysis of projected maps \citep{ser07} and crucial information to constrain gas shape and orientation. All pieces of information were finally combined in a single Bayesian statistical analysis to infer the intrinsic parameters of the halo.

The method was applied to A1689. We proposed the first multi-probe analysis of A1689 combining a data-set spanning from X-ray to lensing to SZe data. We could obtain an unbiased picture of the cluster. A1689 is massive, $M_{200} = (0.9 \pm 0.1)\times 10^{15}M_\odot/h$, and slightly over-concentrated $c_{200} =8\pm1$. The halo is triaxial ($q_1 \sim 0.5 \pm 0.1$) and aligned with the line of sight. The high degree of triaxiality  ($q^\mathrm{ICM}_1 \sim$0.6--0.8) of the gas distribution shows a deviation from hydrostatic equilibrium, which would prefer rounder gas shapes. A significant contribution of non thermal pressure is required for equilibrium, $\sim$20--50 per cent in the center and $\sim$ 20--30 per cent in the outer regions. This level of non-thermal pressure support is consistent with what found by \citet{mol+al10} using a sample of massive relaxed clusters drawn from high resolution cosmological simulations and with recent findings in MS2137.3-2353 by \citet{chi+al12}, who found a 40--50 per cent contribution in the core assuming a spherical model.

The measurement of non-thermal pressure requires an unbiased knowledge of the cluster shape. \citet{chi+al12} found that the effect of the alignment of the major axis with the line of sight is to decrease the non-thermal pressure support required for equilibrium at all radii without changing the distribution qualitatively. Different counterbalancing factors can play a role in the determination of the cluster mass with X-ray methods, $M_\mathrm{X}$. Under the hypothesis of hydrostatic equilibrium, neglecting non-thermal processes, $M_\mathrm{X}$ is usually biased low by $\sim$20-30 per cent \citep{ras+al12}. Non thermal pressure would then make the lensing signal greater than expected given the X-ray derived mass.

On the other hand, if a cluster aligned with the line of sight is considered spherical, $M_\mathrm{X}$ is biased high by $\ls 5-10$ per cent \citep{gav05}. The effect of the elongation on lensing is even more influential, since the central projected mass density of the lens is directly proportional to the extension of the halo along the line of sight.

In this regard, our method seems particularly promising. Shape and inclination are measured and the orientation bias is overcome. The condition for hydrostatic equilibrium is not used to derive the mass and can be employed to determine the non thermal contribution to the pressure. Degeneracies are broken thanks to the joint multi-wavelength data sets giving a reliable picture of the cluster status and properties.

The assessment of the over-concentration depend on the real presence of an upturn in the mass-concentration relation for high mass and redshift clusters \citep{pra+al11}. Massive systems are likely identified when they are substantially out of equilibrium and in a transient stage of high concentration \citep{lud+al12}. The upturn should disappear when only dynamically-relaxed systems are considered. Our results seem to support this picture. A1689 is massive and still not settled in hydrostatic equilibrium which propounds an high value of concentration as observed.

\section*{Acknowledgements}
The authors thank M. Limousin for providing some results of the strong lensing analysis in \citet{lim+al07}.


\setlength{\bibhang}{2.0em}

\end{document}